\documentclass[aip,preprint,groupedaddress]{revtex4-1}

\usepackage{amsfonts}
\usepackage{amsmath}
\usepackage{graphicx}
\usepackage{mathrsfs}

\usepackage{caption}
\usepackage{subfig}
\usepackage{comment}
\usepackage{natbib}
\usepackage[title]{appendix}

\newcommand{\be}{\begin{equation}}
\newcommand{\ee}{\end{equation}}
\newcommand{\bea}{\begin{eqnarray}}
\newcommand{\eea}{\end{eqnarray}}
\newcommand{\bes}{\begin{eqnarray*}}
\newcommand{\ees}{\end{eqnarray*}}

\newcommand*{\shifttext}[2]{%
  \settowidth{\@tempdima}{#2}%
  \makebox[\@tempdima]{\hspace*{#1}#2}%
}
\makeatother

\begin{document}
\title{Normal Modes, Rotational Inertia, and Thermal Fluctuations of Trapped Ion Crystals}
\author{Daniel H.E. Dubin} 

\affiliation{Dept of Physics UCSD, La Jolla CA 92093}

\date{\today}
\begin{abstract}
The  normal modes of a trapped ion crystal are derived using an 
approach based on the Hermitian properties of the system's dynamical matrix. 
This method is equivalent to the standard Bogoliubov method, but for classical 
systems it is arguably simpler and more general in that canonical coordinates 
are not necessary. The  theory is developed for stable, 
unstable, and neutrally-stable systems. The method is then applied to ion crystals in a Penning trap.  
Reduced eigenvalue problems for the case of large applied magnetic field are developed, for which 
the spectrum breaks into ExB drift modes, axial modes, and cyclotron modes. 
Thermal fluctuation levels in these modes are analyzed and shown to be 
consistent with the Bohr-van-Leeuwen theorem, provided that neutrally-stable modes associated with crystal rotations are included in the analysis. An expression for the rotational inertia of the
crystal is derived, and a magnetic contribution to this inertia, which dominates in 
large magnetic fields,  is described. An unusual limit is discovered for the special case 
of spherically-symmetric confinement, in which the rotational inertia does not 
exist and changes in angular momentum leave the rotation frequency unaffected.

\end{abstract}

\maketitle
\newpage

\section{Introduction}

This paper  examines  the normal modes of an ion crystal confined in the static electric and magnetic fields of a Penning trap. Such ion crystals, consisting of anywhere from a few  to thousands of ions, are employed in a variety of applications ranging from fundamental studies of quantum entanglement, quantum simulation and frequency standards\cite{bollinger, bollinger2, Bohnet2016, Gilmore2017, Sawyer2014,Sawyer2020}, to studies of the properties of strongly-coupled plasmas\cite{gilbert, 3Dcrystal, mitchell, dubinschiffer}.   Linear normal modes of oscillation  in these crystals are used as a diagnostic and manipulation tool in some of these studies, and a detailed understanding of the modes is essential to the success of this work.  

	In this paper we lay out a general theory for the normal modes.  In previous work, normal modes of a periodic  correlated Coulomb lattice in a uniform magnetic field were found using Fourier methods,\cite{fukuyama,nagai, bonsal,kalman,bonitz} taking advantage of the periodicity of the system, and in  Refs.~\onlinecite{usov,  chen, baiko} by using a Bogoliubov transformation.\cite{bogoliubov,holstein,fetter}
In  Ref.~\onlinecite{freericks} the normal modes for a  nonuniform 2D planar ion crystal in Penning trap geometry were found, using a method that employed a preliminary diagonalization of the potential matrix and that required the solution of a quadratic eigenvalue problem, but that can nevertheless be connected to a non-canonical version of the approach employed in this paper (see Appendix  B).  Here, we consider the general case of a nonuniform 3-dimensional magnetized ion crystal, from which the previous 2D results can be obtained as a limit. Some of the modes in 3D nonuniform crystals have been previously described\cite{dubinmodesim, dubinschiffer}, but a general theory for all of the modes has not been previously published to our knowledge. 

	In the first half the the paper (Section II), we develop a theory for the normal modes of a general linearized classical Hamiltonian system. The theory differs from the standard Bogoliubov approach mentioned above, focusing on the Hermitian properties of linearized Hamiltonians.  There are some advantages to our approach: canonical coordinates are not required, and extra transformations to/from the creation/annihilation representation of the dynamics (necessary in the Bogoliubov method) are avoided. We include an analysis of neutrally-stable modes, since such modes often occur in trapped ion crystals, associated with rotations in symmetric external trap fields. This requires a discussion of the differing forms of the diagonalized Hamiltonian when constants of the motion are in involution (i.e. their Poisson bracket vanishes), or are not in involution.  For completeness we also consider the modes of an unstable Hamiltonian system.   We find that canonical coordinates remain mixed in pairs of exponentially growing and decaying mode amplitudes in such a way that the system energy remains conserved as mode amplitudes grow and decay.
	
	 In the papers second half (Section III) we apply this theory to determine the modes of an ion crystal, and consider two examples in detail, a Coulomb cluster consisting of two charges, and  a 3D  Coulomb crystal with $N\gg 1$. For the former system, all of the normal modes can be evaluated analytically. In the latter case, the modes are evaluated numerically. Averages over thermal fluctuations are discussed. The fluctuation energy associated with vibrations as well as rotations is analyzed. Expressions for thermal fluctuation amplitudes are shown to be consistent with the Bohr-van-Leewen theorem\cite{bohr} provided that any contributions from zero-frequency rotational modes are included in the averages.  

	In relation to the rotational modes, expressions for rotational inertia of an ion crystal are developed, including a novel magnetic addition to the rotational inertia. This magnetic addition, arising from the vector-potential portion of the angular momentum, is a dominant contibution in many experiments that employ large magnetic fields. A surprising magnetic effect is discussed for the case of a spherically-symmetric trap potential, in which the rotational inertia ceases to exist.  Under these conditions, variations in angular momentum leave the rotation frequency unaffected but instead change the crystals orientation with respect to the magnetic field. This phenomenon is connected to the occurrence of constants of the motion (components of the angular momentum) that are not in involution.
 
	In the large magnetic field limit of interest in many of the experiments, reduced eigenvalue problems are developed that separately describe cyclotron, axial, and $E\times B$ drift eigenmodes. As far as we know, the reduced eigenvalue problems for cyclotron  and $E\times B$ modes have not be written down previously for a general nonuniform crystal structure, although these limits have been considered for periodic lattices\cite{usov, chen}.  We also describe coupling between axial and $E\times B$ modes that occurs in 3-dimensional crystals. These reduced eigenvalue problems provide some intuition as to the form of the eigenmodes and eigenfrequencies in large magnetic fields.

\section{ Normal modes of a linearized Hamiltonian system}

Determination of the normal modes of a system of coupled linear oscillators is a venerable problem in mathematical  physics, with applications in a variety of scientific fields. The standard solution is a textbook problem in classical mechanics, in which  normal modes are found by solving for the eigenvectors and eigenvalues of combined kinetic and  potential energy matrices that arise in a  linearized Lagrangian of the form $L = T-V$, where $T$  and $V$ are kinetic and potential energies respectively. \cite{goldstein}  The coordinate transformation relating particle displacements to a sum over the eigenmode amplitudes (the ``normal coordinates" for the oscillators) is a point transformation, and as such can be easily handled in the context of Lagrangian mechanics. However, for  more general  Lagrangians in which coordinates and velocities are mixed (for instance, through velocity-dependent potentials that arise in the application of a magnetic field),  point transformations are no longer adequate. Problems of this nature arise, for example, in the modes of an ion crystal in a Penning trap, the study of vibrational modes of molecules in applied magnetic fields, and in the normal modes of crystals of point vortices in 2D Euler flow.\cite{fine,liquidHe} A transformation mixing both positions and momenta is now required  in order to diagonalize the Hamiltonian.  

A general approach to the solution of this problem is the Bogoliubov transformation,\cite{bogoliubov,holstein,fetter} a linear transformation  in which the symplectic condition\cite{goldstein2} for canonical transformations is  imposed, and  conditions required to diagonalize the Hamiltonian are  then determined. The standard Bogoliubov transformation was developed with quantum problems in mind, and formulated in terms of creation and annihilation operators which are related to position and momentum by a linear transformation. This approach is reviewed in Appendix D.

In this paper we use a  method that is formulated specifically for the classical problem. To our surprise,  we have not come across a discussion of this method in any previous publication, so we lay out the theory in some detail.  (However, the literature on Hamiltonian methods is vast and it is possible, even likely, that this approach is not novel).  This method focuses on the Hermitian properties of  the matrix operators that appear in linearized Hamiltonian systems, rather than on the symplectic condition. Normal modes are obtained as  eigenvectors and eigenfrequencies of the dynamical matrix $\bf D$ that determines the linear equations of motion. We derive a  Hermitian property of this matrix (with respect to an inner product involving the system Hamiltonian) that is then used to show that  the eigenvectors form a complete orthogonal set under certain conditions involving the systems stability with respect to small perturbations. Diagonalization of the system energy is easily accomplished using these eigenvectors, without imposing the symplectic condition on the linear transformation. 

The added requirement that the transformation to normal mode coordinates be canonical is then satisfied through a specific normalization condition on the eigenvectors. Once this condition is imposed, the  transformation  is equivalent to the Bogoliubov transform. The equivalence to the Bogoliubov method is discussed in detail in Appendix D of the paper. 

Although our diagonalization method is equivalent to the Bogoliubov method, for classical systems it is arguably more straightforward in both its derivation and its application, as extra transformations from the creation/annihilation representation to the position/momentum representation are not required.  Furthermore, it is possible to apply our method to linearized non-canonical Hamiltonian systems, since our method does not require canonical coordinates. We describe an example of this approach to diagonalizing a non-canonical system in Appendix B: the ion crystal in a position-velocity space representation. We  have also used a continuum version of this method in calculations involving the normal modes of a non-canonical conservative system of fluid equations.\cite{dubinfluid1,dubinfluid2}

A linearized Hamiltonian system is  a  dynamical system with $N$ coordinates ${\bf q} = (q_1,...,q_N)$ and associated canonical  momenta ${\bf p} = (p_1,...,p_N)$ whose Hamiltonian has the quadratic form
\begin{equation}\label{a1}
H({\bf z},t)=\frac{1}{2}{\bf z}\cdot{\bf H}\cdot{\bf z} +{\bf f}(t)\cdot {\bf z},
\end{equation} 
where ${\bf z} = ({\bf q},{\bf p})$ is the system phase-space configuration vector,  ${\bf f}(t)$ is a  ``forcing" vector, and  $\bf H$ is the Hamiltonian matrix, a matrix of coefficients independent of $\bf z$ and which for our purposes is also assumed to be time-independent. (There is also considerable interest in the time-dependent problem, particularly in the area of  linear control theory and in  generalizations of parametric resonance to multidimensional systems.\cite{arnold,krein,johnson})  Suitable choices of the off-diagonal coefficients in $\bf H$ allow the  matrix to be of symmetric form, satisfying $H_{i j} = H_{j i}$. 

The linear equations of motion that arise from this Hamiltonian are, in vector form, given in terms of the Poisson bracket $[\cdot, \cdot ]$ as 
\begin{align}
\dot{\bf z} &= [ {\bf z}, H]  \notag \\
     &=[ {\bf z},{\bf z} ]\cdot\frac{\partial H}{\partial{\bf z}}  \notag \\
      &={\bf J}\cdot {\bf H}\cdot{\bf z} + {\bf J}\cdot{\bf f}\notag \\
     & = {\bf D}\cdot {\bf z} + {\bf J}\cdot{\bf f} \label{a2},
      \end{align}
where we introduce the dynamical matrix ${\bf D} = {\bf J}\cdot\bf H$, as well as the  fundamental symplectic matrix\cite{goldstein2} $\bf J \equiv [{\bf z},{\bf z}]$. The fundamental symplectic matrix is  given, in block form, by 
\begin{equation}\label{a2.5}
{\bf J} = \left(
\begin{array}{r r}
{\bf 0}, &{\bf 1} \\
{-\bf 1}, & {\bf 0}
\end{array}
\right),
\end{equation}
and $\bf 1$ and $\bf 0$ are the unit and zero tensors respectively.  The matrix is antisymmetric, and expresses the basic Poisson bracket relations $[q_i, q_j]=[p_i,p_j]=0$, and $[q_i,p_j]=-[p_j, q_i]=\delta_{i j}$ where $\delta_{i j}$ is the Kronecker delta. 

We will consider the normal modes of this linearized Hamiltonian system. The normal modes are unforced (i.e. ${\bf f}=\bf 0$) solutions of Eq.~\eqref{a2}  that are assumed to have a time dependence of the form 
\begin{equation}\label{a3}
{\bf z}(t)= \exp(-i\omega t){\bf u}_{\omega},
\end{equation}
 for some (possibly complex) frequency $\omega$ and  some (time-independent, possibly complex) vector ${\bf u}_\omega$. Substituting Eq.~\eqref{a3} into Eq.~\eqref{a2} and assuming ${\bf f}=\bf 0$ then yields an eigenvalue problem for $\omega$ and ${\bf u}_\omega$,
 \begin{equation}\label{a5}
 -i\omega {\bf u}_\omega = {\bf D}\cdot{\bf u}_\omega.
 \end{equation}

Thanks to the Hamiltonian nature of the linear dynamical equations,  the eigenfrequencies $\omega$ and eigenvectors ${\bf u}_\omega$ of Eq.~\eqref{a5} have the following properties:
\begin{enumerate}
\item The eigenvectors ${\bf u}_\omega$ form an orthogonal set with respect to a generalized inner product  defined for any complex vectors $\bf a$ and $\bf b$ as $({\bf a},{\bf b})\equiv{\bf a}^*\cdot{\bf H}\cdot{\bf b}$:
\end{enumerate}
\begin{center}
$({\bf u}_\omega,{\bf u}_{\bar\omega})= 0 $ provided that $\omega\ne\bar\omega^*$. 
\end{center}
\begin{enumerate}
\setcounter{enumi}{1}
\item A given eigenvalue $\omega$ is real provided that the corresponding eigenvector satisfies $({\bf u}_\omega,{\bf u}_\omega)\ne 0$.

\item For each eigenmode $(\omega,{\bf u}_\omega)$ for which $\omega\ne 0$, there is a second eigenmode $(-\omega^*, {\bf u}_{-\omega^*})$ for which ${\bf u}_{-\omega^*}={\bf u}^*_\omega$. Thus, for real $\omega$ the $\omega\ne 0$ eigenmodes
come in  $\pm\omega$ pairs. 
\end{enumerate}

It is straightforward to prove these properties.  
Property 3 arises from the fact that the dynamical matrix $\bf D$ has real coefficients. Taking the complex conjugate of Eq.~\eqref{a5} then yields 
\begin{equation}-i (-\omega^*){\bf u}^*_\omega = {\bf D}\cdot{\bf u}^*_\omega,
\end{equation}
showing that ${\bf u}_\omega^*$ is also an eigenvector of $\bf D$ with eigenfrequency $-\omega^*$, which completes the proof of property 3.

Properties 1 and 2 follow from the fact that $i \bf D$ is a Hermitian (self-adjoint) matrix with respect to the above-defined inner product, which we prove below. It is well-known that the eigenvalues and eigenvectors of a Hermitian matrix satisfy properties 1 and 2. These properties of Hermitian matrices  are often referred to as the spectral theorem, and a proof may be found in many  linear algebra textbooks. \cite{hermitian} 

 The Hermitian property of the matrix $i\bf D$ is defined by the relation $({\bf a}, i {\bf  D}\cdot {\bf b}) = ({\bf b}, i{\bf D}\cdot {\bf a})^*$, which must be satisfied for all vectors $\bf a$ and $\bf b$. Dividing out the factor of $i$ and using the definition of the inner product, this Hermitian relation  can be expressed as
\begin{equation}\label{a7}
{\bf a}^*\cdot{\bf H}\cdot{\bf D}\cdot{\bf b}=-{\bf b}\cdot{\bf  H}\cdot{\bf D}\cdot{\bf a}^*.
\end{equation}
(A matrix $\bf D$ that satisfies this equation is sometimes referred to as ``anti-Hermitian" with respect to $\bf H$,  due to the negative sign in the equation.) Consider the matrix ${\bf L}\equiv {\bf H}\cdot{\bf D}={\bf H}\cdot{\bf J}\cdot{\bf H}$ that appears in the above expression. This matrix is antisymmetric: $L_{j i}=-L_{i j}$. This follows from the symmetry and antisymmetry respectively of the matrices $\bf H$ and $\bf J$: 

\begin{equation}\label{a8}
L_{j i} = H_{j k} J_{k l} H_{l i}= H_{k j} (-J_{l k}) H_{i l} = - H_{i l} J_{l k} H_{k j} = -L_{i j}.
\end{equation}

The antisymmetry of $\bf L$ proves Eq.~\eqref{a7}, which in turn proves that $i\bf D$ is Hermitian.

\subsection{Stable System}
We can use properties 1-3  in order to analyze the evolution of the solution ${\bf z}(t)$ to the the dynamical equations. First consider the simplest case, of a {\it stable} system for which  $({\bf u}_\omega,{\bf u}_\omega)\ne 0$ for all modes, so that all eigenfrequencies $\omega$ are real (property 2). Assume also (for simplicity) that there are no $\omega = 0$ modes, and that all mode frequencies are different. Cases which have one or more neutrally stable ($\omega=0$) modes introduce certain technical issues that are addressed in Sec. IIb, and  examples of such neutral modes will be considered  in Sec. III. Degeneracies ($\bar\omega=\omega$ for two or more separate eigenvectors) can be handled easily by orthogonalizing degenerate eigenvectors within the subspace created by these vectors; see, for example, Ref.~\onlinecite{goldstein}.

Under these assumptions, the set of $2N$ eigenvectors ${\bf u}_\omega$ then form an orthogonal set in the $2N$ dimensional vector space for phase-space vector $\bf z$, spanning the vector space and thus forming a complete set. We can therefore construct a representation of the vector ${\bf z}(t)$ in terms of the eigenvectors:

\begin{equation}\label{a9}
{\bf z}(t) = \sum_\omega a_\omega(t) {\bf u}_\omega,
\end{equation}
where the complex amplitude $a_\omega(t)$ associated with each eigenvector ${\bf u}_\omega$ can be found by taking an inner product of both sides of Eq.~\eqref{a9}, applying  property 1 (orthogonality)
 of the eigenvectors:
 \begin{equation}\label{a10}
a_\omega(t) =\frac{ ({\bf u}_\omega, {\bf z}(t))}{({\bf u}_\omega,{\bf u}_\omega)}.
\end{equation}
Since eigenmodes come in pairs (property 3), Eq.~\eqref{a9} can also be written as
\begin{equation}\label{a11}
{\bf z}(t) = \sum_{\omega>0} (a_\omega(t) {\bf u}_\omega + a_{-\omega}(t){\bf u}^*_\omega).
\end{equation}
The real nature of the vector $\bf z$ then implies that $a_{-\omega}(t) = a_\omega^*(t)$ and we can then write Eq.~\eqref{a11} as
\begin{equation}\label{a12}
{\bf z}(t) = \sum_{\omega>0} a_\omega(t) {\bf u}_\omega + c.c.,
\end{equation}
where $c.c.$ stands for complex conjugate.

A differential equation for the time evolution of the complex mode amplitude $a_\omega(t)$ follows by substitution of Eq.~\eqref{a9} into Eq.~\eqref{a2}. Taking an inner product with respect to one of the eigenvectors ${\bf u}_\omega$ then yields
\begin{equation}\label{a13}
\dot a_\omega(t) = -i\omega a_\omega(t) + f_\omega(t),
\end{equation}
where $f_\omega(t) = ( {\bf u}_\omega, {\bf J}\cdot{\bf f}(t))/({\bf u}_\omega,{\bf u}_\omega)$. This equation has the solution
\begin{equation}\label{a14}
a_\omega(t) = A_\omega e^{-i\omega t} + \int_0^t e^{-i\omega (t-t')}f_\omega(t')dt',
\end{equation}
where the coefficient $A_\omega$ is determined by the initial conditions via Eq.~\eqref{a10}.  The forcing coefficient $f_\omega$ can also be written as
\begin{equation}\label{a14.5}
f_\omega(t) =  \frac{{\bf u}_\omega^*\cdot{\bf H}\cdot{\bf J}\cdot{\bf f}(t))}{({\bf u}_\omega,{\bf u}_\omega)} = \frac{-i\omega {\bf u}_\omega^*\cdot{\bf f}(t)}{({\bf u}_\omega,{\bf u}_\omega)},
\end{equation}
where we used ${\bf D} = {\bf J}\cdot{\bf H}$ and the complex conjugate of Eq.~\eqref{a5},  along with the symmetry and antisymmetry respectively of ${\bf H}$ and $\bf J$.

The system energy  can be written in terms of the mode amplitudes $a_\omega(t)$ by substituting Eq.~\eqref{a9} into Eq.~\eqref{a1}
\begin{align} \label{a15}
H &= \frac{1}{2} \sum_\omega\sum_{\bar\omega} a_\omega(t) a_{\bar\omega}(t) {\bf u}_{\bar\omega}\cdot{\bf H}\cdot{\bf u}_\omega + \sum_\omega a_\omega(t) {\bf f}\cdot{\bf u_\omega} \notag \\
   & = \frac{1}{2}  \sum_\omega\sum_{\bar\omega} a_\omega(t) a_{- \bar\omega}(t) {\bf u}^*_{\bar\omega}\cdot{\bf H}\cdot{\bf u}_\omega + \sum_\omega a_\omega(t) {\bf f}\cdot{\bf u_\omega} \notag \\
     & =  \frac{1}{2} \sum_\omega\ a_\omega(t) a^*_{\omega}(t) {\bf u}^*_{\omega}\cdot{\bf H}\cdot{\bf u}_\omega + \sum_\omega a_\omega(t) {\bf f}\cdot{\bf u_\omega} \notag \\
     & =   \sum_{\omega>0} H_\omega,
\end{align}
where in the second line we replaced the index $\bar\omega$ with $-\bar\omega$ using property 3,  in the third line we used property 1 (orthogonality) of the modes, and in the last line  we  identified the energy $H_\omega$ in each eigenmode, given by
\begin{equation}\label{a16}
H_\omega = |a_\omega(t)|^2 ({\bf u}_\omega,{\bf u}_\omega) +  2{\bf f}(t)\cdot Re(a_\omega \bf{u}_\omega),
\end{equation}
using property 3  to sum only over positive eigenfrequencies. 

Note also that in many, if not all, applications an unforced stable oscillator  has positive energy compared to the  equilibrium ${\bf z}=0$, in which case Eq.~\eqref{a16} implies that $({\bf u}_\omega,{\bf u}_\omega)>0$.  


Equations~\eqref{a9}, \eqref{a14}, \eqref{a15}, and \eqref{a16} provide a complete description of the dynamics of a stable linearized Hamiltonian system. We have already diagonalized the system energy in Eq.~\eqref{a15}, without consideration of canonical variables in the linear transformation from  $\bf z$ to mode amplitude variables $a_\omega$, and we have found solutions for the evolution of each mode amplitude. 

However, Eq.~\eqref{a16} can also be thought of as a Hamiltonian for mode $\omega$, provided that we introduce the proper canonical variables. These variables can be constructed using the following argument. Consider the Poisson bracket 
$[a_\omega, a_{\bar\omega}^*]$. This bracket can be evaluated using Eq.~\eqref{a10} and the symmetry of the Hamiltonian matrix:
\begin{align}
[a_\omega, a_{\bar\omega}^*] &= \frac{  {\bf u}_{\omega}^*\cdot {\bf H}\cdot [{\bf z}, {\bf z}]\cdot{\bf H}\cdot {\bf u}_{\bar\omega} } { ({\bf u}_\omega,{\bf u_\omega})({\bf u}_{\bar\omega},{\bf u}_{\bar\omega})  }\notag \\
                       &= \frac{  {\bf u}_{\omega}^*\cdot {\bf H}\cdot {\bf J}\cdot{\bf H}\cdot {\bf u}_{\bar\omega} } { ({\bf u}_\omega,{\bf u_\omega})({\bf u}_{\bar\omega},{\bf u}_{\bar\omega})  } \notag \\
                        &= \frac{  {\bf u}_{\omega}^*\cdot {\bf H}\cdot {\bf D}\cdot {\bf u}_{\bar\omega} } { ({\bf u}_\omega,{\bf u_\omega})({\bf u}_{\bar\omega},{\bf u}_{\bar\omega})  } \notag \\   
                        &= \frac{ -i\bar\omega {\bf u}_{\omega}^*\cdot {\bf H}\cdot{\bf u}_{\bar\omega} } { ({\bf u}_\omega,{\bf u_\omega})({\bf u}_{\bar\omega},{\bf u}_{\bar\omega})  } \label{16.5} \\   
                        &= \frac{-i\omega} { ({\bf u}_\omega,{\bf u_\omega})} \delta_{\omega \bar\omega}. \label{16.6}                
\end{align}
 where in the fourth line we used Eq.~\eqref{a5} and in the last line we used orthogonality of the eigenvectors.  Similarly, one can show that
 \begin{equation}\label{16.7}
 [a_\omega, a_{\bar\omega}] = 0
 \end{equation}  for $\omega$ and $\bar\omega$ greater than zero. In this case both eigenvectors in Eq.~\eqref{16.5} are starred. However, recall that ${\bf u}_{\bar\omega}^* = {\bf u}_{-\bar\omega}$, the eigenvector for a negative frequency mode.  This mode is orthogonal to all positive frequency modes by property 1, proving Eq.~\eqref{16.7}.
 
Now, to define canonical variables based on the complex amplitudes $a_\omega$ we find it useful to impose the condition on these amplitudes that, for $\omega>0$ and $\bar\omega>0$,
 \begin{equation}\label{a19}
 [a_\omega, a_{\bar\omega}^*] = - i\delta_{\omega \bar\omega}.
 \end{equation}
  (The reason for this condition will become clear in a moment.)  According to Eq.~\eqref{16.6} we therefore choose normalizations of the eigenvectors such that
 \begin{equation}\label{a20}
  ({\bf u}_\omega,{\bf u_\omega}) = \omega.
  \end{equation}
Since both sides of this equation are positive, a normalization constant for ${\bf u}_\omega$ can be found to satisfy this equation. Note that only the magnitude of this  constant is determined. The phase of the constant can be chosen arbitrarily, which allows a certain degree of latitude in the canonical transformation.   As an aside, note also that Eqs.~\eqref{16.7} and \eqref{a19} are analogous to the commutator relations required for the creation and annihilation operators in the Bogoliubov method; see Appendix D.

We can now introduce $N$ real-valued canonical pairs $(Q_\omega, P_\omega) $, defined by 
\begin{equation} \label{a21}
a_\omega = \frac{1}{\sqrt{2}} (Q_\omega + i P_\omega).
\end{equation} 
 In order to show that these are canonical pairs, invert Eq.~\eqref{a21} (and its complex conjugate) to give
 $Q_\omega = 2^{-1/2}(a_\omega + a^*_\omega)$ and $P_\omega =-i 2^{-1/2}(a_\omega - a^*_\omega)$.  Then 
 \begin{align}
 [Q_\omega, P_{\bar\omega}] &=- \frac{i}{2}( [a_\omega, a_{\bar\omega}]- [a_\omega, a_{\bar\omega}]^* - [a_\omega,a_{\bar\omega}^*]+ [a_{\omega}^*,a_{\bar\omega}] ) \notag \\
  & =- \frac{i}{2}(0-0 +i \delta_{\omega \bar\omega} + i  \delta_{\omega \bar\omega} ) \notag \\
  & =  \delta_{\omega \bar\omega},
 \end{align}
 and similarly $ [P_\omega, P_{\bar\omega}] = 0 =[Q_\omega, Q_{\bar\omega}] $. We now see the point of Eq.~\eqref{a19}: this choice determines that $[Q_\omega, P_\omega] = 1$; a different choice would lead to a value other than $1$ on the right hand side of this Poisson bracket relation.  
  
Applying Eq.~\eqref{a21}  to  Eqs.~\eqref{a15} and \eqref{a16}  and using the normalization condition Eq.~\eqref{a20}, yields the diagonalized system Hamiltonian
\begin{align}\label{a24}
H =& \sum_{\omega>0}  H_\omega, \\
H_\omega =& \frac{\omega}{2}  (Q_\omega^2 + P_\omega^2) + f_{1\omega} Q_\omega + f_{2 \omega}P_\omega,
\end{align}
where  $f_{1\omega}(t) =\sqrt{2} {\bf f}(t)\cdot Re({\bf u}_\omega)$ and  $f_{2\omega}(t) = -\sqrt{2} {\bf f}(t)\cdot Im({\bf u}_\omega)$. Note that for our choice of canonical pairs the $Q_\omega$ and $P_\omega$ variables have the same dimensions of $\sqrt{energy/frequency}$, but other choices are of course possible via a secondary canonical transformation. 

Hamiltons equations of motion applied to Eq.~\eqref{a24} then yield
\begin{align}\label{a25}
\dot Q_\omega &=\frac{\partial H}{\partial P_\omega} =  \omega P_\omega + f_{2\omega}, \notag \\
\dot P_\omega &=-\frac{\partial H}{\partial Q_\omega}= - \omega Q_\omega - f_{1\omega}.
\end{align}
which are seen to be  equivalent to Eqs.~\eqref{a13} and \eqref{a14.5} after application of Eqs.~\eqref{a21}  and Eq.~\eqref{a20}.

When $f_{1\omega}=f_{2\omega} = 0$ the equations for the $\omega$ mode are unforced and the oscillator energy $H_\omega$ is a constant of the motion.

Finally, we note that Eqs.~\eqref{a11} and \eqref{a21} imply that the linear transformation from phase-space variables ${\bf z}=({\bf q},{\bf p})$ to new variables ${\bf Z} = ({\bf Q},{\bf P})$
can be written as a matrix equation 
\begin{equation}\label{28}
{\bf z} = {\bf S}\cdot{\bf Z},
\end{equation}
where the  $2N\times 2N$ symplectic transformation matrix $\bf S$ is given by 
\begin{equation}\label{29}
{\bf S} = \sqrt{2}( \text{Re}{\bf U},-\text{Im}{\bf U}),
\end{equation}
and the $2N\times N$ matrix $\bf U$ has columns consisting of the $N$ eigenvectors ${\bf u}_\omega$ with $\omega>0$, normalized as per Eq.~\eqref{a20}. A similar result holds for the Bogoliubov transformation, although there the phase space variables $\bf z$ and $\bf Z$ are replaced by creation/annihilation pairs (see Appendix D), and extra linear transformations between these pairs and the phase space variables must be performed to obtain Eq.~\eqref{29}.

\subsubsection{Thermal averages}
The diagonalized Hamiltonian simplifies many calculations involving the energy. For example, consider the thermal average of  a phase space function $F({\bf z})$,
\begin{equation}\label{a25.1}
\langle F \rangle = \frac{\int d{\bf z} F({\bf z}) \exp(-H({\bf z})/T)}  {\int d{\bf z}\exp(-H({\bf z})/T)},
\end{equation} 
where $T$ is the temperature. In what follows we assume that the forcing coefficient $\bf f$ is zero.

In many cases transformation to the canonical variables $(Q_\omega,P_\omega)$ can simplify such calculations. Since the transformations are canonical  the Jacobian of the transformation is unity and the averaging integrals become
\begin{equation}\label{a25.2}
\langle F \rangle = \frac{\int F({\bf z}) \prod_{\omega>0} dQ_\omega dP_\omega \exp(-H_\omega/T)  }  {\int \prod_{\omega>0} dQ_\omega dP_\omega\exp(-H_\omega/T)}.
\end{equation} 
For instance, it is easy to show that
\begin{align}
\langle Q_\omega Q_{\bar\omega}\rangle &= \langle P_\omega P_{\bar\omega}\rangle = \frac{T}{\omega}\delta_{\omega\bar\omega},  \label{a25.3}\\
\langle Q_\omega P_{\bar\omega}\rangle &= 0\label{a25.4}
\end{align}
which implies
\begin{align}\label{a25.5}
\langle H_\omega \rangle = T.
\end{align}


\subsection{Neutrally-Stable System}

We now return to Eq.~\eqref{a9} and consider a modification to it that is necessary when there is a neutrally-stable ($\omega=0$) eigenmode of Eq.~\eqref{a5}. Let us assume there is only one such mode, whose eigenvector we label ${\bf u}_0$. This zero-frequency eigenvector satisfies 
\begin{equation}\label{a26}
{\bf D}\cdot{\bf  u}_0 = {\bf 0}.
\end{equation}
Thus, ${\bf u}_0$ is in the null-space of $\bf D$. By assumption it is the only vector in the nullspace. 
This eigenvector is real, since $\bf D$ is real.  

The eigenvector ${\bf u}_0$ is also  in the nullspace of the Hamiltonian matrix $\bf H$. This follows by applying to both sides of Eq.~\eqref{a26} the fundamental symplectic matrix $\bf J$ and using ${\bf D} = {\bf J}\cdot{\bf H}$:
\begin{align}\label{a27}
{\bf J}\cdot{\bf D}\cdot{\bf  u}_0 &= {\bf J}\cdot{\bf J}\cdot{\bf H}\cdot{\bf  u}_0 \notag \\
						&= -{\bf H}\cdot{\bf  u}_0 ={\bf 0},
\end{align}
where in the last step we used the identity ${\bf J}\cdot{\bf J}= -\bf 1$.

Now there are only $2N-1$ independent eigenvectors of the dynamical matrix $\bf D$, the $2(N-1)$ eigenvectors with non-zero frequencies, and the single zero frequency eigenvector. Since the phase space has dimension $2N$ the $2N-1$ eigenvectors no longer form a complete set that can be used to represent general phase-space vectors $\bf z$. However, we require such a representation in order  to fully diagonalize the Hamiltonian. We therefore need one more vector that is orthogonal  to the $2N-1$ eigenvectors. We will refer to this vector as $\bar{\bf u}_0$.  It is not an eigenvector of $\bf D$, but can instead by obtained by consideration of the constants of the motion.
	When there is a zero frequency eigenvector, Eq.~\eqref{a27} implies that the Hamiltonian has a symmetry that produces a new constant of the motion, the momentum $P_0$. This  constant is 
\begin{equation}\label{a28}
P_0 = {\bf u}_0\cdot{\bf J}\cdot{\bf z}.
\end{equation}
The time derivative of $P_0$ can be shown to equal zero using Eq.~\eqref{a2}, if we also assume that ${\bf f}\cdot{\bf u}_0=  0$:
\begin{align}
\dot P_0 &=  {\bf u}_0\cdot{\bf J}\cdot\dot{\bf z} \notag \\
	&= {\bf u}_0\cdot{\bf J}\cdot({\bf D}\cdot{\bf z} +{\bf J}\cdot{\bf f} )  \notag \\
	&= {\bf u}_0\cdot{\bf J}\cdot{\bf J}\cdot({\bf H}\cdot{\bf z} +{\bf f})  \notag \\
	&= -{\bf u}_0 \cdot{\bf H}\cdot{\bf z}-{\bf u}_0\cdot{\bf f}  \notag \\
	&= {\bf 0}\cdot{\bf z} + 0= 0,
\end{align}
	where in the last step we used Eq.~\eqref{a27} and the symmetry of the Hamiltonian matrix.
	
	We can now find the vector $\bar{\bf u}_0$ using the constancy of $P_0$ in the linear dynamics. Each oscillatory eigenmode with $\omega \ne 0$ must satisfy $P_0=constant$, and since $P_0$ is linear in $\bf z$ the only possibility is $P_0=0$. Therefore, the eigenmodes all satisfy
\begin{equation}\label{a30}
0={\bf u}_0\cdot{\bf J}\cdot{\bf u}_\omega.
\end{equation}
Note that this is also true for the $\omega=0$ eigenvector ${\bf u}_0$, since  $0={\bf u}_0\cdot{\bf J}\cdot{\bf u}_0$ by the antisymmetry of $\bf J$. We will construct a vector $\bar{\bf u}_0$ that is orthogonal to all of the eigenmodes by solving the equation
\begin{equation}\label{a31}
	{\bf u}_0\cdot{\bf J}\cdot{\bf u}_\omega =\bar{\bf u}_0\cdot{\bf H}\cdot{\bf u}_\omega
\end{equation}
for all eigenvectors ${\bf u}_\omega$. By Eq.~\eqref{a30}, such a vector will satisfy $(\bar{\bf u}_0,{\bf u}_\omega)=0$ for all $\omega$ including $\omega = 0$. A necessary and sufficient condition for solution of Eq.~\eqref{a31} is that $\bar{\bf u}_0$ satisfy 
\begin{equation}\label{a32}
{\bf H}\cdot\bar{\bf u}_0={\bf u}_0\cdot{\bf J}.
	\end{equation}
However, since $\bf H$ has a vector ${\bf u}_0$ in its nullspace, $\bar{\bf u}_0$  cannot be obtained via a standard matrix inversion solution to Eq.~\eqref{a32} because the inverse of $\bf H$ does not exist. In fact, the solution  to Eq~\eqref{a32} is under-determined. The right-hand-side is perpendicular (in the usual dot-product sense) to  the null-space of $\bf H$  since ${\bf u}_0\cdot{\bf J}\cdot{\bf u}_0 = 0$, and this implies that only $2N-1$ of the equations in Eq.~\eqref{a32} are linearly independent: 
the system satisfies ${\bf u}_0 \cdot{\bf H}\cdot\bar{\bf u}_0 = 0$. The solution to such a problem can be obtained in a number of ways. For example, one can project Eq.~\eqref{a32} onto the subspace that is perpendicular to ${\bf u}_0$, obtaining $2N-1$ independent equations.
A unique particular solution for $\bar{\bf u}_0$, $\bar{\bf u}_p$,  can then obtained by specifying an extra  condition on the solution that, for example, $\bar{\bf u}_0\cdot{\bf u}_0 = 0$. The vector $\bar{\bf u}_p$ is real since all coefficients appearing in the equations are real. The general solution is this particular solution added to the nullspace eigenvector:
\begin{equation}
\bar{\bf u}_0 =\bar{\bf u}_p + C {\bf u}_0
\end{equation}
for any value of the constant $C$. The value of $C$ can be chosen arbitrarily without affecting any of our subsequent results.

We can now use this extended system of vectors to represent a general phase space vector $\bf z$:
\begin{equation}\label{a35}
{\bf z} = \sum_{\omega\ne 0} a_\omega {\bf u}_\omega + a_0 {\bf u}_0 + \bar a_0 \bar{\bf u}_0.
\end{equation}	
The vectors form a complete orthogonal set so the coefficients $a_\omega$  and $\bar a_0$ can be obtained by projection:
\begin{align}
a_\omega &= \frac{({\bf u}_\omega,{\bf z})}{({\bf u}_\omega,{\bf u}_\omega)}, \label{a36n}\\
\bar a_0&= \frac{(\bar {\bf u}_0,{\bf z})}{(\bar{\bf u}_0,\bar{\bf u}_0)}.\label{a37n}
\end{align}
However, $a_0$ cannot be found using the standard projection method because ${\bf u}_0$ is orthogonal to itself: according to Eq.~\eqref{a27},	${\bf u}_0\cdot{\bf H}\cdot{\bf u}_0 = 0$. Instead, $a_0$ can be determined using the properties of the fundamental symplectic matrix. Acting on both sides of Eq.~\eqref{a35} with $\bar{\bf u}_0\cdot{\bf J}$, we obtain
\begin{equation}\label{a35n}
\bar{\bf u}_0\cdot{\bf J}\cdot{\bf z} = \sum_{\omega\ne 0} a_\omega \bar{\bf u}_0\cdot{\bf J}\cdot{\bf u}_\omega + a_0 \bar{\bf u}_0\cdot{\bf J}\cdot{\bf u}_0 + \bar a_0 \bar{\bf u}_0\cdot{\bf J}\cdot\bar{\bf u}_0.
\end{equation}	
However, $\bar{\bf u}_0\cdot{\bf J}\cdot\bar{\bf u}_0 = 0$ due to the antisymmetry of $\bf J$, and also 
$\bar{\bf u}_0\cdot{\bf J}\cdot{\bf u}_\omega =0$ for $\omega\ne 0$. This follows because, for $\omega\ne 0$,
\begin{equation}\label{a39n}
 \bar{\bf u}_0\cdot{\bf J}\cdot{\bf u}_\omega =\frac{ \bar{\bf u}_0\cdot{\bf J}\cdot{\bf D}\cdot{\bf u}_\omega}{-i\omega} = \frac{ \bar{\bf u}_0\cdot{\bf J}\cdot{\bf J}\cdot{\bf H}\cdot{\bf u}_\omega}{-i\omega}
 =\frac{ \bar{\bf u}_0\cdot{\bf H}\cdot{\bf u}_\omega}{i\omega}=0.
 \end{equation}
 We are therefore left with 
 \begin{equation}
 \bar{\bf u}_0\cdot{\bf J}\cdot{\bf z} =  a_0 \bar{\bf u}_0\cdot{\bf J}\cdot{\bf u}_0=- a_0 \bar{\bf u}_0\cdot{\bf H}\cdot\bar{\bf u}_0
 \end{equation}
 where in the second form we employed Eq.~\eqref{a32}. Thus, we obtain for $a_0$
 \begin{equation}\label{a41n}
 a_0 = - \frac{ \bar{\bf u}_0\cdot{\bf J}\cdot{\bf z}} {(\bar{\bf u}_0,\bar{\bf u}_0)}.
 \end{equation}
Returning to Eq.~\eqref{a35},  the complex coefficients $a_\omega$  still come in $\pm\omega$ pairs satisfying $a_{-\omega}=a_\omega^*$ for $\omega\ne 0$. The time-dependence of these coefficients is still given by Eq.~\eqref{a14}. The time-dependence of $a_0$ and $\bar a_0$ follow in the same way,  by substitution of Eq.~\eqref{a35} into the equation of motion, Eq.~\eqref{a2}. The result, after projecting out all the $\omega\ne 0$ eigenvectors, is
\begin{align}\label{a39}
\dot a_0 {\bf u}_0 + \dot{\bar a}_0 \bar {\bf u}_0 &= \bar a_0 {\bf D}\cdot\bar{\bf u}_0 + \Delta {\bf f}, \notag \\
&= \bar a_0 {\bf u}_0 + \Delta {\bf f}
\end{align}
	where we employed ${\bf D}\cdot{\bf u}_0 = \bf 0$, and where $\Delta{\bf f} ={\bf J}\cdot{\bf f} - \sum_{\omega\ne 0}f_\omega {\bf u}_\omega$ is the projection of ${\bf J}\cdot\bf f$ into the $({\bf u}_0,\bar{\bf u}_0)$ subspace. In the second line we used
	\begin{equation}\label{a43n}
	 {\bf D}\cdot\bar{\bf u}_0 ={\bf u}_0,
	 \end{equation}
 which follows from Eq.~\eqref{a32} by applying $\bf J$ to both sides, and using ${\bf u}_0\cdot{\bf J} = - {\bf J}\cdot{\bf u}_0$ and ${\bf J}\cdot{\bf J} = -\bf 1$. Taking an inner product of Eq.~\eqref{a39} with respect to $\bar{\bf u}_0$ then implies that 
\begin{align}\label{a40}
\dot{\bar a}_0 &= \frac{(\bar{\bf u}_0,\Delta{\bf f})}{(\bar{\bf u}_0,\bar{\bf u}_0)} \notag \\
 & = -\frac{{\bf f}\cdot{\bf u}_0}{(\bar{\bf u}_0,\bar{\bf u}_0)},
\end{align}
where in the second line we rewrote the numerator using $(\bar{\bf u}_0,\Delta{\bf f})=(\bar{\bf u}_0,{\bf J}\cdot{\bf f})=({\bf J}\cdot{\bf f},\bar{\bf u}_0)= -{\bf f}\cdot{\bf J}\cdot{\bf H}\cdot\bar{\bf u}_0 =-{\bf f}\cdot{\bf D}\cdot\bar{\bf u}_0=-{\bf f}\cdot{\bf u}_0 $, and in the last step we used Eq.~\eqref{a43n}. 

  When ${\bf f}\cdot{\bf u}_0 = 0$ Eq.~\eqref{a40} is an expression of the conservation  of the momentum $P_0$ in the dynamics. This can be seen by substituting for $\bf z$ from Eq.~\eqref{a35}  into Eq.~\eqref{a28}, yielding
\begin{equation}\label{a51c}
P_0 = \bar a_0 \bar {\bf u}_0\cdot{\bf H}\cdot\bar{\bf u}_0,
\end{equation}
where we employed Eqs.~\eqref{a30} and \eqref{a32}. Taking a time derivative and using Eq.~\eqref{a40} yields $\dot P_0= - {\bf f}\cdot{\bf u}_0$.

  Finally, the dynamics of $a_0$ follows  by  acting on both sides of Eq.~\eqref{a39} with $\bar{\bf u}_0\cdot{\bf J}$, yielding
\begin{equation}
\dot a_0  = \bar a_0  + \frac{\bar{\bf u}_0\cdot{\bf J}\cdot\Delta {\bf f}}{\bar{\bf u}_0\cdot{\bf J}\cdot{\bf u}_0}, 
\end{equation}
Substituting for $\Delta\bf f$, and using ${\bf J}\cdot{\bf J}=-\bf 1$, Eq.~\eqref{a39n}, and Eq.~\eqref{a32}, we are left with
\begin{equation}\label{a48n}
\dot a_0  = \bar a_0  + \frac{\bar{\bf u}_0\cdot {\bf f}}{(\bar{\bf u}_0,\bar{\bf u}_0)}.
\end{equation}
	Thus, for ${\bf f}= \bf 0$, $a_0$ increases linearly in time with a rate given by $\bar a_0$.

We now return to the question of the diagonalization of the system energy and the proper choice of canonical pairs. Substituting  $\bf z$ from Eq.~\eqref{a35} into the system energy Eq.~\eqref{a2},  the same argument as led to Eq.~\eqref{a15} now results  in some new terms involving the zero-frequency modes:
\begin{equation}
H = \sum_{\omega>0} H_\omega  + \frac{1}{2} {\bar a_0}^2 (\bar{\bf u}_0, \bar{\bf u}_0)  + a_0 {\bf f}\cdot{\bf u}_0 + \bar a_0 {\bf f}\cdot\bar{\bf u}_0,
\end{equation}
where $H_\omega$ is still given by Eq.~\eqref{a16}. Poisson brackets involving $a_0$ and $\bar a_0$ follow from Eqs.~\eqref{a41n}, and  \eqref{a37n}:
\begin{align}
[a_0,\bar a_0] &= -\frac{\bar{\bf u}_0\cdot{\bf J}\cdot[{\bf z},\bf z] \cdot{\bf H}\cdot\bar {\bf u}_0}{(\bar{\bf u}_0,\bar{\bf u}_0)^2} \notag \\
& = -\frac{\bar{\bf u}_0\cdot{\bf J}\cdot{\bf J}\cdot{\bf H}\cdot\bar {\bf u}_0}{(\bar{\bf u}_0,\bar{\bf u}_0)^2} = \frac{\bar{\bf u}_0\cdot{\bf H}\cdot\bar {\bf u}_0}{(\bar{\bf u}_0,\bar{\bf u}_0)^2}\notag \\
& = \frac{1}{(\bar{\bf u}_0,\bar{\bf u}_0)}.
\end{align}
This implies that $a_0$ and  the momentum $P_0 = \bar a_0 (\bar{\bf u}_0,\bar{\bf u}_0)$ form a canonical pair (see Eq.~\eqref{a51c}). 

We must also show that $[a_0, a_\omega]= [P_0,a_\omega]=0$ for all $\omega\ne 0$. Using Eqs.~\eqref{a41n} and \eqref{a36n} the first bracket yields
\begin{align}
[a_0,a_\omega] &= -\frac{\bar{\bf u}_0\cdot{\bf J}\cdot[{\bf z},\bf z] \cdot{\bf H}\cdot {\bf u}_\omega^*}{(\bar{\bf u}_0,\bar{\bf u}_0)({\bf u}_\omega,{\bf u}_\omega)} \notag \\
& = -\frac{\bar{\bf u}_0\cdot{\bf J}\cdot{\bf J}\cdot{\bf H}\cdot{\bf u}_\omega^*}{(\bar{\bf u}_0,\bar{\bf u}_0)({\bf u}_\omega,{\bf u}_\omega)} = 
\frac{\bar{\bf u}_0\cdot{\bf H}\cdot{\bf u}_\omega^*}{(\bar{\bf u}_0,\bar{\bf u}_0)({\bf u}_\omega,{\bf u}_\omega)} \notag \\
& = 0,
\end{align}
as required. The second bracket follows from Eqs.~\eqref{a37n} and ~\eqref{a36n}:
\begin{align}
[P_0,a_\omega] &= -\frac{\bar{\bf u}_0\cdot{\bf H}\cdot[{\bf z},\bf z] \cdot{\bf H}\cdot {\bf u}_\omega^*}{({\bf u}_\omega,{\bf u}_\omega)} \notag \\
& = -\frac{\bar{\bf u}_0\cdot{\bf H}\cdot{\bf J}\cdot{\bf H}\cdot{\bf u}_\omega^*}{({\bf u}_\omega,{\bf u}_\omega)} = 
-\frac{\bar{\bf u}_0\cdot{\bf H}\cdot{\bf D}\cdot{\bf u}_\omega^*}{({\bf u}_\omega,{\bf u}_\omega)} \notag \\
& =-\frac{i\omega \bar{\bf u}_0\cdot{\bf H}\cdot{\bf u}_\omega^*}{({\bf u}_\omega,{\bf u}_\omega)} =0.
\end{align}

Finally, we  write the diagonalized Hamiltonian  in terms of the canonical pairs as
\begin{equation}\label{a53}
H = \sum_{\omega>0}  \{ \frac{\omega}{2} (Q_\omega^2 + P_\omega^2) + f_{1\omega} Q_\omega + f_{2\omega} P_\omega \}  + \frac{P_0^2}{2 (\bar{\bf u}_0, \bar{\bf u}_0) }  + a_0 {\bf f}\cdot{\bf u}_0 + P_0 \frac{{\bf f}\cdot\bar{\bf u}_0}{ (\bar{\bf u}_0, \bar{\bf u}_0) }.
\end{equation}
The dynamical equations for $Q_\omega$ and $P_\omega$ remain unchanged from Eqs.~\eqref{a25}.
The equations of motion for $a_0$ and $P_0$ are
\begin{align}
\dot a_0 &=\frac{ \partial H}{\partial P_0} = \frac{P_0}{ (\bar{\bf u}_0, \bar{\bf u}_0) } +  \frac{{\bf f}\cdot\bar{\bf u}_0}{ (\bar{\bf u}_0, \bar{\bf u}_0) }, \\
\dot P_0 &= -\frac{ \partial H}{\partial a_0}= -{\bf f}\cdot{\bf u}_0,
\end{align}
which agree with Eqs.~\eqref{a48n} and \eqref{a40}. In these equations the inner product $ (\bar{\bf u}_0, \bar{\bf u}_0)$ can be interpreted as the  inertia associated with the momentum $P_0$. 

This completes the derivation of the diagonalized Hamiltonian and the canonical coordinates for a neutrally-stable linearized Hamiltonian system with a single zero frequency mode.

Before we move on, a few remarks must be be made regarding the case when there is more than one zero frequency mode. This special case actually arises more often than one might suspect; examples include  systems with spherical symmetry,  systems with translational or cylindrical symmetry that are also undergoing a second order structural phase transition, or  systems with translational symmetry in more than one dimension. 
	Let us consider the case where there are two neutrally stable modes; the case of more than two can be understood by extension of this example.
	
	Now there are two independent eigenvectors ${\bf u}_{0 1}$ and ${\bf u}_{0 2}$ in the nullspace of  both the dynamical matrix $\bf D$ and the Hamiltonian matrix $\bf H$. These eigenvectors produce two constants of the motion, $P_{0 1}$ and $P_{0 2}$, given by
\begin{align}
P_{0 1} &= {\bf u}_{0 1}\cdot{\bf J}\cdot{\bf z}, \label{b62}\\
P_{0 2} &= {\bf u}_{0 2}\cdot{\bf J}\cdot{\bf z}.\label{b63}
\end{align}
	Two possibilities  must be separately considered: (i) ${\bf u}_{0 1}\cdot{\bf J}\cdot {\bf u}_{0 2} = 0$ and (ii) ${\bf u}_{0 1}\cdot{\bf J}\cdot {\bf u}_{0 2} \equiv J_{1 2}\ne 0$. In case (i), the constants of the motion are in involution (their Poisson bracket vanishes), while in case (ii) they are not. This can be seen by evaluating the Poisson bracket $[P_{0 1}, P_{0 2}]$, using Eqs.~\eqref{b62} and \eqref{b63}:
\begin{align}
[P_{0 1}, P_{0 2}] &=-{\bf u}_{0 1}\cdot{\bf J}\cdot[ {\bf z},{\bf z}]\cdot{\bf J}\cdot{\bf u}_{0 2} \notag \\
&= -{\bf u}_{0 1}\cdot{\bf J}\cdot{\bf J}\cdot{\bf J}\cdot{\bf u}_{0 2} \notag \\
& ={\bf u}_{0 1}\cdot{\bf J}\cdot{\bf u}_{0 2} = J_{1 2}.\label{b64.5}
\end{align}

\subsubsection{case (i): constants of the motion in involution}
In case (i), when the constants are in involution, the eigenvectors by themselves are not a complete set and two vectors $\bar {\bf u}_{0 1} $ and $\bar {\bf u}_{0 2}$ are required in order to describe a general phase-space vector $\bf z$ according to
\begin{equation}\label{b65}
{\bf z} = \sum_{\omega\ne 0}  a_\omega {\bf u}_\omega + a_{0 1}{\bf u}_{0 1} +  a_{0 2}{\bf u}_{0 2}  + \bar a_{0 1}\bar {\bf u}_{0 1} +  \bar a_{0 2}\bar{\bf u}_{0 2}.
\end{equation}
The need for the extra vectors $\bar {\bf u}_{0 1} $ and $\bar {\bf u}_{0 2}$ can be seen by acting on the equation with ${\bf u}_{0 1}\cdot{\bf J}$:
\begin{align}\label{b66}
{\bf u}_{0 1}\cdot{\bf J}\cdot{\bf z} = &\sum_{\omega\ne 0}  a_\omega {\bf u}_{0 1}\cdot{\bf J}\cdot{\bf u}_\omega + a_{0 1}{\bf u}_{0 1}\cdot{\bf J}\cdot{\bf u}_{0 1} \notag \\
  &+a_{0 2}{\bf u}_{0 1}\cdot{\bf J}\cdot{\bf u}_{0 2}  + \bar a_{0 1}{\bf u}_{0 1}\cdot{\bf J}\cdot\bar {\bf u}_{0 1} +  \bar a_{0 2}{\bf u}_{0 1}\cdot{\bf J}\cdot\bar{\bf u}_{0 2}.
\end{align}
Conservation of $P_{0 1}$ implies that ${\bf u}_{0 1}\cdot{\bf J}\cdot{\bf u}_\omega =0$, while ${\bf u}_{0 1}\cdot{\bf J}\cdot{\bf u}_{0 1}=0$ by symmetry and ${\bf u}_{0 1}\cdot{\bf J}\cdot{\bf u}_{0 2}=0$ by assumption. Repeating the procedure with ${\bf u}_{0 2}\cdot{\bf J}$, we have
\begin{align}
{\bf u}_{0 1}\cdot{\bf J}\cdot{\bf z} &=\bar a_{0 1}{\bf u}_{0 1}\cdot{\bf J}\cdot\bar {\bf u}_{0 1} +  \bar a_{0 2}{\bf u}_{0 1}\cdot{\bf J}\cdot\bar{\bf u}_{0 2}, \label{b67} \\
{\bf u}_{0 2}\cdot{\bf J}\cdot{\bf z} &=\bar a_{0 2}{\bf u}_{0 2}\cdot{\bf J}\cdot\bar {\bf u}_{0 1} +  \bar a_{0 2}{\bf u}_{0 2}\cdot{\bf J}\cdot\bar{\bf u}_{0 2},\label{b68}
\end{align}
To satisfy these equations  for a general vector $\bf z$ we require nonzero values for both $\bar a_{0 1}$ and $ \bar a_{0 2}$, proving that both vectors $\bar {\bf u}_{0 1} $ and $\bar {\bf u}_{0 2}$ are required for a complete set. 

Let us now consider the solution of these equations for $\bar a_{0 1}$ and $ \bar a_{0 2}$, along with the determination of $a_{0 1}$ and $ a_{0 2}$. Recall that
 these latter two coefficients can be found by applying $\bar {\bf u}_{0 1}\cdot{\bf J}$ and $\bar {\bf u}_{0 2}\cdot{\bf J}$ to Eq.~\eqref{b65}. 
 Noting that $\bar {\bf u}_{0 1}\cdot{\bf J}\cdot{\bf u}_\omega = \bar {\bf u}_{0 2}\cdot{\bf J}\cdot{\bf u}_\omega =0$ (see Eq.~\eqref{a39n}), and 
 that $\bar{\bf u}_{0 1}\cdot{\bf J}\cdot\bar {\bf u}_{0 1}=\bar{\bf u}_{0 2}\cdot{\bf J}\cdot\bar {\bf u}_{0 2}=0$ by symmetry, we are left with 
\begin{align}
\bar {\bf u}_{0 1}\cdot{\bf J}\cdot{\bf z} &=a_{0 1}\bar {\bf u}_{0 1}\cdot{\bf J}\cdot{\bf u}_{0 1} +a_{0 2}\bar {\bf u}_{0 1}\cdot{\bf J}\cdot{\bf u}_{0 2}+  \bar a_{0 2}\bar {\bf u}_{0 1}\cdot{\bf J}\cdot\bar{\bf u}_{0 2}, \label{b69} \\
\bar {\bf u}_{0 2}\cdot{\bf J}\cdot{\bf z} &=a_{0 1}\bar {\bf u}_{0 2}\cdot{\bf J}\cdot{\bf u}_{0 1} +a_{0 2}\bar {\bf u}_{0 2}\cdot{\bf J}\cdot{\bf u}_{0 2}-  \bar a_{0 1}\bar {\bf u}_{0 1}\cdot{\bf J}\cdot\bar{\bf u}_{0 2}. \label{b70}
\end{align}
We can simplify the solution of Eqs.~\eqref{b67} - \eqref{b70}  by recalling that the vectors $\bar {\bf u}_{0 1} $ and $\bar {\bf u}_{0 2}$ are constructed to be orthogonal to all of the eigenvectors by solution of the (under-determined) equations
\begin{align}
{\bf H}\cdot\bar {\bf u}_{0 1} &= {\bf u}_{0 1} \cdot{\bf J}, \label{b71}\\
{\bf H}\cdot\bar {\bf u}_{0 2} &= {\bf u}_{0 2} \cdot{\bf J}.\label{b72}
\end{align}
We first use these two equations to replace Eqs.~\eqref{b67} -\eqref{b70} by the equivalent equations, 
\begin{align}
\bar{\bf u}_{0 1}\cdot{\bf H}\cdot{\bf z} &=\bar a_{0 1}\bar{\bf u}_{0 1}\cdot{\bf H}\cdot\bar {\bf u}_{0 1} +  \bar a_{0 2}\bar{\bf u}_{0 1}\cdot{\bf H}\cdot\bar{\bf u}_{0 2}, \label{b73} \\
\bar{\bf u}_{0 2}\cdot{\bf H}\cdot{\bf z} &=\bar a_{0 2}\bar{\bf u}_{0 2}\cdot{\bf H}\cdot\bar {\bf u}_{0 1} +  \bar a_{0 2}\bar{\bf u}_{0 2}\cdot{\bf H}\cdot\bar{\bf u}_{0 2},\label{b74}\\
\bar {\bf u}_{0 1}\cdot{\bf J}\cdot{\bf z} &=-a_{0 1}\bar {\bf u}_{0 1}\cdot{\bf H}\cdot\bar{\bf u}_{0 1} -a_{0 2}\bar {\bf u}_{0 1}\cdot{\bf H}\cdot\bar{\bf u}_{0 2}+  \bar a_{0 2}\bar {\bf u}_{0 1}\cdot{\bf J}\cdot\bar{\bf u}_{0 2}, \label{b75} \\
\bar {\bf u}_{0 2}\cdot{\bf J}\cdot{\bf z} &=-a_{0 1}\bar {\bf u}_{0 2}\cdot{\bf H}\cdot\bar{\bf u}_{0 1} -a_{0 2}\bar {\bf u}_{0 2}\cdot{\bf H}\cdot\bar{\bf u}_{0 2}-  \bar a_{0 1}\bar {\bf u}_{0 1}\cdot{\bf J}\cdot\bar{\bf u}_{0 2}. \label{b76}
\end{align}
Also, we note that since Eqs.~\eqref{b71} and \eqref{b72}  are under-determined, any linear combination of the null space vectors of $\bf H$  can be added to particular solutions $\bar{\bf u}_{ p 1}$ and $\bar{\bf u}_{ p 2}$  of these equations:
 \begin{align}
 \bar {\bf u}_{0 1} &=  \bar {\bf u}_{p 1} +\alpha_1 {\bf u}_{0 1} + \alpha_2 {\bf u}_{0 2}  \label{b77} \\
 \bar {\bf u}_{0 2} &=  \bar {\bf u}_{p 2} +\beta_1 {\bf u}_{0 1} + \beta_2 {\bf u}_{0 2}. \label{b78}
 \end{align}
It is useful to choose the constants $\alpha_1, \alpha_2,\beta_1, \beta_2$ so that $\bar {\bf u}_{0 1}\cdot{\bf J}\cdot\bar{\bf u}_{0 2}=0$. This can be accomplished, for example,  by taking $\alpha_1=\beta_1=\beta_2 = 0$ and choosing $\alpha_2$ such that $\bar {\bf u}_{p 1}\cdot{\bf J}\cdot\bar{\bf u}_{0 2}+ \alpha_2 {\bf u}_{0 2}\cdot{\bf J}\cdot\bar{\bf u}_{0 2}=0$. Using Eq.~\eqref{b72}, the solution for $\alpha_2$ is
\begin{equation}
\alpha_2 =- \frac{\bar {\bf u}_{p 1}\cdot{\bf J}\cdot\bar{\bf u}_{0 2}}{\bar{\bf u}_{0 2}\cdot{\bf H}\cdot\bar{\bf u}_{0 2}}. \label{b79}
\end{equation}
The condition  $\bar {\bf u}_{0 1}\cdot{\bf J}\cdot\bar{\bf u}_{0 2}=0$ implies that we can drop the $\bar a_{0 1}$ and $\bar a_{0 2}$ terms in Eqs.~\eqref{b75} and \eqref{b76}, so they become
\begin{align}
\bar {\bf u}_{0 1}\cdot{\bf J}\cdot{\bf z} &=-a_{0 1}\bar {\bf u}_{0 1}\cdot{\bf H}\cdot\bar{\bf u}_{0 1} -a_{0 2}\bar {\bf u}_{0 1}\cdot{\bf H}\cdot\bar{\bf u}_{0 2}, \label{b80} \\
\bar {\bf u}_{0 2}\cdot{\bf J}\cdot{\bf z} &=-a_{0 1}\bar {\bf u}_{0 2}\cdot{\bf H}\cdot\bar{\bf u}_{0 1} -a_{0 2}\bar {\bf u}_{0 2}\cdot{\bf H}\cdot\bar{\bf u}_{0 2}. \label{b81}
\end{align}

Equations~\eqref{b80} and \eqref{b81} can be solved for $a_{0 1}$ and $a_{0 2}$, while Eqs.~\eqref{b73} and \eqref{b74} can be solved for $ \bar a_{0 1}$ and $\bar a_{0 2}$. The solutions are
\begin{align}
a_{0 1} &= \frac{ h_{12}j_{2z}-h_{22} j_{1z}}{h_{11}h_{22}-h_{12}^2 },\ \ \ a_{0 2} = \frac{ h_{12}j_{1z}-h_{11} j_{2z}}{h_{11}h_{22}-h_{12}^2 }, \label{b82}\\
\bar a_{0 1} &= \frac{ h_{22} h_{1z}-h_{12}h_{2z}}{h_{11}h_{22}-h_{12}^2 },\ \ \ \bar a_{0 2} = \frac{ h_{11} h_{2z}-h_{12}h_{1z}}{h_{11}h_{22}-h_{12}^2 },\label{b83}
\end{align}
 where $h_{i j} = \bar{\bf u}_{0 i}\cdot{\bf H}\cdot\bar{\bf u}_{0 j}$,  $h_{ i z} =  \bar{\bf u}_{0 i}\cdot{\bf H}\cdot{\bf z}$, and $j_{ i z} =  \bar{\bf u}_{0 i}\cdot{\bf J}\cdot{\bf z}$. We assume throughout that the Hamiltonian satisfies $h_{11}h_{22}-h_{12}^2\ne 0$. (This requirement has an origin similar to the requirement that the determinant of the inertia tensor of a rigid body must be nonzero. It is a requirement on any physical Hamiltonian system. )
 
 The system energy can be evaluated by substituting Eq.~\eqref{b65} into Eq.~\eqref{a1}, yielding
 \begin{align}
 H= \sum_{\omega>0} H_\omega +& \frac{1}{2} h_{11} \bar a_{0 1}^2 + \frac{1}{2} h_{22} \bar a_{0 2}^2 + h_{12} \bar a_{0 1}\bar a_{0 2}  \notag \\
 &+ a_{0 1} {\bf f}\cdot {\bf u}_{0 1} + a_{0 2} {\bf f}\cdot {\bf u}_{0 2}+ \bar a_{0 1} {\bf f}\cdot \bar {\bf u}_{0 1}+  \bar a_{0 2} {\bf f}\cdot \bar {\bf u}_{0 2}.
 \end{align}
 
 Canonical variables must be found in order to use this expression as a Hamiltonian. The Poisson brackets  of the zero frequency amplitudes can be found using Eqs.~\eqref{b82} and \eqref{b83}. We obtain
 $[a_{0 1}, a_{0 2}]=[\bar a_{0 1}, \bar a_{0 2}] = 0$  and the nontrivial brackets
 \begin{align}
 [a_{0 1}, \bar a_{0 1}] &= \frac{h_{22}}{h_{11}h_{22}-h_{12}^2}, \notag\\
  [a_{0 2}, \bar a_{0 2}] &= \frac{h_{11}}{h_{11}h_{22}-h_{12}^2}, \label{b85}\\
   [a_{0 1}, \bar a_{0 2}] &=- \frac{h_{12}}{h_{11}h_{22}-h_{12}^2}.\notag
   \end{align}
 
 Canonical pairs can be found by noting that the constants of the motion $P_{0 1}$ and $P_{0 2}$ are related to $\bar a_{0 1}$ and $\bar a_{0 2}$ via
 \begin{align}
 P_{0 1} &= h_{11}\bar a_{0 1} + h_{1 2} \bar a_{0 2}, \label{b86} \\
 P_{0 2} &= h_{2 2}\bar a_{0 2} + h_{1 2} \bar a_{0 1}, \label{b87}
 \end{align}
 where we substituted Eq.~\eqref{b65} into Eqs~\eqref{b62} and \eqref{b63} and applied Eqs.~\eqref{b71} and \eqref{b72}.
 
 Using these variables along with Eqs.~\eqref{b85} it is then an exercise to show   that
 $[a_{0 1},P_{0 1}]= [a_{0 2}, P_{0 2}] = 1$ while $[a_{0 2}, P_{0 1}] = [a_{0 1},P_{0 2}]=[P_{0 1}, P_{0 2}]= 0$.
 Thus, the canonical pairs are $(a_{0 1}, P_{0 1})$ and $(a_{0 2}, P_{0 2})$.
 We now need only invert Eqs.~\eqref{b86} and \eqref{b87},

\begin{align}
 \bar a_{0 1} & = \frac{ h_{22} P_{0 1} - h_{1 2} P_{0 2}}{ h_{11}h_{22}-h_{12}^2 }, \label{b88}\\
  \bar a_{0 2} & = \frac{ h_{11} P_{0 2} - h_{1 2} P_{0 1}}{ h_{11}h_{22}-h_{12}^2 }, \label{b89}
 \end{align} 
 and employ these results in the Hamiltonian, which becomes
 \begin{align}
 H= \sum_{\omega>0} H_\omega +& \frac{1}{2} \frac{ h_{22} P_{0 1}^2 + 2h_{12} P_{01}P_{02}+h_{11}P_{02}^2 } {h_{11}h_{22}-h_{12}^2   } \notag \\
 &+ \frac{ h_{22} P_{0 1} - h_{1 2} P_{0 2}}{ h_{11}h_{22}-h_{12}^2 } {\bf f}\cdot \bar{\bf u}_{0 1} +   \frac{ h_{11} P_{0 2} - h_{1 2} P_{0 1}}{ h_{11}h_{22}-h_{12}^2 }   {\bf f}\cdot \bar{\bf u}_{0 2}+  a_{0 1} {\bf f}\cdot  {\bf u}_{0 1}+   a_{0 2} {\bf f}\cdot  {\bf u}_{0 2}.
 \end{align}
  The equations of motion for $P_{0 1}$ and $P_{0 2}$ are then
 \begin{align}
  \dot P_{0 1} = -\frac{\partial H}{\partial a_{0 1}} = -{\bf f}\cdot{\bf u}_{0 1}, \label{b93.1}\\
 \dot P_{0 2} = -\frac{\partial H}{\partial a_{0 2}} = -{\bf f}\cdot{\bf u}_{0 2}, \label{b93.2}
  \end{align}
 and the equations of motion for the amplitudes $a_{0 1}$ and $a_{0 2}$ are
\begin{align}
\dot  a_{0 1} &= \frac{\partial H}{\partial P_{0 1}} = \frac{h_{22} P_{0 1}+  h_{12} P_{0 2}}{  h_{11}h_{22}-h_{12}^2  }+ \frac{h_{2 2} {\bf f}\cdot \bar{\bf u}_{0 1}-h_{1 2} {\bf f}\cdot \bar{\bf u}_{0 2}}{ h_{11}h_{22}-h_{12}^2  }, \label{b94.1}\\
\dot  a_{0 2} &= \frac{\partial H}{\partial P_{0 2}} = \frac{h_{11} P_{0 2} + h_{12} P_{0 1}}{  h_{11}h_{22}-h_{12}^2  }+ \frac{h_{1 1} {\bf f}\cdot \bar{\bf u}_{0 2}-h_{1 2} {\bf f}\cdot \bar{\bf u}_{0 1}}{ h_{11}h_{22}-h_{12}^2  },.\label{b94.2}
\end{align}
   When the forcing $\bf f$ is zero, the Hamiltonian is independent of $a_{0 1}$ and $a_{0 2}$ so the canonical momenta $P_{0 1}$ and $P_{0 2}$ are constant, as expected, and $a_{0 1}$ and $a_{0 2}$ both have uniform rates of change.

 \subsubsection{case (ii): constants of the motion not in involution}
 
 When the two zero-frequency modes satisfy $J_{1 2} ={\bf u}_{0 1}\cdot{\bf J}\cdot{\bf u}_{0 2} \ne 0$, the constants of the  motion $P_{0 1}$ and $P_{0 2}$ are not in involution (Eq.~\eqref{b64.5}). This case is easier to deal with than the previous case of constants in involution. Now the eigenvectors of $\bf D$ by themselves form a complete set for any phase space vector ${\bf z}$, allowing us to write
 \begin{equation}\label{b94}
 {\bf z} = \sum_{\omega\ne 0} a_\omega {\bf u}_\omega + a_{0 1} {\bf u}_{0 1} + a_{0 2}{\bf u}_{0 2}.
 \end{equation}
No extra vectors $\bar {\bf u}_{0 1}$ or $\bar {\bf u}_{0 2}$ are needed.   If such vectors were needed, they would satisfy  Eqs.~\eqref{b72} and \eqref{b73}; but these equations no longer have solutions. This can be seen by, for example, taking a dot product of ${\bf u}_{0 1}$ with Eq.~\eqref{b73}:
\begin{equation}
0 = {\bf u}_{0 1}\cdot{\bf H}\cdot\bar{\bf u}_{0 2} = {\bf u}_{0 2}\cdot{\bf J}\cdot{\bf u}_{0 1}=-J_{1 2}\ne 0,
\end{equation}
a contradiction.

The amplitude coefficients $a_{01}$ and $a_{02}$ in Eq.~\eqref{b94} can now be obtained by acting with ${\bf u}_{02}\cdot\bf J$ and ${\bf u}_{01}\cdot\bf J $ respectively,  yielding
\begin{align}
a_{01} =-&\frac{ {\bf u}_{02}\cdot{\bf J}\cdot{\bf z}}{J_{12}} =-\frac{P_{0 2}}{J_{12}}, \label{b96}\\
a_{02} =&\frac{ {\bf u}_{01}\cdot{\bf J}\cdot{\bf z}}{J_{12}} = \frac{P_{0 1}}{J_{12}}, \label{b97}
\end{align}
 where we used $  {\bf u}_{01}\cdot{\bf J}\cdot{\bf u}_\omega={\bf u}_{02}\cdot{\bf J}\cdot{\bf u}_\omega= 0$ (see Eq.~\eqref{a39n}), and where the second forms in terms of the constants of the motion follow from Eqs.~\eqref{b62} and \eqref{b63}. The coefficients $a_\omega$ are still obtained with the usual inner product, see Eq.~\eqref{a10}.
 
 The system energy is found by applying Eq.~\eqref{b94} to Eq.~\eqref{a1}, yielding
 \begin{align}
 H &= \sum_{\omega>0} H_\omega + a_{01}{\bf f}\cdot{\bf u}_{0 1} + a_{0 2}{\bf f}\cdot{\bf u}_{02} \notag \\
     &= \sum_{\omega>0} H_\omega -\frac{ P_{02}}{J_{12}}{\bf f}\cdot{\bf u}_{0 1} + \frac{P_{0 1}}{J_{12}}{\bf f}\cdot{\bf u}_{02},
 \end{align}
where in the second line we employed Eqs.~\eqref{b96} and \eqref{b97}. 
 Equations of motion for $P_{0 1}$ and $P_{0 2}$ then follow from Eq.~\eqref{b64.5}:
 \begin{align}
 \dot P_{0 1} = [P_{0 1}, H] &= J_{12}\frac{\partial H}{\partial P_{0 2}} = -{\bf f}\cdot{\bf u}_{0 1}, \\
 \dot P_{0 2} = [P_{0 2}, H] &= -J_{12}\frac{\partial H}{\partial P_{0 1}} = -{\bf f}\cdot{\bf u}_{0 2}. 
  \end{align}
which are the same as when $P_{01}$ and $P_{02}$  are in involution (see Eqs.~\eqref{b93.1} and \eqref{b93.2}). However, now these two variables form a canonical set. This implies that when ${\bf f}=0$, the amplitudes $a_{01}$ and $a_{02}$ are time-independent (see Eqs.~\eqref{b96} and \eqref{b97}). This behavior differs from the previous case, where $a_{01}$ and $a_{02}$ continued to evolve at a fixed rate when ${\bf f}=0$ (Eqs.~\eqref{b94.1} and \eqref{b94.2}).

\subsection{Unstable System}
We now consider the normal modes in an unstable Hamiltonian system. Unstable conservative systems  are of importance in several contexts, such as in the study of ideal fluid and plasma instabilities. Here we will consider the case of a system with an unstable mode with complex frequency $\omega = \Omega \equiv \Omega_r + i \gamma$, where $\Omega_r>0$ and $\gamma>0$ are real frequency and growth rate respectively, with both assumed to be greater than zero.   In addition to this complex mode, a second mode with complex frequency $\Omega^* = \Omega_r -i\gamma$ must also occur. This follows because the dynamical matrix in the eigenmode problem,   Eq.~\eqref{a5}, has only real coefficients, which implies that, in order to solve the characteristic polynomial in $\omega$,  all complex mode frequencies must come in pairs, $\Omega_r\pm i\gamma$.

 Note that the above mode with frequency $\Omega^*$ has  negative growth rate $-\gamma$, and  is independent of the mode required by property 3, with frequency $-\Omega^*$. The latter mode  has positive growth rate, and is needed (when $\Omega_r\ne 0$) in order to construct a real solution for the phase space configuration $\bf z$, in analogy to the argument accompanying Eq.~\eqref{a12}. A fourth mode with frequency $-\Omega$ and negative growth rate is also required by property 3, and is the complex conjugate of the mode with frequency $\Omega^*$, in order to produce a real solution for $\bf z$. (If $\Omega_r=0$ only two of these four modes are required, as the others are redundant. In what follows we assume that $\Omega_r> 0$.  The $\Omega_r = 0$ case will be briefly discussed at the end of the subsection.)
 
 The modes of the unstable system form an orthogonal set according to property 1, and we use the eigenvectors associated with the modes to describe the phase space vector $\bf z$ via
 \begin{equation}\label{a56nn}
 {\bf z}(t) = \sum_{\omega>0} a_\omega(t) {\bf u}_\omega + a_\Omega(t) {\bf u}_\Omega + a_{\Omega^*}(t) {\bf u}_{\Omega^*} + c.c.
 \end{equation}
  just as was done for a stable system. Some differences become apparent however. 
According to property 2, a mode with complex eigenfrequency $\Omega$ must be orthogonal to itself: $({\bf u}_\Omega,{\bf u}_\Omega) = 0$. It is therefore not possible to determine $a_\Omega$ in terms of $\bf z$ using the standard projection formula, Eq.~\eqref{a10}. Fortunately, however, the complex mode  with frequency $\Omega^*$ can be used to determine $a_\Omega$ via projection. According to property 1, this mode is orthogonal to all other eigenmodes (as well as itself), {\it except} for the mode with complex frequency $\Omega$. Therefore, for modes with complex frequencies Eq.~\eqref{a10} is replaced by
\begin{equation}\label{a58}
a_\Omega(t) = \frac{({\bf u}_{\Omega^*}, {\bf z}(t))}{({\bf u}_{\Omega^*},{\bf u}_\Omega)},
\end{equation}
with an analogous expression for $a_{\Omega^*}$.
 The dynamics of a mode with complex frequency then follows by substitution of Eq.~\eqref{a56nn} into the equation of motion Eq.~\eqref{a2}, followed by projection, just as for the stable modes: :
 \begin{equation}\label{a59}
 \dot a_\Omega(t) = -i\Omega a_\Omega(t)+ f_\Omega(t),
 \end{equation}
 where here $f_\Omega(t) = ({\bf u}_{\Omega^*},{\bf J}\cdot {\bf f})/({\bf u}_{\Omega^*},{\bf u}_\Omega)$. This forcing coefficient can also be written as
 \begin{equation}\label{a59.5}
 f_\Omega(t) =-i\Omega \frac{{\bf u}_{\Omega^*}^*\cdot {\bf f}}{({\bf u}_{\Omega^*},{\bf u}_\Omega)},
 \end{equation}
 using the same algebraic steps as led to Eq.~\eqref{a14.5}.
 When $f_\Omega = 0$ the differential equation \eqref{a59} is unforced and the solution grows exponentially with time at the growth rate $\gamma$, $a_\Omega = A_\Omega \exp(-i\Omega t)\propto \exp(\gamma t)$, where $A_\Omega$ is an integration constant determined by initial conditions. The analogous equation for the mode with frequency $\Omega^*$ implies a decaying mode amplitude $a_{\Omega^*} = A_{\Omega^*}\exp(-i \Omega^* t)\propto\exp(-\gamma t)$.
 
 	The system energy can be found in terms of the mode amplitudes by substitution of Eq.~\eqref{a56nn} into Eq.~\eqref{a1}, just as for a stable system. However, when orthogonality of the eigenmodes is applied, the energy is no-longer perfectly diagonalized:
\begin{equation}
H = \sum_{\omega>0} H_\omega + H_u,
\end{equation}
where $H_u$, the unstable mode contribution to the energy, is 
\begin{equation}
H_u = 2Re\left\{ a_\Omega a_{\Omega^*}^* ({\bf u}_{\Omega^*},{\bf u}_\Omega) \right\} + 2 Re\left\{a_\Omega {\bf f}\cdot{\bf u}_\Omega  +a_{\Omega^*} {\bf f}\cdot{\bf u}_{\Omega^*}\right\}, 
\end{equation}
and where the stable mode contribution $H_\omega$ is unchanged, given by Eq.~\eqref{a16}.  The non-diagonal form of $H_u$ is required by energy conservation. For an unforced system, one can see that although $a_\Omega(t)$ and $a_{\Omega^*}(t)$ have differing time dependences $\exp(-i\Omega t)$ and $\exp(-i\Omega^* t)$ respectively (the former growing and the latter decaying), the combination
$a_\Omega a_{\Omega^*}^*$ is time-independent, as required for an energy-conserving system.

	Just as for a stable system, the energy can be formulated as a  Hamilitonian when the proper canonical coordinates are introduced. First, we consider the Poisson bracket $[a_\Omega, a_\omega^*]$ for $Re\omega >0$. Using Eq.~\eqref{a58} and the same series of steps as led to Eq.~\eqref{16.6}, we obtain
	\begin{equation}
	[a_\Omega, a_\omega^*] =\left\{
	\begin{array}{c c}
	\frac{-i\Omega}{({\bf u}_{\Omega^*},{\bf u}_\Omega)}, & \omega = \Omega^*, \notag \\
	0  & \text{otherwise}. \notag
	\end{array}
	\right.,
	\end{equation}
and similarly, $[a_\Omega, a_\omega]=0$ for all eigenfrequencies $\omega$ with $Re \omega>0$. For canonical coordinates, we therefore choose normalization  ${\bf u}_{\Omega^*}$ and ${\bf u}_{\Omega}$ such that 
\begin{equation}\label{a63nn}
({\bf u}_{\Omega^*},{\bf u}_\Omega) = \Omega.
\end{equation}
It is possible for the inner product in Eq.~\eqref{a63nn} to evaluate to the  complex frequency $\Omega$ because the vectors appearing in the inner product are different, with different normalization coefficients. We then introduce real-valued canonical pairs $(Q_\Omega, P_\Omega)$ and $(Q_{\Omega^*}, P_{\Omega^*})$ via the linear transformation
\begin{align}
a_\Omega &= \frac{Q_\Omega + i P_{\Omega^*}}{\sqrt{2}}, \label{a64nn}\\
a_{\Omega^*} &= \frac{Q_{\Omega^*} + i P_{\Omega}}{\sqrt{2}}. \label{a65nn}
\end{align}
With these choices one can easily show  that $[Q_\Omega,P_\Omega]=[Q_{\Omega^*},P_{\Omega^*}]=1$ and $[Q_\Omega,Q_\omega]=[P_\Omega,P_\omega]=0$.  When written in terms of these variables the unstable mode Hamiltonian is 
\begin{align}
H_u &= \Omega_r  (Q_\Omega Q_{\Omega^*} + P_\Omega P_{\Omega^*} ) + \gamma (Q_\Omega P_\Omega - Q_{\Omega^*} P_{\Omega^*}), \notag \\
   & + \sqrt{2}{\bf f}\cdot(Q_\Omega Re{\bf u}_\Omega - P_{\Omega^*}Im {\bf u}_\Omega + Q_{\Omega^*} Re{\bf u}_{\Omega^*} - P_\Omega Im{\bf u}_{\Omega^*}),
\end{align}
where we have also employed Eq.~\eqref{a63nn} and have taken $\Omega = \Omega_r + i\gamma$. Hamiltons equations for the complex-frequency modes then yield
\begin{align}
\dot Q_\Omega &=\frac{\partial H_u}{\partial P_\Omega} = \Omega_r P_{\Omega^*} + \gamma Q_\Omega - \sqrt{2}{\bf f}\cdot Im{\bf u}_{\Omega^*}, \notag \\
\dot P_{\Omega^*} &=-\frac{\partial H_u}{\partial Q_{\Omega^*}} = -\Omega_r Q_{\Omega} + \gamma P_{\Omega^*} - \sqrt{2}{\bf f}\cdot Re{\bf u}_{\Omega^*}, \notag \\
\dot Q_{\Omega^*} &=\frac{\partial H_u}{\partial P_{\Omega^*}} = \Omega_r P_{\Omega} - \gamma Q_{\Omega^*} - \sqrt{2}{\bf f}\cdot Im{\bf u}_\Omega, \notag \\
\dot P_{\Omega} &=-\frac{\partial H_u}{\partial Q_{\Omega}} = -\Omega_r Q_{\Omega^*} - \gamma P_{\Omega} - \sqrt{2}{\bf f}\cdot Re{\bf u}_{\Omega}. \notag 
\end{align}
The first two equations represent the dynamics of the unstable mode, and may be seen to agree with Eqs.~\eqref{a59}  and \eqref{a59.5}  once one applies Eqs.~\eqref{a63nn}, \eqref{a64nn} and \eqref{a65nn}. Similarly, the last two equations determine the dynamics of the exponentially decaying mode and may be seen to agree with Eq.~\eqref{a59} upon replacing $\Omega\rightarrow\Omega^*$ in this equation.

Finally, we briefly mention a few salient points regarding the special case  $\Omega_r = 0$. In this case the eigenvalue problem for the unstable mode with frequency $\Omega = i \gamma$ can be written $\gamma {\bf u}_{i \gamma} = {\bf D}\cdot{\bf u}_{i \gamma}$. Since $\gamma$ and $\bf D$ are  real, 
the unstable eigenvector ${\bf u}_{i \gamma}$ is also real. The corresponding mode with frequency $\Omega^* = -i\gamma$  also has a real eigenvector ${\bf u}_{-i \gamma}$. Since $\Omega = -\Omega^*$ these two modes are already paired according to property 3, and there are no other associated complex eigenmodes.  Thus, Eq.~\eqref{a56nn} becomes
 \begin{align}\label{a56n1}
 {\bf z}(t) =& \sum_{\omega>0} a_\omega(t) {\bf u}_\omega + cc. + a_{i \gamma}(t) {\bf u}_{i \gamma} + a_{-i \gamma}(t) {\bf u}_{-i \gamma},
 \end{align}
 with $({\bf u}_{i \gamma},  {\bf u}_{i \gamma})=({\bf u}_{-i \gamma},  {\bf u}_{-i \gamma})=0$ according to property 2, but $({\bf u}_{-i \gamma},  {\bf u}_{i \gamma})\ne 0$ according to property 1. 
Following through with the rest of the algebra we find that $ a_{i \gamma(t)}$ and $ a_{-i \gamma}(t)$ are real; that the  normalization condition for canonical coordinates is
$({\bf u}_{-i \gamma},  {\bf u}_{i \gamma}) = \gamma$;  that the canonical coordinates can be chosen as $Q_\gamma = a_{i \gamma}, P_{\gamma}=a_{-i \gamma}$; and that for this choice the  unstable mode Hamiltonian is
\begin{align}
H_u &= \gamma Q_\gamma P_\gamma  + {\bf f}\cdot(Q_\gamma {\bf u}_{i \gamma}  +P_\gamma {\bf u}_{-i \gamma}).
\end{align}
This Hamiltonian leads to the equations of motion
\begin{align}
\dot Q_\gamma &=\frac{\partial H_u}{\partial P_\gamma} = \gamma Q_\gamma + {\bf f}\cdot{\bf u}_{-i \gamma}, \notag \\
\dot P_\gamma &=-\frac{\partial H_u}{\partial Q_\gamma} = -\gamma P_\gamma - {\bf f}\cdot{\bf u}_{i \gamma}, \notag 
\end{align}
which agree with Eq.~\eqref{a59} and \eqref{a59.5} when $\Omega = \pm i\gamma$. The first equation represents the dynamics of the unstable growing mode,  with the second equation corresponding to the exponentially-decaying mode.

This completes our discussion of the modes of an unstable linearized Hamiltonian system.

\section{Normal modes of an ion crystal. }

As an example of the Hamiltonian approach outlined in the previous section, consider the dynamics of $N$ positive charges confined in the fields of a Penning trap:  a uniform magnetic field ${\bf B}= -B \hat z$, with $B>0$, 
and  an electrostatic trap potential $\phi_0(r,z)$ that is confining in the $z$ direction for positive charges. In some experiments this potential is nearly a pure quadrupole, $\phi_0(r,z) =(1/2)  E_0 (z^2 - r^2/2)$, where $E_0>0$, and this form will be used in our examples. However, this quadrupole form  is not necessary in the general theory described below. Each particle has mass $m_i$ and charge $q_i>0$, and position ${\bf r}_i$, $i=1,...,N$.  (For negative charges in the trap, remove the $-$ sign from $\bf B$ and add a $-$ sign to  $\phi_0$ so that $B$ and $E_0$ remain positive, and treat $q_i>0$ as the magnitude of each  charge. This preserves the signs for all the subsequent coefficients and formulas used in this section.) 

The charges are assumed to rotate about the $z$~axis with some mean rotation frequency $\omega_r>0$,  (i.e.  the rotation is in the positive $\phi$ direction).
In a frame rotating with the charges, the system Hamiltonian is
\begin{equation}
H = \sum_i \frac{ ({\bf p_i} - m_i \Omega_i {\bf A}({\bf r}_i))^2}{2 m_i} + \Phi({\bf r}_1,...,{\bf r}_N),
\end{equation}
Here the canonical momentum for particle $i$ is ${\bf p}_i = m_i\dot{\bf r}_i + m_i \Omega_i {\bf A}({\bf r}_i)$ where $\Omega_i = q_i B/(m_i c) -2\omega_r$ is the ``vortex frequency" for particle $i$ (the cyclotron frequency shifted by Coriolis effects)  and ${\bf A}({\bf r})$ is the scaled magnetic vector potential, defined so that $\nabla\times{\bf A} = -\hat z$, where $\hat z$ is the unit vector in the z direction. A useful gauge choice for $\bf A$ is the cylindrically-symmetric gauge ${\bf A} = -(1/2)\hat z\times{\bf r}$. This choice of gauge makes
\begin{equation}\label{a57nn}
{\bf p}_i =  m_i\dot{\bf r}_i - \frac{1}{2} m_i \Omega_i \hat z\times{\bf r}_i.
\end{equation}
The function $\Phi$ is the total electrostatic potential energy  of the system (as seen in the rotating frame) given by 
\begin{equation}\label{a57n}
\Phi = \sum_{i>j} \phi_{i j} + \sum_i q_i\phi_i(r_i,z_i),
\end{equation}
 where  $\phi_{i j} = q_i q_j/|{\bf r}_i - {\bf r}_j| $
 is the electrostatic Coulomb  potential between particles $i$ and $j$ (neglecting for simplicity image charge effects in the surrounding electrodes), and  
 \begin{equation} \label{a58n}
 q_i \phi_i(r_,z)= q_i \phi_0 (r,z) + \frac{\omega_r }{2}\left(\frac{q_i B}{c} - m_i \omega_r\right)r^2
 \end{equation}
 is the effective external  potential energy for charge $i$ as seen in the rotating frame, including both the force from rotation through the magnetic field and centrifugal force. For a quadrupole trap potential this can be written as 
  $q_i \phi_i(r_,z)= (1/2) q_i E_0 (z^2 + \beta_i r^2)$, where the trap parameter $\beta_i \equiv \omega_r (B/c - m_i\omega_r/q_i)/E_0 - 1/2$.  

\begin{figure}[!tbp]

{\includegraphics[width = 2.12in]{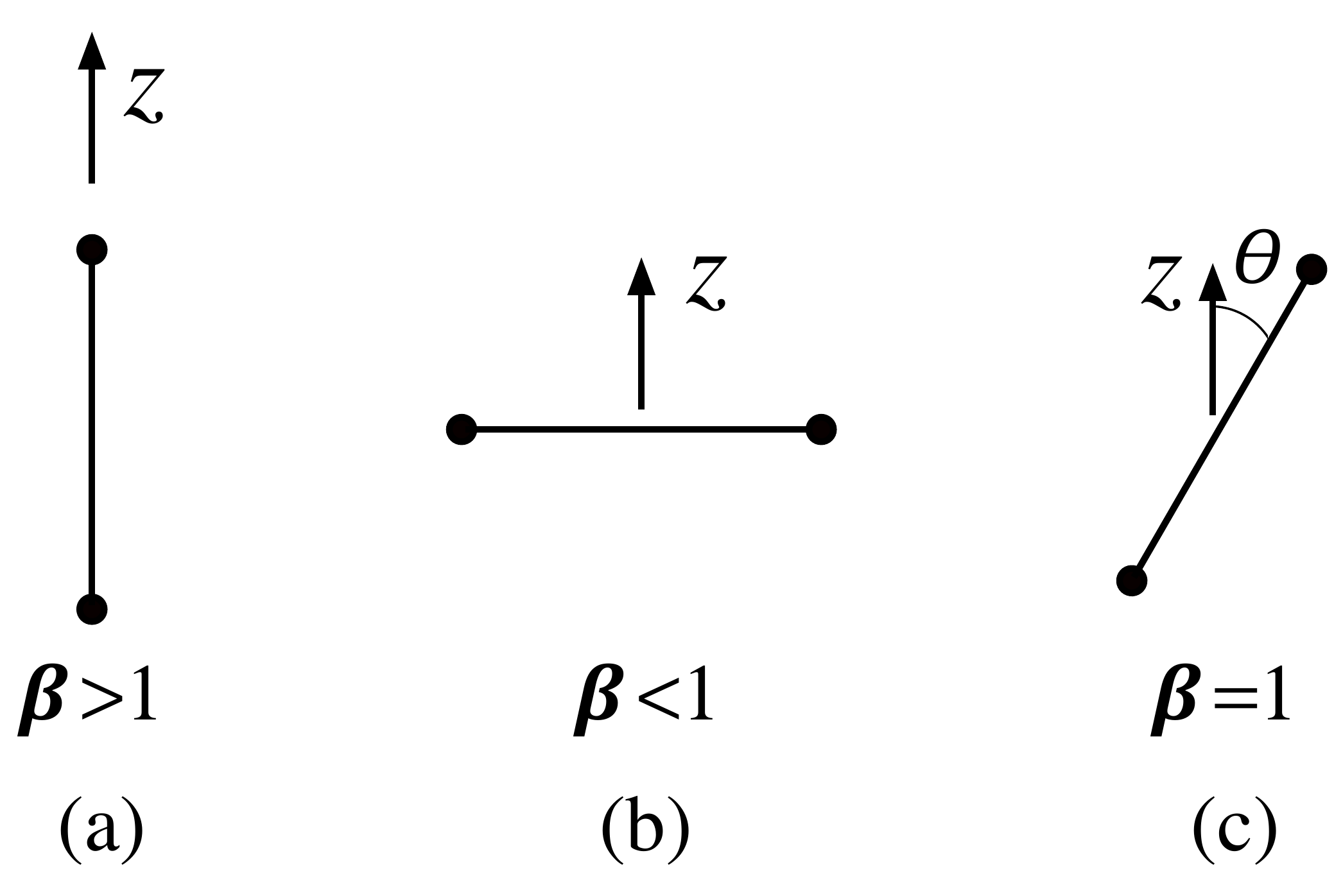}}

\caption{\label{fig0} Equilibrium of 2 identical charges in a quadrupole trap}
  \end{figure}

	For positive values of the trap parameters $\beta_i$ and $E_0$ (or, more generally, for a potential $q_i \phi_i(r,z)$ that increases from the trap center with both increasing  $r$ and $z$) a (neutrally) stable ion crystal equilibrium exists with $\dot{\bf r}_i = \bf 0$ and   ${\bf r}_i={\bf R}_i$ for equilibrium positions ${\bf R}_i$ satisfying $\partial\Phi/\partial {\bf R}_i = {\bf 0}, i=1, ..., N$. (In fact for $N\gg 1$ there are typically many such equilibria corresponding to different crystalline configurations with slightly different arrangements of the charges.) These crystal equilibria have been discussed in some detail in several previous publications.\cite{dubinoneil,schiffer,dubinRMP,mitchell} We will consider two examples in detail. Figure~\ref{fig0} displays the simplest nontrivial crystal consisting of two identical charges in a quadrupole trap, a 2-ion Coulomb cluster\cite{wineland,rafac,cornell}. When the trap parametrer $\beta$ is  greater than one, the equilibrium has the charges on the $\pm z$ axis, each at a distance $d=(q/ 4E_0)^{1/3}$ from the origin. When $\beta<1$, the ions are on opposite sides of the origin in the $x-y$ plane, each a distance $d/\beta^{1/3}$ from the origin, and for $\beta = 1$ the charges can be at any angle $\theta$ with respect to the $z$ axis. This Coulomb cluster can be thought of as a classical version of a symmetric molecule such as $H_2$ or $N_2$, in which the electrons are replaced by a neutralizing background charge (the ``plum-pudding" model of J.J. Thompson). Later in this section we will analytically evaluate the normal modes for this system, including the effect of the magnetic field. 
	
\begin{figure}[!tbp]

{\includegraphics[width = 2.12in]{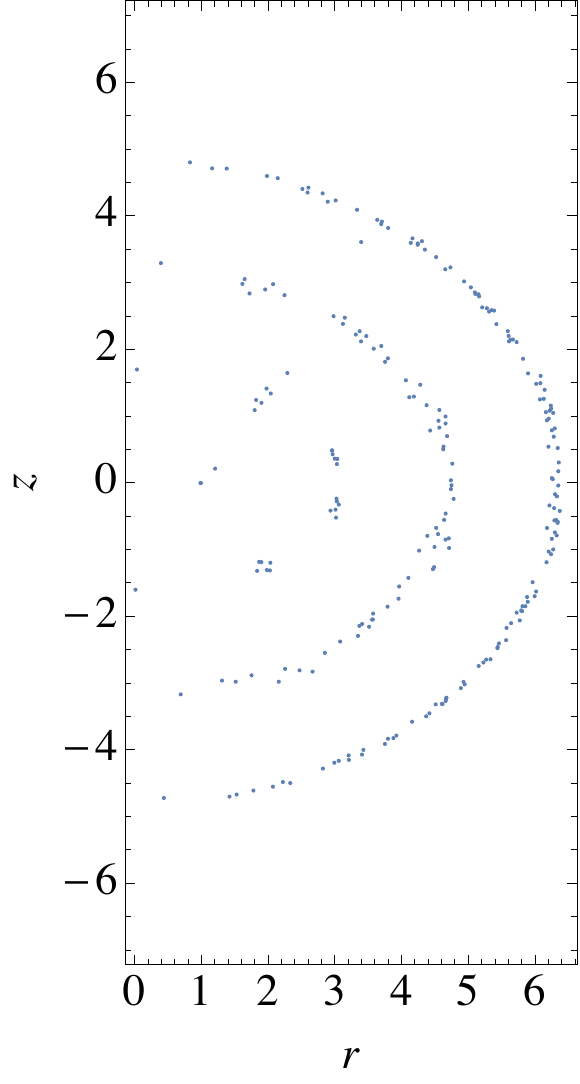}}

\caption{\label{fig1} $r-z$ positions in a spheroidal Coulomb crystal of $N=236$ identical charges in a quadrupolar Penning trap with trap parameter $\beta = 3/4$. Distances are  in terms of the distance $ (q/E_0)^{1/3}$. }
  \end{figure}

	An equilibrium configuration with larger $N$ is displayed in   Fig.~\ref{fig1}, which shows the $r-z$ positions of a crystal (local minimum energy state)  consisting of $N=236$ identical charges in a quadrupole trap with trap parameter $\beta = 3/4$. The charges tend to arrange themselves in spheroidal shells\cite{schiffer,dubinoneil} with an average density that is determined by the rotation rate and the external trap fields. For $\beta<1$ the system tends to form an oblate spheroid, which for sufficiently small $\beta$ collapses into the $z=0$ plane. This particular regime of a single-plane plasma crystal is currently of interest as a useful system for the purposes of quantum simulation.\cite{bollinger, bollinger2, ball} For  $\beta>1$ the system forms a prolate spheroid and for sufficiently large $\beta$ the system forms a one-dimensional Coulomb string of charges distributed along the $z$ axis.\cite{dubinstruc,schifferstruc}
	
	These equilibria are all neutrally stable with respect to rotations about the $z$ axis.  When $\beta = 1$ the spherical symmetry of the effective trap potential implies that rotations about $x$ and $y$ axes are also neutral modes. 

	In what follows, we assume that particles are displaced only slightly from one such equilibrium, with positions ${\bf r}_i = {\bf R}_i +\delta {\bf r}_i$ and momenta  ${\bf p}_i =  m_i \Omega_i {\bf A}({\bf R}_i)+ \delta{\bf p}_i $. Taylor expansion to second order in the small displacements from equilibrium then results in a linearized Hamiltonian system, with Hamiltonian
	\begin{equation}
H = \frac{1}{2} \left( \sum_i \frac{ (\delta {\bf p_i} + m_i \Omega_i \hat z\times\delta {\bf r}_i/2)^2}{m_i} + \sum_{i j}\delta {\bf r}_i\cdot{\bf V}_{i j}\cdot\delta{\bf r}_j\right),
\end{equation}  
where ${\bf V}_{i j} = \partial^2\Phi/\partial{\bf R}_i\partial{\bf R}_j$. The form of ${\bf V}_{i j}$ is given in Appendix A.

This Hamiltonian can be put in the matrix form of Eq.~\eqref{a1} with ${\bf f} = \bf 0$, with phase space vector ${\bf z} =({\bf r},{\bf p})$ where ${\bf r} =  (\delta{\bf r}_1,...,\delta{\bf r}_N)$ and ${\bf p}= (\delta{\bf p}_1,...,\delta{\bf p}_N)$,   and with the symmetric  Hamiltonian matrix given in block form by
\begin{equation}\label{a119n1}
{\bf H} =\left(
\begin{array}{c c}
{\bf V + C} &{\bf \Omega} \\
{\bf \Omega}^{tr} &{\bf M}^{-1}
\end{array}
\right).
\end{equation}
Here  ${\bf V}$,  ${\bf C}$, ${\bf \Omega}$ and ${\bf M}^{-1}$ are $3N\times 3N$ matrices.  The matrix ${\bf M}^{-1}$ is the inverse of the diagonal  mass matrix $\bf M$ for the system, with the diagonal elements $M^{-1}_{ i i}$ given by the vector $(m_1^{-1}, m_1^{-1}, m_1^{-1}, ..., m_N^{-1}, m_N^{-1}, m_N^{-1})$. The matrix $\bf \Omega$  is the Lorentz matrix coupling positions and momenta in the Hamiltonian. For our symmetric choice of vector potential the Lorentz matrix is antisymmetric, and is zero everywhere except in $3\times 3$ blocks along the diagonal:
\begin{equation}\label{a120n1}
{\bf \Omega} = \left(
\begin{array} {c c c}
{\bf \Omega}_1 &  &0 \\
  & \ddots &   \\
  0&  & {\bf\Omega}_N
 \end{array}
 \right)
 \end{equation}
Each diagonal block is, in dyadic notation, given by  
\begin{equation}\label{a63}
{\bf \Omega}_i = \frac{1}{2}\Omega_i (\hat x \hat y - \hat y\hat x),\ \  i=1,...,N.
\end{equation}
The symmetric matrix ${\bf V}=\partial^2\Phi/\partial{\bf R}\partial{\bf R}$  is the potential energy matrix given in block form by
\begin{equation}
{\bf V} = \left(
\begin{array}{c c c c}
{\bf V}_{1 1} & {\bf V}_{1 2} & ...&{\bf V}_{1 N} \\
{\bf V}_{2 1} & {\bf V}_{2 2} & ... &{\bf V}_{2 N}\\
... &  &  & \\
{\bf V}_{N 1} & {\bf V}_{N 2} & ... &{\bf V}_{N N}\\
\end{array}
\right),
\end{equation}
The matrix $\bf C$  is a magnetic potential contribution whose elements are  zero everywhere except along the diagonal. The  vector of diagonal elements $C_{i i}$ is given by 
$$\frac{1}{4} (m_1\Omega_1^2,m_1\Omega_1^2,0,...,m_N\Omega_N^2,m_N\Omega_N^2, 0).$$

The dynamical matrix ${\bf D}={\bf J}\cdot{\bf H}$ corresponding to this Hamiltonian matrix is
\begin{equation} \label{a65}
{\bf D} = \left(
\begin{array}{c c}
-{\bf \Omega} &{\bf M}^{-1} \\
-{\bf V}-{\bf C} & -{\bf \Omega}
\end{array}
\right)
\end{equation}
where we used the antisymmetry of the Lorentz matrix to write ${\bf \Omega}^{tr} = - \bf \Omega$. The eigenvalues and eigenvectors of $\bf D$ provide us with the normal modes of the system, as per Eq.~\eqref{a5}.  As discussed in Sec. II these modes diagonalize the system energy. 

It is well-known that a few of the eigenmodes have simple analytic descriptions. In a pure quadrupole trap with a single species, there are three ``center of mass" (COM) modes that consist of a displacement of the entire crystal. The axial COM mode consists of an oscillation in the $z$ direction and has frequency $\omega_z$ where $\omega_z\equiv \sqrt{q E_0/m}$ is  referred to in the literature as the single particle axial frequency; it is the frequency at which a single trapped particle oscillates in $z$ when displaced from the origin.  The cyclotron and $E\times B$ COM modes consist of rotational motions of the center of mass on the $x-y$ plane, with frequencies $\omega_{+}$ and $\omega_{-}$ respectively,  where
 \begin{equation}
 \omega_{\pm} = \sqrt{\Omega^2 + 2\omega_\bot^2 \pm\Omega\sqrt{\Omega^2+4\omega_\bot^2}}/\sqrt{2}
  \end{equation}
 and where $\omega_\bot\equiv\sqrt{\beta}\omega_z$
is the single particle transverse frequency in an unmagnetized trap. All three COM mode frequencies are independent of the number of charges in the trap. 

In addition, in a cylindrically-symmetric trap potential (quadrupolar or not) there is  a zero frequency eigenmode which is a pure rigid rotation about the z axis, with eigenvector ${\bf u}_{0 z} = ({\bf r}_{0 z},{\bf p}_{0 z})$
where 
\begin{align}
{\bf r}_{0 z} &= (\hat z\times {\bf R}_1,...,\hat z\times {\bf R}_N), \notag \\  
{\bf p}_{0 z} &=  -\frac{1}{2}(m_1\Omega_1 \hat z\times\hat z\times  {\bf R}_1, ...,m_N \Omega_N \hat z\times\hat z\times {\bf R}_N)\label{eig0}\\
 & = \frac{1}{2}(m_1\Omega_1 R_1\hat r,..., m_N \Omega_N R_N\hat r),\notag 
\end{align} 
and where $R_j$ is the cylindrical radius of equilibrium position ${\bf R}_j$ for the $j$th ion.  As was discussed previously in more general terms, this eigenmode corresponds to a  constant of the motion, the momentum $P_{0 z}$ given by Eq.~\eqref{a28}:
\begin{align}\label{a67}
P_{0 z} &= {\bf u}_{0 z}\cdot{\bf J}\cdot{\bf z} = {\bf r}_{0 z}\cdot{\bf p}-{\bf p}_{0 z}\cdot{\bf r} \notag \\
& = \sum_i  \left(R_i \hat\phi\cdot\delta{\bf p}_i - \frac{1}{2} m_i \Omega_i R_i \hat r\cdot \delta {\bf r}_i\right) \notag \\
& = \sum_i \left( m_i R_i ^2\delta\dot\phi_i -  m_i R_i \Omega_i\delta r_i\right),
\end{align}
where in the last step we substituted for $\delta {\bf p}_i$ using Eq.~\eqref{a57nn}.  The constant $P_{0 z}$ is, of course, the perturbed total canonical angular momentum associated with rotations about the $z$ axis. 

The corresponding vector $\bar{\bf u}_{0 z}$, required for the rotational inertia $(\bar{\bf u}_{0 z}, \bar{\bf u}_{0 z})$ (see Eq.~\eqref{a53}),  is the solution of Eq.~\eqref{a32}. In general this equation requires a numerical solution but it might also be of interest to note that there is a case where
$\bar {\bf u}_{0 z}$ can be evaluated analytically: when the equilibrium consists of identical charges trapped  in a quadrupolar trap with trap parameter $\beta$ chosen to be sufficiently small so that the crystal equilibrium is a planar crystal confined to  the $z=0$ plane. In this case (see Appendix C) 
\begin{equation}\label{a128nn}
\bar{\bf u}_{0 z} = \left( -\frac{2{\bf p}_{0 z}}{3 m\omega_\bot^2}, (1 + \frac{\Omega^2}{6\omega_\bot^2}){{\bf r}_{0 z}}  \right)
\end{equation}
 and  the rotational inertia $(\bar{\bf u}_{0 z}, \bar{\bf u}_{0 z})$  is then given by the expression
\begin{equation}\label{a129nn}
(\bar{\bf u}_{0 z}, \bar{\bf u}_{0 z}) =m  \sum_i R_i^2\left(1 + \frac{ \Omega^2}{3 \omega_\bot^2 }\right)
\end{equation}
The first term in the parenthesis gives  the usual kinetic inertia associated with rigid rotation, while the second term (which can dominate in strong magnetic fields) arises from electrostatic energy associated with compression/expansion of the crystal as the rotation rate changes.

A few other cases allow analytic solution  for all of the modes. One example is  the $N=2$  Coulomb cluster shown in Fig.~\ref{fig0}. Assuming that the particles are aligned in equilibrium in the $x-z$ plane, with particle $1$ above the $z=0$ plane and particle $2$ below the plane, the potential matrix $\bf V$ evaluates to
\begin{equation}
\frac{{\bf V}}{m\omega_z^2} =
\left(
\begin{array}{c c c c c c}
\beta + \frac{1}{4}(1-3\cos2\theta)& 0 & \frac{3}{4}\sin2\theta & -\frac{1}{4}(1-3\cos2\theta)& 0  &-\frac{3}{4}\sin2\theta \\
 & \beta-1/2 & 0 & 0 &\beta - 1/2 & 0 \\
&  &    \frac{1}{4}(5+3\cos2\theta) & -\frac{3}{4}\sin2\theta & 0  & -\frac{1}{4}(1+3\cos2\theta) \\
&  & & \beta + \frac{1}{4}(1-3\cos2\theta)& 0  &\frac{3}{4}\sin2\theta \\
& &  & &\beta - 1/2 & 0 \\
& &  & &  & \frac{1}{4}(5+3\cos2\theta)  \\
\end{array}
\right).
\end{equation}
The frequencies and corresponding eigenvectors are provided in Table 1. These eigenvectors are not normalized. 

We first consider $\beta>1$, where the charges align along the $z$ axis in equilibrium, with $\theta = 0$. In this case there is no zero frequency rotational mode. In addition to the three center of mass modes there are three other modes in which the  charges perform opposite motions, $\delta{\bf r}_1 = -\delta{\bf r}_2$. One of these is an axial stretch mode only along the $z$ axis, with frequency $\sqrt{3}\omega_z$. The other two modes consist of circular motion in $x$ and $y$ at the frequencies $\omega_{r+}$ and $\omega_{r-}$ where
\begin{equation}\label{a125n1}
\omega_{r\pm}=\sqrt{\Omega^2 + 2(\beta-1)\omega_z^2 \pm\Omega\sqrt{\Omega^2+4(\beta-1)\omega_z^2}}/\sqrt{2}.
\end{equation}

As $\beta\rightarrow 1, \omega_{r -}$ approaches zero frequency and the eigenvector corresponds to a sum of rotations about the $x$ and $y$ axes, which are neutral modes in the spherically-symmetric $\beta = 1$ limit.

\begin{table}[h!]
  \begin{center}
    \caption{Eigenmodes for two identical charges in a quadrupole trap}
    \label{tab:table1}
    \begin{tabular}{c c c}
      $\omega $ & ${\bf u}_\omega$ &  $({\bf u}_\omega,{\bf u}_\omega)$ \\
      \hline
                    & COM modes: &  \\
      \hline
      $\omega_z$ & \ \ $(0,0,1,0,0,1,0,0,-i m\omega,0,0,-i m\omega)$ \ \ & $4 m\omega^2$\\
       $\omega_\pm$ & \ \ {\tiny $(1,\pm i,0,1,\pm i,0,-i m\omega \pm i \frac{m\Omega}{2},\pm m\omega - \frac{m\Omega}{2},0,-i m\omega \pm i \frac{m\Omega}{2},\pm m\omega - \frac{m\Omega}{2},0)$} \ \ & $4m(\omega^2+\omega_\bot^2)$\\
       \hline
                           & other modes for $\beta>1$: &  \\
                           \hline
      $\sqrt{3}\omega_z$ & $(0,0,1,0,0,-1,0,0,-im\omega,0,0,im\omega)$ &  $4 m \omega^2$    \\ 
      $\omega_{r \pm}$ & \ \ {\tiny $(1,\pm i,0,-1,\mp i,0,-i m\omega \pm i \frac{m\Omega}{2},\pm m\omega - \frac{m\Omega}{2},0,i m\omega \mp i \frac{m\Omega}{2},\mp m\omega + \frac{m\Omega}{2},0)$} \ \ & $4m(\omega^2+(\beta-1)\omega_z^2)$\\

       \hline
                           & other modes for $\beta<1$: &  \\
                           \hline
$\omega_z\sqrt{1-\beta}$ & $(0,0,1,0,0,-1,0,0,-i \omega,0,0,i\omega) $ & $4 m \omega^2$ \\

     $\sqrt{\Omega^2+3\omega_\bot^2} $ &{\tiny $(1,\frac{i\Omega}{\omega},0,-1,-i\frac{\Omega}{\omega},0, -im\omega+i\frac{m\Omega^2}{2\omega}, \frac{m\Omega}{2},0, im\omega-i\frac{m\Omega^2}{2\omega}, -\frac{m\Omega}{2},0   )$} &  $4m\omega^2$    \\ 

$ 0$ & $ (0,1,0,0,-1,0,-\frac{m \Omega}{2},0,0,\frac{m \Omega}{2},0,0)$ & $0$ \\
       \hline
                           & other modes for $\beta=1$: &  \\
                           \hline
 $\omega_{\theta \pm}$ & $({\bf r}_\theta,-{\bf r}_\theta,{\bf p}_{\theta},-{\bf p}_{\theta})$ &   $\frac{4m}{3} \omega^2 \frac{\Omega^4 -\omega^2(\Omega^2- 3\omega_z^2)-3\Omega^2\omega_z^2\cos2\theta}{\omega_z^2\sin^2\theta}$ \\


 & $ { \bf r}_\theta = ( -i\omega, \Omega,-2i\omega\frac{\omega^2-\Omega^2-3\omega_z^2\sin^2\theta}{3\omega_z^2\sin2\theta} ) $ &  \\
& ${\bf p}_{\theta} = ( \frac{\Omega^2}{2} -\omega^2,-i \frac{\omega \Omega}{2}, -2  \omega^2 \  \frac{\omega^2-\Omega^2-3\omega_z^2\sin^2\theta}{3\omega_z^2 \sin2\theta})$& \\

 $ 0$ & ${\bf u}_{0 z} = (0,1,0,0,-1,0,-\frac{m \Omega}{2},0,0,\frac{m \Omega}{2},0,0)$ & $0$ \\
$ 0 $ &  ${\bf u}_{0 y}= (\cos\theta,0,-\sin\theta,-\cos\theta,0,\sin\theta,0,-\frac{m\Omega}{2}\cos\theta,0,0,\frac{m\Omega}{2}\cos\theta,0)$ &  0 
     \end{tabular}
  \end{center}
\end{table}
The energy takes the standard form for a stable system (see Eq.~\eqref{a15}):
\begin{equation}\label{ham}
H=\sum_{\omega>0} |a_\omega|^2 ({\bf u}_\omega,{\bf u}_\omega),
\end{equation}
with the coefficients $({\bf u}_\omega,{\bf u}_\omega)$ given in the Table.

On the other hand, for $\beta<1$ the equilibrium is in the $x-y$ plane.  Now there is a zero frequency rotation about the $z$ axis in addition to the three center of mass modes. The associated constant of the motion is the angular momentum given by Eq.~\eqref{a67}.  For the two ion system it is convenient to divide out the radii $R=d/\beta^{1/3}$ of the ions, defining $P'_{0 z}=P_{0 z}/R=  \delta p_{1 y}-\delta p_{2 y} - m \Omega(\delta x_1 -\delta x_2)$. 
In addition we introduce the normalized vector $\bar{\bf u}'_0=\bar{\bf u}_0/R$ with $\bar{\bf u}_0$ given by Eq.~\eqref{a128nn}.  
The two other modes are a tilt mode consisting only of axial motion with $\delta z_1 = -\delta z_2$, at frequency $\omega_z\sqrt{1-\beta}$, and an ``upper-hybrid" mode consisting of elliptical motion of the charges in the $x-y$ plane at frequency $\sqrt{\Omega^2 + 3\omega_\bot^2}$. 
Due to the neutral mode the energy takes the form
\begin{equation}\label{a131new}
H=\sum_{\omega>0} |a_\omega|^2 ({\bf u}_\omega,{\bf u}_\omega) +\frac{1}{2} \frac{P_{0 z}'^2}{(\bar{\bf u}'_0,\bar{\bf u}'_0)}
\end{equation}
where the scaled moment of inertia  $(\bar{\bf u}'_0,\bar{\bf u}'_0)=2m (1+\Omega^2/3\omega_\bot^2)$, see Eq.~\eqref{a129nn}.

Finally, when $\beta = 1$ the system has spherical symmetry and there are now equilibria oriented at any angle $\theta$ with respect to the $z$ axis.  In addition to the center of mass modes, there are two zero frequency modes consisting of free rotations asbout the $y$ and $z$ axes, and two modes whose frequencies $\omega_{\theta+}$ and $\omega_{\theta -}$ depend
 on $\theta$:
 \begin{equation}
 \omega_{\theta \pm} = \sqrt{\frac{\Omega^2 + 3\omega_z^2 \pm\sqrt{(\Omega^2-3\omega_z^2)^2+12\Omega^2\omega^2_z\sin^2\theta}}{2}}.
 \end{equation}
As $\theta\rightarrow 0$, these frequencies approach $\omega_{r +}=\Omega$ and the stretch mode frequency $\sqrt{3}\omega_z$. As $\theta\rightarrow\pi/2$, $\omega_{\theta -} \rightarrow 0$ and  $\omega_{\theta +} \rightarrow \sqrt{\Omega^2+3\omega_z^2}$. 

The two zero frequency modes produce two constants of the motion, the angular momentum $P'_{0 z}$ due to rotation of the cluster about the $z$ axis (again dividing the radii $d \sin\theta$), and  angular momentum $P'_{0 y}$ due to rotations about the $y$ axis (also dividing out distance $d$  from the axis): 
\begin{equation}
P'_{0 y}=  \cos\theta(\delta p_{x 1} -\delta p_{x 2} +  \frac{1}{2}m\Omega (\delta y_1-\delta y_2))-\sin\theta(\delta p_{z1}-\delta p_{z2}).
\end{equation} 
These two constants are not in involution, $J'_{x y}= [P'_{0 y},P'_{0 z}]= 2m\Omega \cos\theta$.  Therefore the constants do not appear in the energy, so it takes the form given in Eq.~\eqref{ham}.

It is instructive to compare the evolution of the $\beta=1$ system to that of the $\beta<1$ system when the perturbed axial angular momentum $P'_{0 z}$ is nonzero. This comparison illustrates a physical difference between  systems with constants of the motion in involution and those for which the constants are not in involution. For $\beta<1$ Hamiltonian ~\eqref{a131new} implies that the angle variable $\delta\phi = a_0$ changes linearly with time according to 
\begin{equation}
\delta{\dot\phi} = \frac{P_{0 z}'}{(\bar{\bf u}'_0,\bar{\bf u}'_0)},
\end{equation}
  because a change in angular momentum corresponds to a change in the rotation frequency of the cluster. There is a finite scaled moment of inertia $(\bar{\bf u}'_0,\bar{\bf u}'_0)$ which relates the scaled angular momentum change $P_{0 z}'$ to the rotation frequency change $\delta{\dot\phi}$ . 

However, for $\beta = 1$, neither  $P'_{0 z}$  nor $P'_{0 y}$ appear in the Hamiltonian. The phase space configuration $\bf z$  evolves according to
\begin{equation}
{\bf z} =d\  \delta\theta {{\bf u}_{0 y}} + R\delta\phi{{\bf u}_{0 z}} + \sum_{\omega\ne 0}a_\omega {\bf u}_\omega,
\end{equation}
where ${\bf u}_{0 z}$ and ${\bf u}_{0 y}$ are given in Table 1, and where $ \delta\theta = -\frac{P'_{0 z}}{J'_{x y}d } = constant$ and $ \delta\phi=\frac{P'_{0 y}}{J'_{x y}R} =constant$ (see Eqs.~\eqref{b94}, \eqref{b96} and \eqref{b97}). The angle $\delta\phi$ does not evolve in time as it did for $\beta<1$.
This is because the axial angular momentum perturbation $P'_{0 z}$ does not change the rotation frequency  of the cluster. The moment of inertia $(\bar{\bf u}_{0 z},\bar{\bf u}_{0 z})$  is undefined, indeed, the vector $\bar{\bf u}_{0 z}$ does not exist, as discussed in Sec. IIbii. 
The angular momentum perturbation is instead accomplished by a rotation  $\delta\theta $  of the cluster about the $y$ axis, because for $\Omega\ne 0$ the
 canonical angular momentum depends on the cylindrical radius $R$ of the charges (through the vector potential term) , which varies as $\theta$ varies - see Fig.~\ref{fig0}(c). 
 
This surprising feature of the rotational inertia in magnetized spherically-symmetric crystals also occurs for larger Coulomb clusters, provided that the clusters canonical angular momentum depends on its orientation for spherical confinement. A few cases were discussed in Ref. ~\onlinecite{dubinRMP} in the context of a study of the configurations of minimum energy as angular momentum is varied.
 
\begin{figure}[!tbp]
{\includegraphics[width = 3.12in]{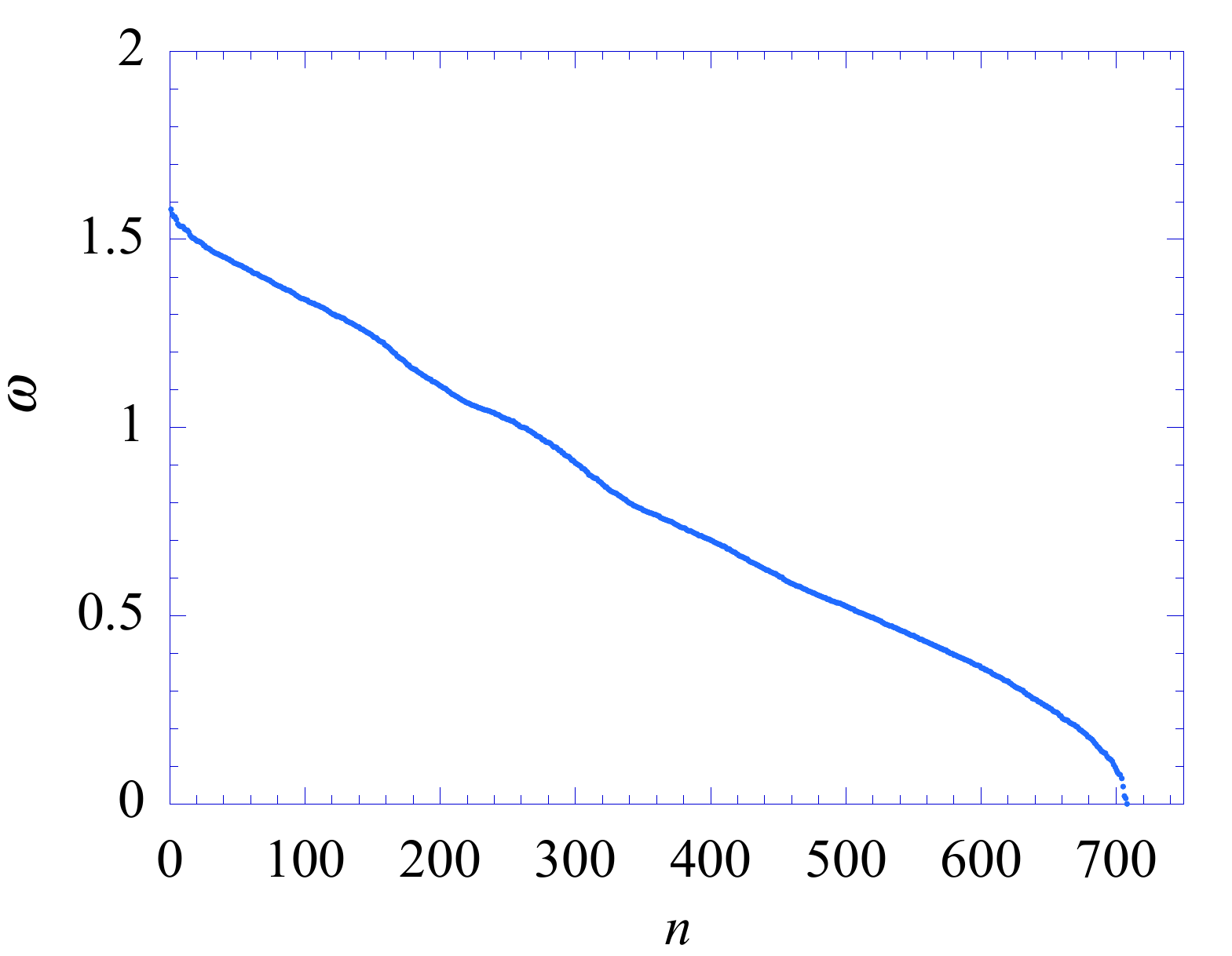}}
\caption{\label{fig2} Eigenfrequencies (in units of $\omega_z$)  for  the spheroidal Coulomb crystal  shown in Fig.~\ref{fig1} for vortex frequency $\Omega = 0$, counting in order from highest frequency to lowest (positive) frequency. }
  \end{figure}

\begin{figure}
{\includegraphics[width = 3.12in]{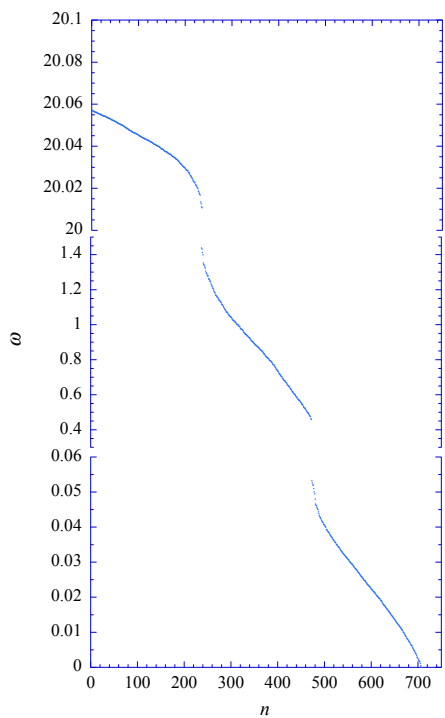}}
\caption{\label{fig2.5} Eigenfrequencies (in units of $\omega_z$)  for  the spheroidal Coulomb crystal  shown in Fig.~\ref{fig1} for vortex frequency $\Omega = 20\omega_z$, counting in order from highest frequency to lowest (positive) frequency. Note the breaks in the frequency axis. }
  \end{figure}

Turning now to an example with more charges, in  Figs.~\ref{fig2}  and \ref{fig2.5} we display the mode frequencies for the case of the spheroidal Coulomb crystal with $N=236$, shown in Fig~\ref{fig1}. We choose $\Omega=0$ in Fig. ~\ref{fig2}  and  $\Omega = 20 \omega_z$, in Fig.~\ref{fig2.5}. 
There is one zero frequency mode corresponding to  a rotation about the $z$ axis, with an associated constant of the motion corresponding to the perturbed angular momentum.
There are also several  modes with quite low frequencies, corresponding to torsional motions of the crystal with weak restoring forces. 

 For $\Omega = 20\omega_z$,  the mode frequencies condense into three groups: a group of $N$ cyclotron modes with $\omega>\Omega$; a group of $N$ axial modes with $\omega\approx 1$, and  a group of $N$ $E\times B$ modes with low frequencies. Variation of $\Omega$ indicates that these latter mode frequencies scale with $\Omega$ as $1/\Omega$.   These mode groupings have been identified in previous publications.\cite{chen, dubinmodes, dubinschiffer, freericks} There is one COM mode in each group, and there in one neutral mode in the $E\times B$ group associated with rotation about the $z$ axis.

Aside from a few special cases such as the center of mass and neutral modes, the eigenmodes have numerical forms the details of which vary depending on the precise crystal structure. Nevertheless, some of these modes are close to the sorts of modes predicted in a cold fluid theory of the normal mode oscillations.\cite{dubinmodes} In this theory, which treats the system as a uniform charged fluid, modes for a system trapped in a quadrupolar trap were worked out analytically in terms of Legendre functions. An initial condition consisting of displacements associated with a given cold fluid normal mode can be described as a superposition of  exact crystal eigenmodes, with only a small number of these eigenmodes dominating, provided that the mode is of low order (with a wavelength large compared to the interparticle spacing). \cite{dubinschiffer}

\begin{figure}[!tbp]

{\includegraphics[width = 3.12in]{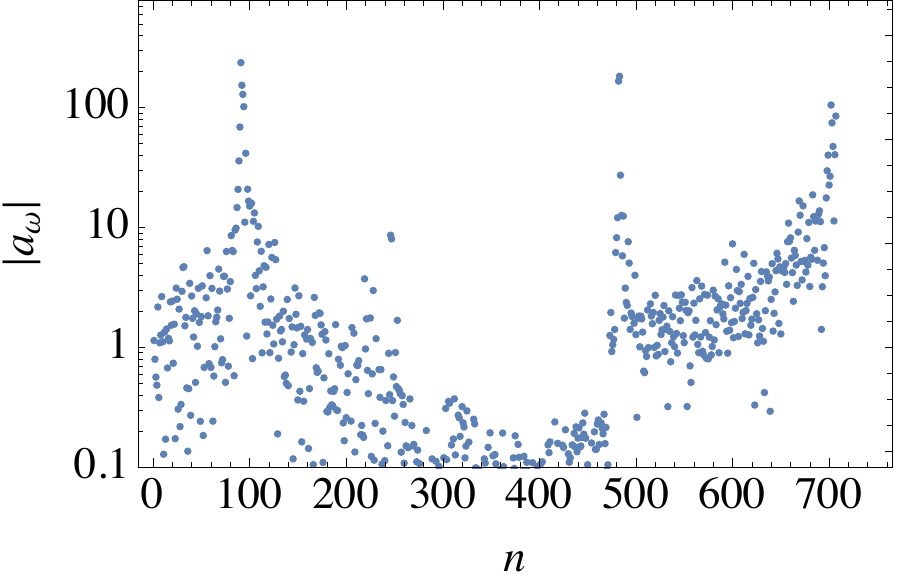}}

\caption{\label{fig3} The magnitude of eigenmode amplitudes $a_\omega$  plotted on a logarithmic scale versus mode number $n$, for an initial condition described in the text.}
  \end{figure}

A single example is displayed in Fig.~\eqref{fig3}. Here we consider a fluid displacement of  the form $\delta{\bf r}_{fluid} = (\delta x,\delta y, \delta z)_{fluid}\propto(x,-y,0)$. Such a displacement creates an ellipsoidal distortion of the crystal that, in fluid theory, is associated with two modes: a cyclotron frequency mode and a $E\times B$ diocotron mode, in which this distortion propagates around the plasma in the $\phi$ direction. For the case of a spheroidal plasma with $\beta = 3/4$  the frequencies of these two modes is predicted to 
be $\omega_{fluid} = \sqrt{\Omega^2/4+0.925253\omega_z^2}\pm\Omega/2 $.
In the figure, we determine the mode amplitude $a_\omega$ for each mode using Eq.~\eqref{a10}, with ${\bf z}$ given by $\delta{\bf r}_{fluid}$ for each particle, along with the associated canonical momentum $-m\Omega\hat z\times\delta{\bf r}_{fluid}/2$:
${\bf z} = (x_1,-y_1,0,...,x_N,-y_N,0,-m\Omega y_1/2,-m\Omega x_1/2,0,...,,-m\Omega y_N/2,-m\Omega x_N/2,0)$. This phase space configuration corresponds to an initially-stationary elliptical distortion.

The resulting plot of the magnitude of the amplitude of each eigenmode displays three strong peaks near  $n=90, 250, 480$. These frequency peaks dominate the dynamics as the system evolves from this initial condition. The frequencies corresponding to each peak are, respectively, $\omega/\omega_z = 20.0464, 1.3063, 0.0461$. These may be compared to $\omega_{fluid}$ which evaluates to $\omega_{fluid}/\omega_z = 20.0462, 0.0462$ for the cyclotron and $E\times B$ mode respectively.  The weaker of the three peaks at $\omega = 1.3063\omega_z$ does not correspond to either of these fluid modes, but is instead close to  the cylindrically-symmetric axial fluid mode (the ``(2,0)" mode)\cite{dubinmodes,bollingermodes} with fluid frequency $\omega_{fluid} = 1.3145\omega_z$ for these conditions. In the fluid theory this cylindrically-symmetric mode has no overlap with the $\theta$-dependent fluid displacement used here. Evidently a weak coupling to this  mode occurs due to the finite number of charges in the crystal. There also appears to be strong coupling to some of the very low frequency $E \times B$ modes, which is also not predicted in fluid theory.
  
\subsubsection{thermal averages and  zero frequency modes}
Thermal averages over various functions are of importance in many applications. For example, consider the average  $\langle \delta z_j \delta z_k\rangle$ for two particles $j$ and $k$. The $jth$ and $kth$ axial displacements appearing here correspond to the $\alpha$ and $\beta$ components of the phase space vector ${\bf z}$ where $\alpha = 3j$ and $\beta = 3k$. Then, remembering that there is a zero-frequency mode in this system, the position $z_j$ can be written in terms of the eigenmodes via
 Eqs.~\eqref{a35}, \eqref{a21}, and \eqref{a30}, 
 \begin{equation}
 \delta z_j = \sum_{\omega>0} (Q_\omega + i P_\omega) u_{\omega \alpha}/\sqrt{2}+ c. c. + a_0 u_{0\alpha} + P_0\frac{ \bar u_{0 z \alpha}}{(\bar{\bf u}_{0 z},\bar{\bf u}_{0 z})},
 \end{equation}
  
 and similarly for $\delta z_k$. Using Eqs.~\eqref{a25.3} and \eqref{a25.4}, the thermal averages then yield
\begin{equation}\label{a65.5}
\langle \delta z_i \delta z_j\rangle = T \sum_{\omega>0} \frac{u_{\omega\alpha}u_{\omega\beta}^* }{({\bf u}_\omega,{\bf u}_\omega)} + c.c + \frac{T}{(\bar{\bf u}_{0 z},\bar{\bf u}_{0 z})}  \bar u_{0 z \alpha} \bar u_{0 z \beta},
\end{equation}
where we used Eq.~\eqref{a20} to replace $\omega$ by $({\bf u}_\omega,{\bf u}_\omega)$, so that the expression is valid for unnormalized eigenvectors.
The last term arises from the average $\langle P_0^2\rangle$, which we evaluated using Hamiltonian \eqref{a53}. There is no contribution from  $a_0$  because the neutrally-stable eigenmode is a pure rotation about the $z$ axis by angle $\delta\phi= a_0$, with eigenvector given by Eq.~\eqref{eig0}. This eigenvector has no axial component (i.e. $u_{0\alpha}=0$).

We can evaluate $\langle \delta z_j\delta z_k\rangle$ for the $2$ ion system using the information in Table 1. For $\beta>1$ only the axial COM mode and the stretch mode contribute. There is no contribution from the zero frequency mode because it does not exist for $\beta>1$ since charges  in equilibrium are aligned along the $z$ axis. We thus obtain, for $\beta>1$,
\begin{align}
  \langle \delta z_1^2\rangle=\langle \delta z_2^2\rangle &= \frac{2T}{4m\omega_z^2} +  \frac{2T}{12m\omega_z^2} =\frac{2T}{3m\omega_z^2}, \label{a137nn}\\
  \langle \delta z_1 \delta z_2 \rangle &= \frac{2T}{4m\omega_z^2} -  \frac{2T}{12m\omega_z^2} = \frac{T}{3m\omega_z^2},
  \end{align}
where the first term is from the axial COM mode and the second term is from the stretch mode. 

For $\beta<1$ only the axial COM and the tilt mode contribute, and again the rotational mode does not contribute because there is no axial component of $\bar{\bf u}_0$ when the charges are trapped in the $x-y$ plane, see Eq.~\eqref{a128nn}. Then Eq.~\eqref{a65.5} yields
\begin{align}
  \langle \delta z_1^2\rangle = \langle \delta z_2^2\rangle &= \frac{2T}{4m\omega_z^2} +  \frac{2T}{4m\omega_z^2(1-\beta)} = \frac{T}{2m\omega_z^2}\frac{2-\beta}{1-\beta},  \label{a140n}\\
  \langle \delta z_1 \delta z_2 \rangle &= \frac{2T}{4m\omega_z^2} -  \frac{2T}{4m\omega_z^2(1-\beta)} = -\frac{T}{2m\omega_z^2}\frac{\beta}{1-\beta}, \label{a141n}
  \end{align}
where the first term is again from the axial COM mode and now the second term is from the tilt mode. These averages diverge when $\beta\rightarrow 1$, as the tilt mode becomes zero frequency, allowing large fluctuations in the axial displacements of the charges. Note that the equations imply that $\langle (\delta z_1 + \delta z_2)^2\rangle/4 =  T/(2 m\omega_z^2)$, independent of $\beta$. This is the mean square fluctuation in the axial center of mass axial position, with the expected form $T/(N m \omega_z^2)$ for a particle of mass $N m$ in a harmonic well. 

Also note that none of these averages depend on the magnetic field strength (i.e. on $\Omega$), as expected from the Bohr-Van Leeuwen theorem\cite{bohr}. Of course, in this example none of the modes contributing to the averages depended on
 $\Omega$. A less trivial application of the theorem arises in the evaluation of a different average, $\langle \delta x_j   \delta x_k \rangle$. This average is finite for an $N=2$ cluster with equilibrium positions in the $x-z$ plane because zero-frequency rotations through $\phi$ are in the $y$ direction and do not affect the average. The formula for this average is the same as Eq.~\eqref{a65.5}, except that now,  $\alpha = 3j-2$ and $\beta = 3k-2$.
 For the two particle Coulomb cluster, and taking $\beta>1$,  we obtain
 \begin{align}
  \langle \delta x_1^2   \rangle = \langle \delta x_2^2   \rangle &=\frac{T}{2m}\left( \frac{1}{\omega_{-}^2+\omega_\bot^2} +  \frac{1}{\omega_{+}^2+\omega_\bot^2} + \frac{1}{\omega_{r-}^2+(\beta-1)\omega_z^2} +  \frac{1}{\omega_{r+}^2+(\beta-1)\omega_z^2}\right), \label{a65.61} \\
 \langle \delta x_1 \delta x_2   \rangle & =\frac{T}{2m}\left( \frac{1}{\omega_{-}^2+\omega_\bot^2} +  \frac{1}{\omega_{+}^2+\omega_\bot^2} - \frac{1}{\omega_{r-}^2+(\beta-1)\omega_z^2} -  \frac{1}{\omega_{r+}^2+(\beta-1)\omega_z^2}\right). \label{a65.62}
 \end{align}
Each term on the right hand side depends explicitly on $\Omega$,  but their sum does not. In fact, $ {1}/({\omega_{-}^2+\omega_\bot^2}) +  1/({\omega_{+}^2+\omega_\bot^2}) = 1/\omega_\bot^2$, and ${1}/({\omega_{r-}^2+(\beta-1)\omega_z^2}) +  
{1}/({\omega_{r+}^2+(\beta-1)\omega_z^2})=1/((\beta-1)\omega_z^2)$. When these formulae are applied to Eqs.~\eqref{a65.61} and \eqref{a65.62}, we obtain the $\Omega$-independent result
\begin{align}
  \langle \delta x_1^2   \rangle = \langle \delta x_2^2   \rangle &=\frac{T}{2m\omega_z^2}\left( \frac{1}{\beta} +  \frac{1}{\beta-1}\right), \\
 \langle \delta x_1 \delta x_2   \rangle  &=\frac{T}{2m\omega_z^2}\left( \frac{1}{\beta}-  \frac{1}{\beta-1}\right),
 \end{align}
where we used the relation $\omega_\bot^2 = \beta \omega_z^2$.

Re-evaluating the averages for $\beta<1$ we obtain
 \begin{align}
  \langle \delta x_1^2   \rangle = \langle \delta x_2^2   \rangle &=\frac{T}{2m}\left( \frac{1}{\omega_{-}^2+\omega_\bot^2} +  \frac{1}{\omega_{+}^2+\omega_\bot^2} +
  \frac{1}{\Omega^2+3\omega_\bot^2} +\frac{(\Omega/3\omega_\bot^2)^2}{1+\Omega^2/3\omega_\bot^2} \right),  \\
 \langle \delta x_1 \delta x_2   \rangle & =\frac{T}{2m}\left( \frac{1}{\omega_{-}^2+\omega_\bot^2} +  \frac{1}{\omega_{+}^2+\omega_\bot^2} -\frac{1}{\Omega^2+3\omega_\bot^2} -\frac{(\Omega/3\omega_\bot^2)^2}{1+\Omega^2/3\omega_\bot^2}  \right).
 \end{align}
The first two terms are from the COM modes, the third term is from the upper hybrid mode, and the last term is from the zero-frequency rotational mode, where we used Eq.~\eqref{a128nn} for $\bar{\bf u}_0$ and Eq.~\eqref{a129nn} for $(\bar{\bf u}_0, \bar{\bf u}_0)$. Again, each term  depends on $\Omega$, but when summed the results are $\Omega$-independent:
\begin{align}
  \langle \delta x_1^2   \rangle = \langle \delta x_2^2   \rangle &=\frac{2T}{3m\omega_\bot^2},  \label{a147nn}\\
 \langle \delta x_1 \delta x_2   \rangle & =\frac{T}{3m\omega_\bot^2}.
 \end{align}

As another test of the Bohr- Van Leeuwen theorem we consider the average $\langle(\delta{\bf r}_1\cdot\hat R_1)^2\rangle = \langle(\delta x_1 \sin\theta + \delta z_1 \cos\theta)^2\rangle$ for the $N=2$ cluster at $\beta = 1$, oriented at angle $\theta$ with respect to the $z$-axis. Here $\hat R_1=(\sin\theta,0,\cos\theta)$ is the unit vector in the direction of ${\bf R}_1$, the equilibrium position of charge 1.  This average is finite because zero-frequency rotations of the cluster about the $y$  or $z$ axes have no effect- see Table 1. Only the COM modes and the modes at frequencies $\omega_{\theta \pm}$ enter the average:
\begin{align}
  \langle(\delta{\bf r}_1\cdot\hat R_1)^2\rangle =& \langle \delta x_1^2\rangle \sin^2\theta + \langle \delta z_1^2\rangle \cos^2\theta + 2 \langle \delta x_1 \delta z_1\rangle 
   \sin\theta \cos\theta \notag \\
   =&\sin^2\theta \frac{T}{2m} \sum_{\omega = \omega_\pm} \frac{1}{\omega^2 +\omega_\bot^2}+\cos^2 \theta  \frac{T}{2m\omega_z^2} \notag \\
   &+ 2T \sum_{\omega = \omega_{\theta_\pm}}\frac{1}{({\bf u}_\omega,{\bf u}_\omega)}\left(\omega^2\sin^2\theta +  \Delta^2\cos^2\theta +2  \omega\Delta \sin\theta\cos\theta 
  \right),
   \end{align}
where $\Delta = 2\omega\frac{\omega^2-\Omega^2-3\omega_z^2\sin^2\theta}{3\omega_z^2\sin2\theta}$. The first two terms on the right hand side of the second line arise 
from the three COM modes, and for $\beta=1$ they sum to $T/(2m\omega_z^2)$, the expected radial thermal fluctuation for a single particle of mass $2m$  in a spherically-symmetric harmonic well of frequency $\omega_z$. Surprisingly, perhaps, considering its complexity, the last term sums  to $T/(6 m \omega_z^2)$. Thus we obtain
\begin{align}
 \langle(\delta{\bf r}_1\cdot\hat R_1)^2\rangle =\frac{2T}{3m\omega_z^2},
 \end{align}
which agrees with Eq.~\eqref{a147nn}  when $\theta=\pi/2$ and $\beta=1$, and with Eq.~\eqref{a137nn} when $\theta=0$.

\begin{figure}[!tbp]

{\includegraphics[width = 3.12in]{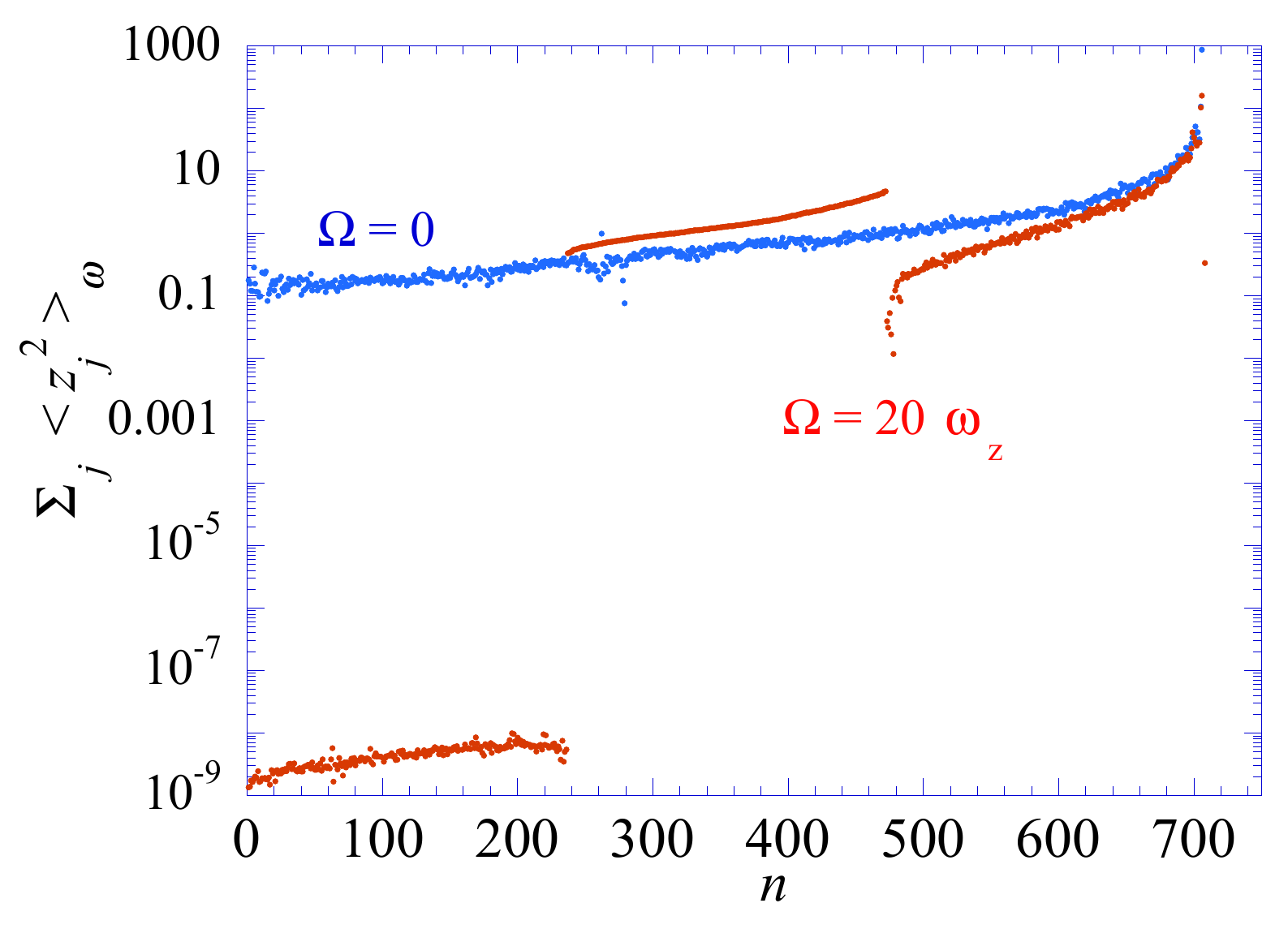}}

\caption{\label{figtherm} The thermal average $\sum_{j=1}^N \langle \delta z_j^2\rangle_\omega$ for each  eigenmode,  in units of $T/(m\omega_z^2)$, for the spherioidal crystal shown in Fig.~\ref{fig1}, and  for two  vortex frequencies, $\Omega = 0 $ and  $\Omega=20\omega_z$. }
  \end{figure}

  As an example of thermal averages in a larger $N$ system, in  Fig.~\ref{figtherm} we numerically evaluate the following thermal average  for the $N=236$ crystal: $\sum_{j=1}^N \langle \delta z_j^2\rangle_\omega$, 
  where  $\langle \delta z_j^2\rangle_\omega = 2T |u_{\omega\ 3j }|^2 /{({\bf u}_\omega,{\bf u}_\omega)}$ is the mean square fluctuation in axial position $z_j$ caused by mode $\omega$, as per Eq.~\eqref{a65.5}. 
  
 We use this average to evaluate each mode's contribution to $ \langle \delta z^2\rangle=N^{-1} \sum_{j=1}^N \langle z_j^2\rangle$ via
\begin{equation}\label{a129n}
N  \langle \delta z^2\rangle =\sum_{\omega>0} \sum_{j=1}^N   \langle \delta z_j^2\rangle_\omega + 
\frac{T}{(\bar{\bf u}_{0 z},\bar{\bf u}_{0 z})}  \sum_{j=1}^N \bar u_{0 z \ 3 j}^2,
\end{equation}
see Eq.~\eqref{a65.5}.

 From the figure one can see that for $\Omega=0$ the lowest frequency torsional modes dominate the thermal average, as one would expect (in the figure the mode number $n$ is ordered from highest to lowest frequency as in Fig.~\ref{fig2}). For the large magnetic field case $\Omega = 20\omega_z$, cyclotron modes make a negligible contribution to the average, as one might also expect; we obtain $\sum_{\omega_\text{cyclotron}} \sum_{j=1}^N   \langle \delta z_j^2\rangle_\omega = 1.096\times 10^{-6} T/(m \omega_z^2)$. Axial modes make a larger contribution, contributing to the average an amount $\sum_{\omega_\text{axial}} \sum_{j=1}^N   \langle \delta z_j^2\rangle_\omega = 385.0969 T/(m \omega_z^2)$.  Surprisingly, however, the low-frequency $E\times B$ drift modes dominate by a factor of 10, contributing $\sum_{\omega_\text{axial}} \sum_{j=1}^N   \langle \delta z_j^2\rangle_\omega = 3575.0989 T/(m \omega_z^2)$.  One normally thinks of $E\times  B$ drift modes as  motions in the $x-y$ plane, but there can also be substantial axial motion in these modes, as we will see; and low-frequency torsional $E\times B$ motions make a large contribution to the axial fluctuations. Finally, the zero-frequency rotational mode (the last term in Eq.~\ref{a129n}) contributes $0.2463$ to the right hand side of the equation, for a total mean square fluctuation of $\langle \delta z^2\rangle = 3690.4421/N = 16.7815$ in units of $T/m \omega_z^2$. This fluctuation is independent of magnetic field strength as expected from the Bohr-van-Leeuwen theorem; we have checked that precisely the same value can be computed by summing over the $\Omega=0$  mode contributions shown in the figure. Note that for 
 $\Omega=0$ there is no contribution to this thermal average from the zero frequency rotational mode, since for $\Omega = 0$,  the axial components of the vector $\bar{\bf u}_{0 z}$ vanish, i.e. $\bar {\bf u}_{0 z\  3 j} =  0$; see Appendix C.

  Calculations like that displayed in Fig.~\ref{figtherm} may impact the prospects of extending sensitive techniques\cite{bollinger2, Sawyer2014} developed for the spectroscopy and thermometry of normal modes with single-plane ion crystals to large 3-dimensional ion crystals.  The basic technique uses an oscillating spin-dependent force with a frequency $\mu$ to map motion parallel to the magnetic field and with the same frequency $\mu$ onto precession of an internal spin-degree of freedom of the trapped ion.  The spin precession can then be read out with high signal-to-noise ratio.  Technically the spin-dependent force is an optical dipole force created at the intersection of two laser beams and is characterized by a wavelength $\lambda_{eff}$.  Accurate spectroscopy and thermometry require the so-called Lamb-Dicke confinement criteria where the ion axial fluctuations $\langle \delta z^2\rangle$ are small compared to $\lambda_{eff}$.  This could open the door for trapped-ion quantum simulation and sensing work \cite{bollinger, Bohnet2016, Gilmore2017} with large 3-dimensional ion crystals.  The large axial excursions of the low-frequency $E\times B$ modes should improve the prospects for laser Doppler cooling of these modes relative to that possible for single-plane crystals.

  
{

\subsubsection{limiting cases}
We now consider several limiting cases for which the eigenmode problem for charges in a trap simplifies. We first consider the unmagnetized case $\Omega_i = 0$. 
In this case the dynamical matrix Eq.~\eqref{a65} reduces to 
\begin{equation}
{\bf D} = \left(
\begin{array}{c c}
{\bf 0} &{\bf M}^{-1} \\
-{\bf V} & {\bf 0}
\end{array}
\right)
\end{equation}
and the eigenvalue problem Eq.~\eqref{a5} corresponds to the coupled equations $-i\omega{\bf r} ={\bf M}^{-1}\cdot{\bf p}$, $-i\omega{\bf p} =-{\bf V}\cdot{\bf r}$ which can be combined into a reduced eigenvalue problem for $\omega^2$,
$\omega^2{\bf r} = {\bf M}^{-1}\cdot{\bf V}\cdot{\bf r}$. This is the standard Hermitian eigenvalue problem for coupled unmagnetized oscillators presented in many physics textbooks, and referred to in the introduction to Sec. II.  The real eigenvectors ${\bf r}$  are found to form an orthogonal set with respect to the reduced inner product $({\bf a},{\bf b}) ={\bf a}\cdot{\bf M}\cdot{\bf b}$ for real vectors ${\bf a}$ and ${\bf b}$. The mode frequencies $\omega$ are found to be real provided that the crystal equilibrium is stable (or neutrally stable), just as in the more general problem discussed in Sec. II. Other features of the unmagnetized eigenmodes for an ion crystal in a trap have been examined in  previous papers and we will not comment further on this special case. 

We next turn to the case of a large magnetic field, such that $\Omega_i\gg \omega_r$. As we saw in Fig.~\ref{fig2.5}  the normal modes now separate into three frequency groupings. There are  $N$ $E\times B$ drift modes with low  frequencies that scale with magnetic field strength $B$ as $1/B$; $N$ intermediate frequency axial modes with frequencies that are  independent of $B$ (for large $B$), and $N$ high frequency cyclotron modes with frequencies close to (but slightly larger than) $\Omega_i$. If there are several species of charge in the trap  with different values of the vortex frequency $\Omega_i$, there are modes near each value. The number of modes is the number of particles with that vortex frequency. In general, the frequency difference  
$\omega - \Omega_i$  for these modes  scales as $1/B$. 

Many characteristics of these strongly-magnetized modes have been considered in previous publications\cite{usov, nagai,  baiko, chen, dubinmodesim, dubinschiffer, freericks}. Here we present reduced eigenvalue problems for each mode type, and consider a few special cases in more detail. 

The eigenmodes in a large magnetic field can be evaluated using degenerate perturbation theory applied to Eq.~\eqref{a5}. We break up the dynamical matrix $\bf D$ and the Hamiltonian matrix $\bf H$ into zeroth-order and first order parts. The zeroth-order parts are 
\begin{equation}\label{a84}
{\bf H}^{(0)}= \left(
\begin{array}{c c}
{\bf C} & {\bf \Omega} \\
{\bf -\Omega} & {\bf M}^{-1}
\end{array}
\right).
\end{equation}
and 
\begin{equation}
{\bf D}^{(0)}= {\bf J}\cdot{\bf H}^{(0)} = \left(
\begin{array}{c c}
{\bf -\Omega} & {\bf M}^{-1} \\
{\bf -C} & {-\bf \Omega}
\end{array}
\right).
\end{equation}
while the first-order parts are
\begin{equation}\label{a86}
{\bf H}^{(1)}= \left(
\begin{array}{c c}
{\bf V} & {\bf 0} \\
{\bf 0} & {\bf 0}
\end{array}
\right).
\end{equation}
and 
\begin{equation}\label{a87}
{\bf D}^{(1)}= {\bf J}\cdot{\bf H}^{(1)} = \left(
\begin{array}{c c}
{\bf 0} & {\bf 0} \\
-{\bf V} & {\bf 0}
\end{array}
\right).
\end{equation}

The zeroth-order matrices describe the dynamics of non-interacting charges in a magnetic field. Because the charges are non-interacting the eigenvectors and eigenvalues can be worked  out for each particle separately. The  $6N$ dimensional eigenvector ${\bf u}_\omega$ for one of these modes consists of zeros in every element except for those elements corresponding to particle $j$. We label this particular eigenvector ${\bf u}_\omega = {\bf u}^{(0)}_{j,\alpha}$ where the superscript $(0)$ indicates that it is a zeroth order eigenvector, the subscripts $j$ and $\alpha$ label the particle and the mode type respectively. There are five different mode types as we will see in a moment. This eigenvector has the form
\begin{equation}\label{a88}
\begin{array}{l l l l  l}
{\bf u}^{(0)}_{j \alpha} &= ({\bf 0},...,&{\bf r}_{j \alpha}, {\bf 0},...&\ \ \ {\bf p}_{j \alpha}, {\bf 0},...&{\bf 0}).  \\
 & i=1,...,&j,...,&N+j,...& 2N
\end{array}
\end{equation}
The vector $({\bf r}_{j  \alpha},{\bf p}_{j \alpha})$ solves the six-dimensional single-particle eigenvalue problem corresponding to the particle $j$ elements in ${\bf D}^{(0)}$:
\begin{equation}
-i\omega ({\bf r}_{j  \alpha},{\bf p}_{j \alpha}) = \left(
\begin{array}{cccccc}
0  &-\Omega_j/2 & 0 & 1/m_j & 0   & 0 \\
\Omega_j/2  & 0 & 0 & 0 & 1/m_j   & 0 \\
0  & 0 & 0 & 0 & 0   & 1/m_j  \\
-m \Omega_j^2/4 & 0 & 0 & 0 & -\Omega_j/2 & 0 \\
0 & -m \Omega_j^2/4 & 0 &\Omega_j/2 & 0 &  0 \\
0  & 0 & 0 & 0 & 0  & 0 \\
\end{array}
\right)\cdot ({\bf r}_{j  \alpha},{\bf p}_{j \alpha}).
\end{equation}
This eigenvalue problem has five independent eigenvectors, three of which correspond to zero-frequency modes, and two of which are cyclotron modes with frequencies $\pm\Omega_j$.   The positive and negative frequency cyclotron eigenvectors for particle $j$  are 
\begin{align}
({\bf r}_{j +},{\bf p}_{j +}) &=\frac{1}{2\sqrt{2} m_j \Omega_j} ( 2,2i,0,-i m_j \Omega_j,m_j\Omega_j,0), \ \ &\omega=\Omega_j, \label{a89}\\
({\bf r}_{j -},{\bf p}_{j -}) &= ({\bf r}_{j +},{\bf p}_{j +})^*, \ \  &\omega=-\Omega_j.\label{a90}
\end{align}
The negative frequency cyclotron eigenvector is the complex conjugate of the positive frequency eigenvector, as expected from property 3. Both eigenvectors are needed to describe the real phase space vector $\bf z$ corresponding to cyclotron motion for particle $j$. Both modes correspond to rotation of the particle position and momentum vectors in the counter-clockwise sense (the positive $\hat\phi$ direction). 

The three independent zero-frequency eigenvectors  for particle $j$ are
\begin{align}
({\bf r}_{j X},{\bf p}_{j X}) &= ( 1,0,0,0,-m_j\Omega_j/2,0), \label{a91}\\
({\bf r}_{j Y},{\bf p}_{j Y}) &=( 0,1,0,m_j\Omega_j/2,0,0),  \label{a92}\\
({\bf r}_{j Z},{\bf p}_{j Z}) &=( 0,0,1,0,0,0).  \label{a93}
\end{align}
These eigenvectors correspond to displacements in the $x,y$ and $z$ directions respectively, and are labelled as such. The nonzero momentum components arise  because canonical momentum depends on position. There are corresponding constants of the motion $P_{j X},P_{j Y},P_{j Z}$. For example, $P_{j X} =({\bf r}_{j X},{\bf p}_{j X})\cdot{\bf J}\cdot(\delta x_j,\delta y_j,\delta z_j,\delta p_{x j},\delta p_{y j},\delta p_{z j})$, and so on.  For $\Omega_j\ne 0$, one can check that $P_{j X}$ and $P_{j Y}$ are not in involution,
\begin{equation}\label{a93.11}
[P_{j X},P_{j Y}]=m_j \Omega_j,
\end{equation} 
but $[P_{j Y},P_{j Z}]=[P_{j X},P_{j Z}]=0$.

These five eigenvectors are {\it not} sufficient to form a complete set to describe the motion of particle $j$; we require a sixth vector. This problem should be familiar as it was covered in section IIb in the discussion of neutrally-stable systems. We require a vector that is orthogonal to the other five eigenvectors. It could be found by solution of the linear algebra problem given by Eq.~\eqref{a32}, but we can identify the solution without any algebra:
\begin{align}\label{a94}
(\bar {\bf r}_{j Z}, \bar {\bf p}_{j Z}) = (0,0,0,0,0,0,1)
\end{align}
corresponding to constant velocity in the $z$ direction. There is one of these vectors for each particle. We refer to the  corresponding $6N$ dimensional vector as $\bar {\bf u}_{j, Z}$, with zeroes in all elements except for those corresponding to the $jth$ particle:
\begin{equation}\label{a95}
\begin{array}{l l l  l}
\bar{\bf u}_{j Z} &= ({\bf 0},... &,{\bf 0},\bar{\bf p}_{j Z}, {\bf 0},...&{\bf 0}).  \\
 & i=1,...&\,N+j,...& 2N
\end{array}
\end{equation}
These vectors are not eigenvectors, but instead satisfy 
\begin{equation}\label{a96}
{\bf D}^{(0)}\cdot \bar {\bf u}_{j, Z} = {\bf u}_{j, Z}^{(0)}/m_j.
\end{equation}
 The vectors are orthogonal  to all of the zeroth-order eigenvectors ${\bf u}_{k \alpha}^{(0)}$, both with respect to a standard dot product as well as  with respect to the inner product defined by the zeroth order Hamiltonian matrix as
 \begin{equation}\label{a96.5}
 ({\bf a},{\bf b})^{(0)} = {\bf a}^*\cdot{\bf H}^{(0)}\cdot{\bf b}.
 \end{equation} 
 
We now employ the zeroth-order eigenvectors in order to construct a cyclotron-frequency eigenmode that includes the effect of particle interactions to first order. Let us assume that there are $N_a$ particles of species $a$, all of which have identical vortex frequency $\Omega_a$, and we will assume there are no other particles with this vortex frequency. We will then use degenerate perturbation theory to solve for the perturbed eigenmode ${\bf u}_\omega$, writing it as a superposition of the degenerate zeroth order modes with frequency $\Omega_a$:
\begin{equation}\label{a97}
{\bf u}_\omega = \sum_{j=1}^{N_a} c_{j\omega} {\bf u}_{j,+}^{(0)} + {\bf u}^{(1)},
\end{equation}
where  the sum over $j$ sums only over particles of species $a$, the coefficients $c_{j\omega}$ are to be determined, and where ${\bf u}^{(1)}$ is a small correction to the eigenmode. By assumption this correction is orthogonal to the zeroth-order eigenmodes. (It will turn out that we do not need to calculate ${\bf u}^{(1)}$.) We substitute this eigenvector into Eq.~\eqref{a5} and write ${\bf D} = {\bf D}^{(0)} + {\bf D}^{(1)}$ to obtain
\begin{equation}
-i\omega\sum_{j=1}^{N_a} c_{j\omega} {\bf u}_{j,+}^{(0)}-i\omega  {\bf u}^{(1)} = \sum_{j=1}^{N_a} c_{j\omega}  {\bf D}^{(0)}\cdot {\bf u}_{j,+}^{(0)} +  \sum_{j=1}^{N_a} c_{j\omega}  {\bf D}^{(1)}\cdot {\bf u}_{j,+}^{(0)} +  {\bf D}^{(0)}\cdot {\bf u}^{(1)}+ {\bf D}^{(1)}\cdot {\bf u}^{(1)},
\end{equation}

We drop the last term in this equation because it is second-order, use the fact that ${\bf D}^{(0)}\cdot {\bf u}_{j,+}^{(0)} =-i\Omega_a {\bf u}_{j,+}^{(0)}$, multiply through by $i$,  and take a zeroth-order inner product with respect to one of the eigenmodes ${\bf u}_{k,+}^{(0)}$. Orthogonality  and the Hermitian nature of the matrix $i {\bf D}^{(0)}$ then annihilates several terms in the equation, leaving us with
\begin{equation}\label{a99}
\omega c_{k\omega} ({\bf u}_{k,+}^{(0)}, {\bf u}_{k,+}^{(0)})^{(0)}=\Omega_a c_{k\omega} ({\bf u}_{k,+}^{(0)}, {\bf u}_{k,+}^{(0)})^{(0)} +  \sum_{j=1}^{N_a} c_{j\omega} ({\bf u}_{k,+}^{(0)}, i {\bf D}^{(1)}\cdot {\bf u}_{j,+}^{(0)} )^{(0)},
\end{equation}
The inner product $ ({\bf u}_{k,+}^{(0)}, {\bf u}_{k,+}^{(0)})^{(0)}$ involves only particle $k$ and, using Eqs. \eqref{a84}, \eqref{a88}, and \eqref{a89}, evaluates to 
\begin{equation}\label{a99.5}
 ({\bf u}_{k,+}^{(0)}, {\bf u}_{k,+}^{(0)})^{(0)}=\frac{1}{m_a}
 \end{equation}
 (recall that all particles in Eq.~\eqref{a99} are of species $a$).  The inner product $({\bf u}_{k,+}^{(0)},i {\bf D}^{(1)}\cdot {\bf u}_{j,+}^{(0)} )^{(0)}$ involves only particles $j$ and $k$ in species $a$ and evaluates to 
\begin{align}
({\bf u}_{k,+}^{(0)}, i{\bf D}^{(1)}\cdot {\bf u}_{j,+}^{(0)} )^{(0)} &={\bf u}_{k,+}^{(0) *}\cdot\ {\bf H}^{(0)}\cdot i{\bf D}^{(1)}\cdot {\bf u}_{j,+}^{(0)} \notag \\
&={\bf u}_{k,+}^{(0) *}\cdot\ {\bf H}^{(0)}\cdot {\bf J}\cdot i{\bf H}^{(1)}\cdot {\bf u}_{j,+}^{(0)} \notag\\
& = -i \Omega_a {\bf u}_{k,+}^{(0) *}\cdot i {\bf H}^{(1)}\cdot {\bf u}_{j,+}^{(0)} \label{a100.5} \\
&  = \Omega_a {\bf r}_{k +}^*\cdot{\bf V}_{k j} \cdot {\bf r}_{j +} \notag \\
&  = \frac{1}{2 m_a^2\Omega_a} (1,-i,0)\cdot{\bf V}_{k j} \cdot (1,i,0)\notag \\
&= \frac{1}{2 m_a^2\Omega_a}  ({ V}_{k j xx} + { V}_{k j yy}),\label{a100}
\end{align}
where we used the conjugate transpose of Eq.~\eqref{a5} in the third line, Eqs.~\eqref{a86} and \eqref{a88} in the fourth line,  and Eq.~( \ref{a89}) in the fifth line, and where ${ V}_{k j xx}$ is the $\hat x\hat x$ component of the potential matrix ${\bf V}_{k j}=\partial^2\Phi/\partial{\bf R}_j\partial{\bf R}_k$ and similarly for ${ V}_{k j yy}$. Equation \eqref{a99} can then be written in vector form as
\begin{equation}\label{a101}
(\omega-\Omega_a){\bf c}_\omega = {\bf F}\cdot{\bf c}_\omega,
\end{equation}
where the  matrix $\bf F$ has components 
\begin{align}\label{a102}
F_{j k} &=   \frac{{V}_{ j k x x}+{V}_{ j k y y}  }{2m_a\Omega_a}.
\end{align}
Thus, the frequency shift $\omega-\Omega_a$ is  an eigenvalue of the matrix ${\bf F}$. The matrix  is real and symmetric, so its eigenvalues are real and the eigenvectors are also real, and they form a complete orthogonal set. 

Equations ~\eqref{a101} and \eqref{a102} are the reduced eigenvalue problem for the cyclotron modes of species $a$. Note that only particles of this species enter into the matrix $\bf F$ so the matrix is $N_a\times N_a$, yielding $N_a$ eigenfrequencies that differ from $\Omega_a$ by a small shift, proportional to $1/\Omega_a$. 

For a stable system the shift is positive; the extra restoring force from the oscillator potentials tends to increase the mode frequencies.
  One can see this  from the following argument:  any displacement ${\bf r}$ of the charges from equilibrium must increase the potential energy: ${\bf r}\cdot{\bf V}\cdot{\bf r}>0$ for any real nonzero vector $\bf r$. Consider now a complex displacement  ${\bf r} $ created by a superposition of cyclotron eigenvectors for species $a$ particles only. From Eq.~\eqref{a89}, each particles position change is complex, of the form $v_k (1,i,0)$ for some complex amplitude $v_k$. Then consider the quantity $ {\bf r}^*\cdot{\bf V}\cdot{\bf r} =2m_a\Omega_a {\bf v}^*\cdot{\bf F}\cdot{\bf v}$ where $\bf v$ is the vector of complex coefficients $v_k$. However, ${\bf r}^*\cdot{\bf V}\cdot{\bf r}>0$ because, for any complex vector ${\bf r} = {\bf a}+ i{\bf b}$ (for real $\bf a$ and $\bf b$), ${\bf r}^*\cdot{\bf V}\cdot{\bf r}= {\bf a}\cdot{\bf V}\cdot{\bf a}+ {\bf b}\cdot{\bf V}\cdot{\bf b}$ (since $\bf V$ is symmetric) and both terms on the right hand side are positive.  Therefore the quantity ${\bf v}^*\cdot{\bf F}\cdot{\bf v}$ must also be greater than zero for any vector $\bf v$, which implies that the eigenvalues of $\bf F$ must be positive. This follows by writing $\bf v$ as a superposition of the real eigenvectors of $\bf F$, and evaluating ${\bf v}^*\cdot{\bf F}\cdot{\bf v}$.   


In Fig~\eqref{fig4} we compare the eigenfrequencies evaluated using Eq.~\eqref{a101} to the exact eigenfrequencies obtained using Eq.~\eqref{a5}, as a test of the perturbation theory. For the large magnetic field $\Omega = 20 \omega_z$, the fractional error between the exact mode frequencies and the approximate frequencies is quite small.
		\begin{figure}[!tbp]

{\includegraphics[width = 3in]{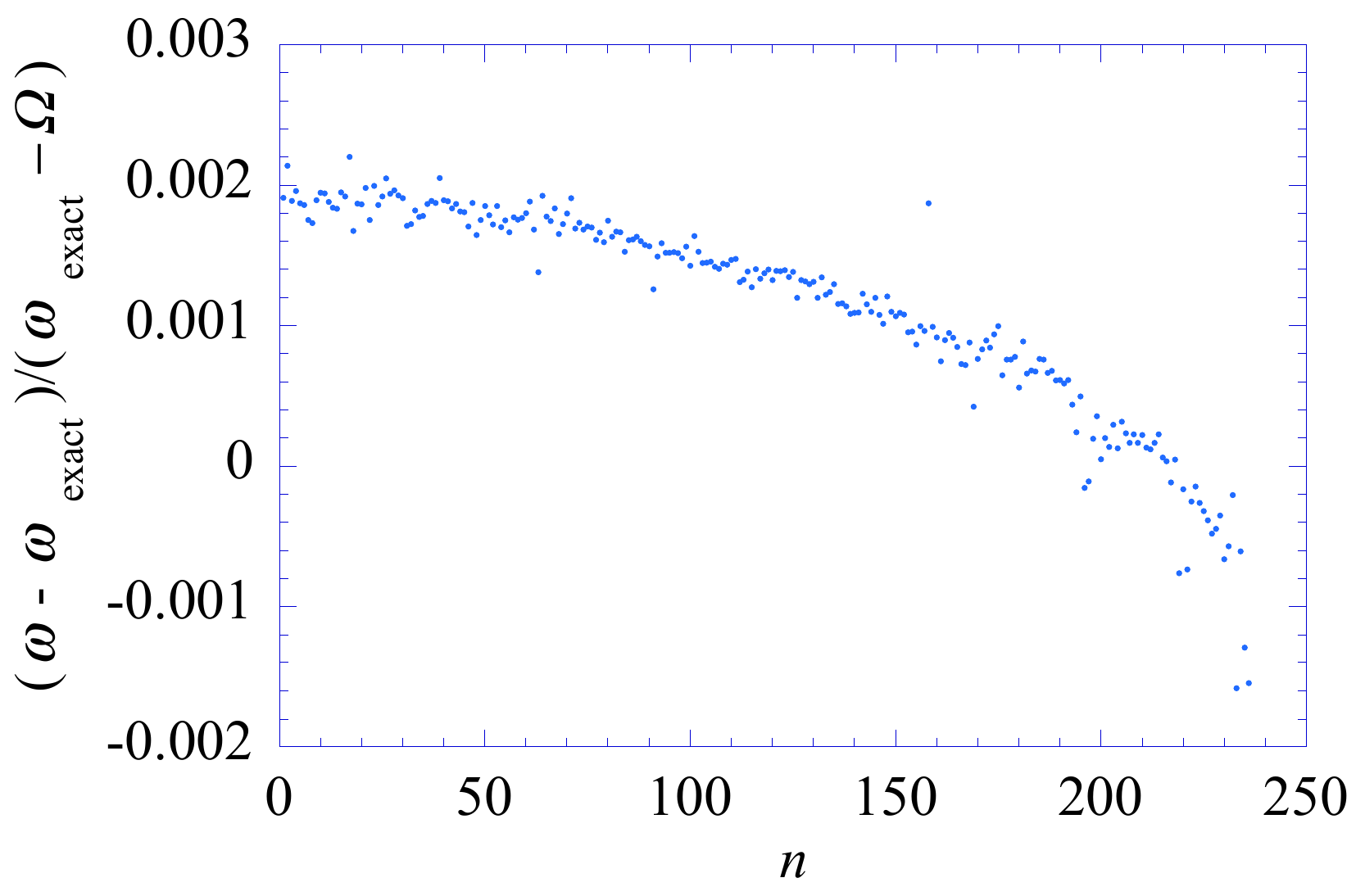}}

\caption{\label{fig4}  Fractional difference between cyclotron frequencies evaluated using Eq.~\eqref{a101}  and the exact frequencies evaluated using Eq.~\eqref{a5} and displayed in Fig.~\ref{fig2.5}, for the same parameters as in that figure. }
  \end{figure}

The energy in cyclotron modes can be determined in terms of the approximate cyclotron frequencies and eigenvectors. (Of, course, the energy of any given mode is also determined exactly via Eq.~\eqref{a16}.)  Consider a  phase space displacement due to species $a$ cyclotron modes,  ${\bf z} = (1/\sqrt{2})\sum_{\omega>0}  a_\omega \sum_{j=1}^{N_a} c_{j\omega} {\bf u}^{(0)}_{j +} + c.c.$, where $a_\omega$ is the amplitude of each mode. This definition, with the extra factor of $1/\sqrt{2}$,  allows us to identify $|c_{k\omega}|$ as $m_a$ multiplied by the speed of particle $k$ in a given mode, for unit amplitude $a_\omega = 1$, because unit amplitude results in a cyclotron radius for this particle of $|c_{k\omega}|/m_a\Omega_a$, according to  Eqs.~\eqref{a89} and \eqref{a90}.  Applying this phase space displacement to the energy, Eq.~\eqref{a1}, breaking the Hamilitonian matrix into zeroth and first order parts, and using orthogonality of the cyclotron modes along with Eqs.~\eqref{a99.5}, \eqref{a100.5} and \eqref{a101}, the cyclotron energy diagonalizes:
\begin{align}
H&=  \frac{1}{2}\sum_{\omega,\bar\omega>0} a_\omega a_{\bar\omega}^* \sum_{j, k=1}^{N_a} c_{k\omega} c_{j\bar\omega} \left(  {\bf u}_{k +}^{(0)*}\cdot{\bf H}^{(0)}\cdot {\bf u}_{j +}^{(0)} + {\bf u}_{k +}^{(0)*}\cdot{\bf H}^{(1)}\cdot {\bf u}_{j +}^{(0)}   \right) \notag \\
&=   \frac{1}{2}\sum_{\omega,\bar\omega>0} a_\omega a_{\bar\omega}^* \sum_{j, k=1}^{N_a} c_{k\omega} c_{j\bar\omega} \left( \frac{1}{m_a}\delta_{j k} +   \frac{1}{m_a\Omega_a} F_{k j}  \right) \notag \\
&=  \sum_{\omega,\bar\omega>0} a_\omega a_{\bar\omega}^* {\bf c}_{\omega}\cdot {\bf c}_{\bar\omega} \left( \frac{1}{2m_a} +   \frac{1}{2m_a\Omega_a} (\bar\omega-\Omega_a) \right) \label{a102.3} \\
&= \frac{1}{2m_a \Omega_a} \sum_{\omega>0} \omega | a_\omega |^2 {\bf c}_\omega\cdot{\bf c}_\omega. \label{a102.5}
\end{align}
The first term in the parenthesis in Eq.~\eqref{a102.3} gives the total kinetic energy associated with free particle cyclotron motion: a sum of the squares of the particle kinetic momenta in a given mode, $|a_\omega|^2{\bf c}_\omega\cdot{\bf c}_\omega$, also summed over the modes, and divided by two times the particle mass. The second term in the parenthesis, proportional to $\bar\omega-\Omega_a$, is the small positive correction to the kinetic energy due to interactions between the charges.

Let us now turn to the axial modes and the $E\times B$ modes. These modes can also be described using a reduced eigenvalue problem that stems from degenerate perturbation theory applied to Eq.~\eqref{a5}. Now, however, we expand an eigenvector ${\bf u}_\omega$ in terms of the zero-frequency eigenvectors along with the extra vectors $\bar{\bf u}_{j Z}$:
\begin{equation}\label{a103}
{\bf u}_\omega = \sum_{j=1}^N (X_j {\bf u}^{(0)}_{j X} + Y_j {\bf u}^{(0)}_{j Y}+Z_j {\bf u}^{(0)}_{j Z}+ P_j  \bar{\bf u}^{(0)}_{j Z} ) +{\bf u}^{(1)},
\end{equation}
where $X_j, Y_j$ and $Z_j$ are displacement amplitudes for particle $j$ in the $x,y$, and $z$ direction respectively,  $P_j$ is the axial momentum of particle $j$, and ${\bf u}^{(1)}$ is a small correction (which we will avoid having to evaluate in what follows). This correction is assumed to satisfy orthogonality conditions
 \begin{equation}\label{a104}
 {\bf u}^{(0)}_{j  X} \cdot{\bf J}\cdot{\bf u}^{(1)}= {\bf u}^{(0)}_{j  Y} \cdot{\bf J}\cdot{\bf u}^{(1)}=  {\bf u}^{(0)}_{j  Z} \cdot{\bf J}\cdot{\bf u}^{(1)}= \bar{\bf u}^{(0)}_{j  Z} \cdot{\bf J}\cdot{\bf u}^{(1)}=0.
 \end{equation}
 We substitute Eq.~\eqref{a103} into Eq.~\eqref{a5} and again break $\bf D$ into zeroth-order and first-order parts:
\begin{align}\label{a104.5}
-i\omega\sum_{j=1}^{N}((X_j {\bf u}^{(0)}_{j X} + Y_j {\bf u}^{(0)}_{j Y}+Z_j {\bf u}^{(0)}_{j Z} &+ P_j  \bar{\bf u}^{(0)}_{j Z} ) -i\omega  {\bf u}^{(1)} = \notag \\
&{\bf D}^{(0)}\cdot \sum_{j=1}^N (X_j {\bf u}^{(0)}_{j X} + Y_j {\bf u}^{(0)}_{j Y}+Z_j {\bf u}^{(0)}_{j Z}+ P_j  \bar{\bf u}^{(0)}_{j Z} ) \notag\\
+&{\bf D}^{(1)}\cdot \sum_{j=1}^N (X_j {\bf u}^{(0)}_{j X} + Y_j {\bf u}^{(0)}_{j Y}+Z_j {\bf u}^{(0)}_{j Z}+ P_j  \bar{\bf u}^{(0)}_{j Z} ) \notag\\
+&{\bf D}^{(0)}\cdot{\bf u}^{(1)} + {\bf D}^{(1)}\cdot{\bf u}^{(1)}. 
\end{align}
We drop the last term on the right hand side since it is second order. Also, ${\bf D}^{(0)}\cdot {\bf u}_{j\alpha}^{(0)}=0$ for all the zero-frequency eigenmodes, so Eq.~\eqref{a104.5} simplifies to
\begin{align}\label{a105}
-i\omega\sum_{j=1}^{N}&((X_j {\bf u}^{(0)}_{j X} + Y_j {\bf u}^{(0)}_{j Y}+Z_j {\bf u}^{(0)}_{j Z} + P_j  \bar{\bf u}^{(0)}_{j Z} ) -i\omega  {\bf u}^{(1)} = \notag \\
& \sum_{j=1}^N P_j  {\bf u}^{(0)}_{j Z}/m_j  
+ {\bf D}^{(1)}\cdot \sum_{j=1}^N (X_j {\bf u}^{(0)}_{j X} + Y_j {\bf u}^{(0)}_{j Y}+Z_j {\bf u}^{(0)}_{j Z}+ P_j  \bar{\bf u}^{(0)}_{j Z} ) \notag\\
+&{\bf D}^{(0)}\cdot{\bf u}^{(1)}. 
\end{align}
where we also applied Eq.~\eqref{a96}.

Now, recall from Section IIb that we can project out zero-frequency modes using the fundamental symplectic matrix $\bf J$ rather than the inner product, since such modes are orthogonal to themselves. Acting on Eq.~\eqref{a105} with $-\bar{\bf u}^{(0)}_{k Z} \cdot{\bf J} = {\bf u}^{(0)}_{k Z}$  and using the orthogonality conditions  \eqref{a104} yields
\begin{align}\label{a106}
-i\omega Z_k = &P_k/m_k 
+  \bar{\bf u}^{(0)}_{k Z}\cdot{\bf H}^{(1)}\cdot \sum_{j=1}^N (X_j {\bf u}^{(0)}_{j X} + Y_j {\bf u}^{(0)}_{j Y}+Z_j {\bf u}^{(0)}_{j Z}+ P_j  \bar{\bf u}^{(0)}_{j Z} ), 
\end{align}
where we used the identity ${\bf J}\cdot{\bf D}={\bf J}\cdot{\bf J}\cdot{\bf H}=-{\bf H}$.  However, one can  use Eqs.~\eqref{a86} and \eqref{a95} to check that $\bar{\bf u}^{(0)}_{k Z}\cdot{\bf H}^{(1)}=\bf 0$, which annihilates the sum on the right hand side so Eq.~\eqref{a106} simplifies to
\begin{align}\label{a107}
-i\omega Z_k  = &P_k/m_k, 
\end{align}
the standard relation between axial position and momentum.
Now  act  on Eq.~\eqref{a105} with ${\bf u}_{k Z}^{(0)}\cdot{\bf J}=\bar{\bf u}_{k Z}^{(0)}$, which results in
\begin{align}\label{a108}
-i\omega P_k =  &-{\bf u}^{(0)}_{k Z}\cdot{\bf H}^{(1)}\cdot \sum_{j=1}^N (X_j {\bf u}^{(0)}_{j X} + Y_j {\bf u}^{(0)}_{j Y}+Z_j {\bf u}^{(0)}_{j Z}+ P_j  \bar{\bf u}^{(0)}_{j Z} ) \notag \\
 = & -   \sum_{j=1}^N (V_{k j z x} X_j + V_{k j z y} Y_j+ V_{k j z z} Z_j ).
\end{align}
where the second line applied Eqs.~\eqref{a86}, \eqref{a88} and (\ref{a91} - \ref{a95}). The right hand side is the axial force on particle $k$ caused by other particle displacements from equilibrium.

Next,  act  on Eq.~\eqref{a105} with ${\bf u}_{k Y}^{(0)}\cdot{\bf J}$. Here we will use ${\bf u}_{k Y}^{(0)}\cdot{\bf J}\cdot{\bf u}_{k X}^{(0)}= -m_k \Omega_k $, which follows from Eq.~\eqref{a93.11} and \eqref{b64.5} or can be checked directly using Eqs.~\eqref{a88}, \eqref{a91} and \eqref{a92}. This projection results in 
\begin{align}\label{a109}
i\omega m_k\Omega_k X_k =  &-{\bf u}^{(0)}_{k Y}\cdot{\bf H}^{(1)}\cdot \sum_{j=1}^N (X_j {\bf u}^{(0)}_{j X} + Y_j {\bf u}^{(0)}_{j Y}+Z_j {\bf u}^{(0)}_{j Z}+ P_j  \bar{\bf u}^{(0)}_{j Z} ) \notag \\
 = &  -   \sum_{j=1}^N (V_{k j y x} X_j + V_{k j y y} Y_j+ V_{k j y z} Z_j ).
\end{align}
 The right hand side is the force in the $y$ direction on particle $k$ caused by displacements of other particles. The force produces  an $E\times B$ drift  velocity $-i\omega X_k$ in the $-x$ direction. 
 Finally, act  on Eq.~\eqref{a105} with ${\bf u}_{k X}^{(0)}\cdot{\bf J}$.  This projection results in 
\begin{align}\label{a110}
- i\omega m_k\Omega_k  Y_k =   &  -   \sum_{j=1}^N (V_{k j x x} X_j + V_{k j x y} Y_j+ V_{k j x z} Z_j ).
\end{align}
 The right hand side is the force in the $x$ direction on particle $k$. This force produces  an $E\times B$ drift  velocity $-i \omega Y_k$ in the $y$ direction. 

Equations (\ref{a107}-\ref{a110}) constitute a reduced eigenvalue problem for axial and $E\times B$ modes that has projected out the cyclotron modes. However, the problem still mixes the $E\times B$ modes with the axial plasma modes. One can see from Eqs.~\eqref{a109} and \eqref{a110}  that  axial displacements $Z_j$  are coupled to $x$ and $y$ drifts through the $x$ and $y$ forces such axial displacements can produce. Also,  axial accelerations can be caused by $X$ and $Y$ displacements, as seen  in Eq.~\eqref{a108}.

 Now, there are circumstances where this coupling vanishes.  For example, when the crystal equilibrium is a single lattice plane  in the $z=0$ plane, symmetries of this equilibrium imply that $V_{k j x z}=V_{k j y z}=0$. The coupling between  axial and transverse motions also vanishes when the crystal is a one-dimensional line of charges along the $z$ axis. However, for  more general crystal equilibria the coupling is nonzero.

In order to further decouple the $E\times B$ modes from the axial modes, we must, in general,  resort to an asymptotic two-timescale analysis based on the different frequencies of these modes.  The $E\times B$ modes, with frequencies scaling as $1/B$, are low frequency compared to the axial modes provided that $B$ is sufficiently large. This regime implies that we may write  $Z_j$ as a sum of a slowly evolving and a rapidly evolving contribution, $Z_j = Z^{slow}_j + Z^{fast}_j$. The slow evolution, on the $E\times B$ timescale, is conditioned on the axial particle positions being in axial force balance:
\begin{equation}
0 =   -   \sum_{j=1}^N (V_{k j z x} X_j + V_{k j z y} Y_j+ V_{k j z z} Z_j ^{slow}).
\end{equation}
We can regard this force balance condition as a set of coupled linear equations for the slow axial displacements, written in vector form as ${\bf V}_{z x}\cdot {\bf X} + {\bf V}_{ z y}\cdot {\bf Y}+ {\bf V}_{z z}\cdot {\bf Z}^{slow} =\bf 0$, where the tensor ${\bf V}_{x z} = \partial\Phi/\partial{\bf X}\partial{\bf Z}$ and similarly for the other terms. This equation can be solved by matrix inversion,
\begin{equation}\label{a114}
{\bf Z}^{slow} = -{\bf V}^{-1}_{z z}\cdot({\bf V}_{z x}\cdot {\bf X} + {\bf V}_{z y}\cdot {\bf Y}).
\end{equation}
The fast motion in $z$ is then evaluated using Eqs.~\eqref{a107} and \eqref{a108} after cancelling the slow terms:
\begin{align}\label{a115}
-i\omega Z_k^{fast}  = &P_k/m_k,  \\
- i\omega P_k = & -\sum_{j=1}^N  V_{k j z z} Z_j^{fast}.\notag
\end{align}
This is the reduced eigenvalue problem for the axial modes.\cite{dubinschiffer}
When the approximations used in its derivation are poor, the full eigenvalue problem is available for exact results, or the reduced eigenvalue problem consisting of Eqs.~\eqref{a107} - \eqref{a110} could be employed. 

Equations~\eqref{a115} can be combined into a standard generalized eigenvalue problem for $\omega^2$, of the familiar type encountered in the textbook Lagrangian theory of coupled oscillators, 
\begin{align}\label{a116}
\omega^2 {\bf M}_1\cdot{\bf Z}^{fast}_\omega = {\bf V}_{ z z} \cdot {\bf Z}^{fast}_\omega
\end{align}
where ${\bf Z}^{fast}_\omega$ is an eigenvector of axial  displacements and the matrix ${\bf M}_1$ is the diagonal mass matrix of dimension $N$ with diagonal elements $m_j, j=1,...,N$. The $N$ eigenvalues $\omega^2$ are the squares of the axial mode frequencies, and can be shown to be real and positive for a stable equilibrium, using standard arguments. The eigenvectors ${\bf Z}^{fast}_\omega$ are real and orthogonal with respect to the inner product $({\bf a, \bf b})_z={\bf a}\cdot{\bf M}_1\cdot{\bf b}$. 
In Fig.~\ref{fig5} we compare the frequencies determined using Eq.~\eqref{a116} to those obtained from the exact analysis, as a test of the theory. Just as for the cyclotron modes, the errors are small when the magnetic field is large.
		\begin{figure}[!tbp]

{\includegraphics[width = 3in]{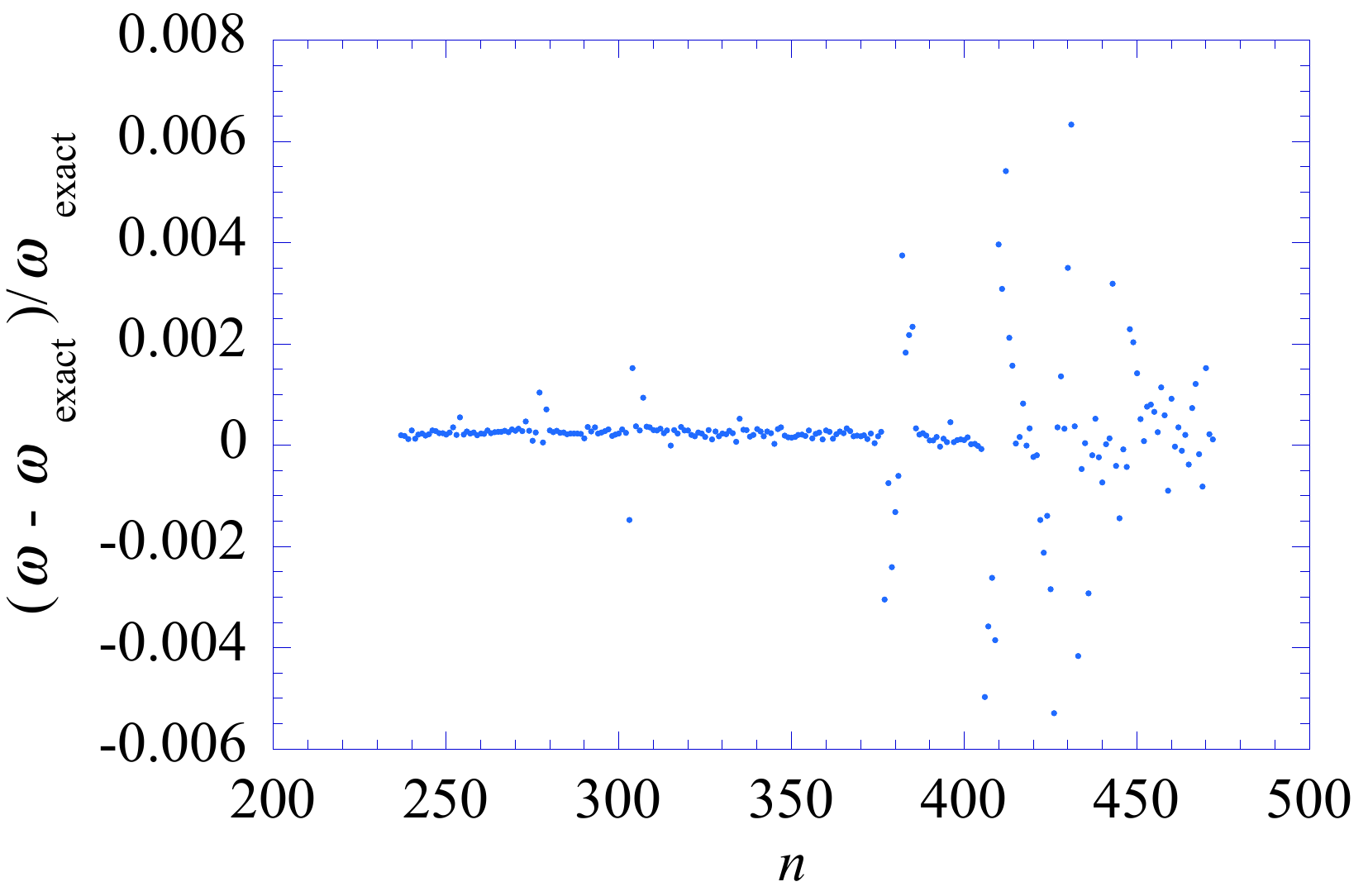}}

\caption{\label{fig5}  Fractional difference between axial frequencies $\omega$ evaluated using Eq.~\eqref{a116}  and the exact frequencies $\omega_{exact}$ evaluated using Eq.~\eqref{a5} and displayed in Fig.~\ref{fig2.5}, for the same parameters as in that figure. }
  \end{figure}

The axial eigenvectors diagonalize the axial energy $H_z$ where
\begin{equation}\label{a117}
H_z = \frac{1}{2}\sum_{j=1}^N P_j^2 /m_j + \frac{1}{2} \sum_{j,k=1}^NZ_j V_{j k zz} Z_k.
\end{equation}
Writing $Z_j = \sum_{\omega } a_\omega Z^{fast}_j$, where $a_\omega$ is the amplitude of mode $\omega$,  and applying Eqs.~\eqref{a115} and \eqref{a116} to Eq.~\eqref{a117}, the kinetic and potential terms contribute equally, yielding for the axial energy the expression
\begin{equation}
H_z = \sum_{\omega} a^2_\omega \omega^2 ({\bf Z}^{fast},{\bf Z}^{fast})_z.
\end{equation}

Turning to the $E\times B$ modes, we apply Eq.~\eqref{a114} for the slow axial motion to Eqs.~\eqref{a109} and \eqref{a110}. These equations can then be combined into a vector form. Defining a $2N$ dimensional transverse displacement eigenvector  ${\bf R}_{\perp\omega} = ({\bf X},{\bf Y})$, the equations become
\begin{equation}\label{a122}
-i\omega  {\bf R}_{\perp\omega} = {\bf D}_\perp\cdot {\bf R}_{\perp\omega},
\end{equation}
where ${\bf D}_\perp={\bf G}\cdot{\bf J}\cdot{\bf V}_\perp$ is the dynamical matrix for $E\times B$ drift modes,  ${\bf G}$ is a diagonal  $2N\times 2N$ matrix with diagonal elements 
$((m_1\Omega_1)^{-1},...,(m_N\Omega_N)^{-1}, (m_1\Omega_1)^{-1},...,(m_N\Omega_N)^{-1})$, 
${\bf J}$ is the $2N\times2N$ fundamental symplectic matrix (see Eq.~\eqref{a2.5}), and ${\bf V}_\perp$ is the following symmetric $2N\times 2N$ potential energy tensor,
\begin{equation}
{\bf V}_\perp = \left(
\begin{array}{c c}
{\bf V}_{x x} & {\bf V}_{x y} \\
{\bf V}_{y x} & {\bf V}_{y y}
\end{array}
\right) - 
\left (
\begin{array}{c c}
{\bf V}_{x z}\cdot{\bf V}^{-1}_{z z}\cdot{\bf V}_{z x} &\ \  {\bf V}_{x z}\cdot{\bf V}^{-1}_{z z}\cdot{\bf V}_{z y} \\
{\bf V}_{y z}\cdot{\bf V}^{-1}_{z z}\cdot{\bf V}_{z x}  &\ \  {\bf V}_{y z}\cdot{\bf V}^{-1}_{z z}\cdot{\bf V}_{z y}
\end{array} 
\right). 
\end{equation}
This tensor determines the potential energy $V$ in an $E\times B$ displacement of the form $({\bf X},{\bf Y},{\bf Z}^{slow})$  through the expression
\begin{equation}
V = \frac{1}{2}({\bf X},{\bf Y})\cdot{\bf V}_\perp\cdot({\bf X},{\bf Y}).
\end{equation}
This can be proven by writing $V$ as
\begin{equation}
V =\frac{1}{2} ({\bf X},{\bf Y},{\bf Z}^{slow})\cdot\left(
\begin{array}{c c c}
{\bf V}_{x x} & {\bf V}_{x y} & {\bf V}_{x z} \\
{\bf V}_{y x} & {\bf V}_{y y} & {\bf V}_{y z} \\
{\bf V}_{z x} & {\bf V}_{z y} & {\bf V}_{z z} 
\end{array}
\right)\cdot ({\bf X},{\bf Y},{\bf Z}^{slow}),
\end{equation}
and applying Eq.~\eqref{a114} along with the symmetry of the matrix elements under interchange of the $x,y,z$ subscripts.

Equation~\eqref{a122} is the reduced eigenvalue problem for $E\times B$ drift modes. The frequencies all scale as $1/B$ since the dynamical matrix ${\bf D}_\perp$ is proportional to $1/B$ through its dependence on $\bf G$.  The dynamical matrix  has properties that mirror the general Hamiltonian matrices discussed in Sec. II, which in turn determine properties of the eigenmodes. First, the matrix $i  {\bf D}_\perp$ is Hermitian with respect to the inner product 
$({\bf a},{\bf b})_\perp = {\bf a}^*\cdot{\bf V}_\perp\cdot{\bf b}$ for any vector ${\bf a}$ and ${\bf b}$. This follows from the fact that the matrix ${\bf L} = {\bf V}_\perp\cdot{\bf G}\cdot{\bf J}\cdot{\bf V}_\perp$ is antisymmetric, which in turn follows from the symmetry and antsymmetry respectively of ${\bf V}_\perp$ and ${\bf J}$, along with the fact that ${\bf G}\cdot{\bf J}={\bf J}\cdot{\bf G}$.   The proof follows along the same path as in Eq.~\eqref{a8}.

As discussed in relation to properties 1 and 2 in Sec. II, the Hermitian nature of the dynamical matrix implies that the eigenfrequencies are real provided that the system is at least neutrally-stable, and that non-degenerate complex eigenvectors are orthogonal with respect to the above inner product. Also, since the matrix elements of ${\bf D}_\perp$ are real, the nonzero frequency modes come in $\pm\omega$ pairs, as per property 3. 

In Fig.~\ref{fig6} we compare the $E\times B$ frequencies determined using Eq.~\eqref{a122} to those obtained from the exact analysis, as a test of the theory. Just as for the cyclotron and axial modes, the errors are small.
		\begin{figure}[!tbp]

{\includegraphics[width = 3in]{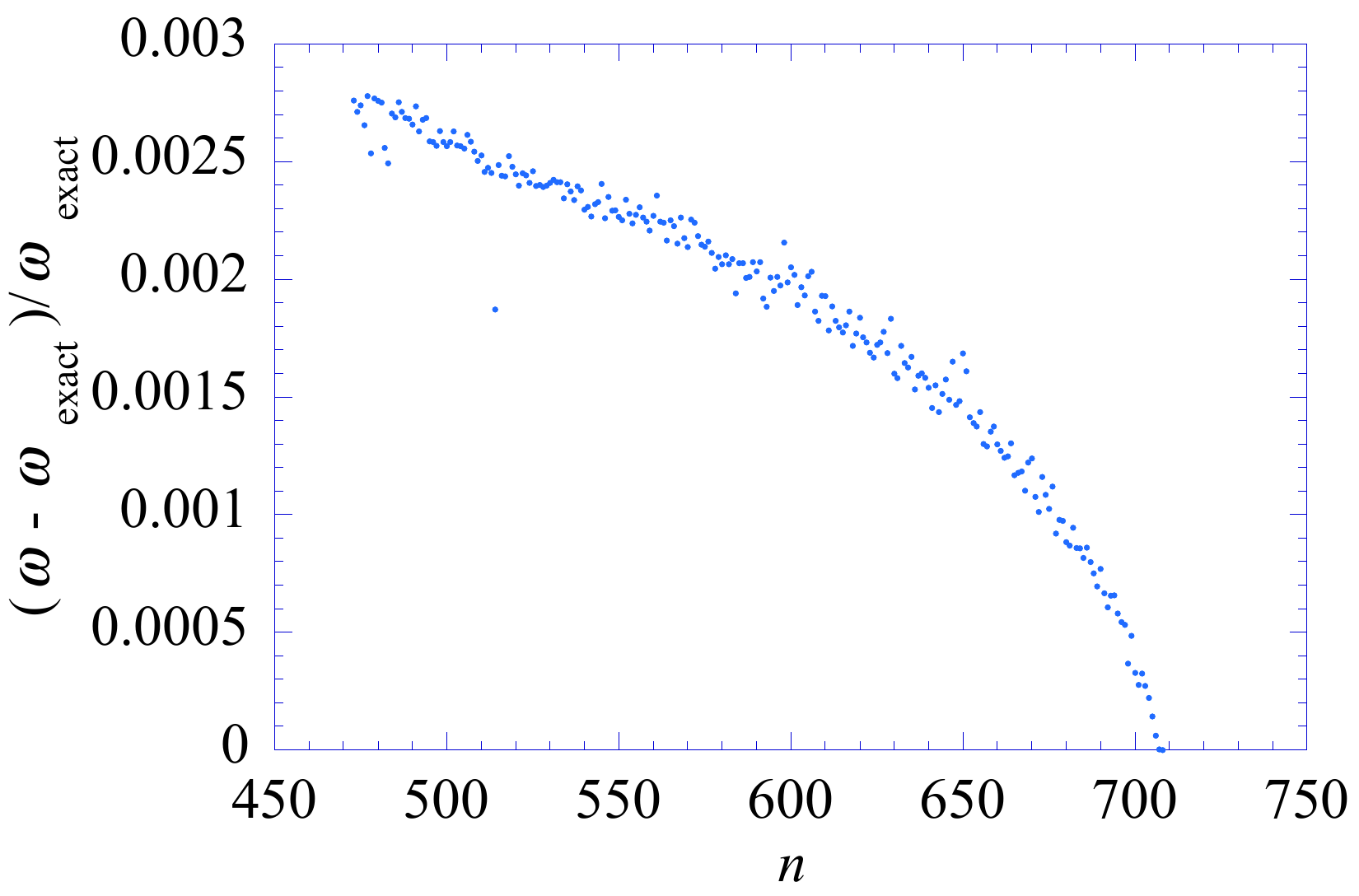}}

\caption{\label{fig6}  Fractional difference between $E\times B$ frequencies $\omega$ evaluated using Eq.~\eqref{a122}  and the exact frequencies $\omega_{exact}$ evaluated using Eq.~\eqref{a5} and displayed in Fig.~\ref{fig2.5}, for the same parameters as in that figure. }
  \end{figure}

The energy of an $E\times B$ mode can be written in terms of the eigenvectors ${\bf R}_{\perp\omega}$. Since the $E\times B$ modes form an orthogonal set, a general $E\times B$ displacement of the form $\sum_{\omega>0} a_\omega ({\bf R}_{\perp\omega},{\bf Z}^{slow}) + c.c. $, where $a_\omega$ is the amplitude of a given mode, produces a diagonalized potential energy $V$ given by
\begin{equation}
V = \sum_{\omega>0} |a_\omega|^2  {\bf R}_{\perp\omega}^*\cdot{\bf V}_\perp\cdot {\bf R}_{\perp\omega}.
\end{equation}
The kinetic energy contribution to $E\times B$ modes is negligible, scaling as $1/B^2$. For a cylindrically-symmetric trap potential, there is an additional contribution to the potential energy from a zero frequency mode. The analysis of this contribution follows the same procedure as was developed in Sec. IIb. The extra energy from this mode is  the potential energy change from radial compression of the plasma due to a change in the rotation rate. 

\section{Discussion}

In this paper we have detailed a  method of diagonalizing the Hamiltonian for a general linearized Hamiltonian system.  The method relies on the Hermitian properties of the dynamical matrix $\bf D$ and, for a stable system, requires only the evaluation of the eigenmodes of this matrix. For a neutrally-stable system, we found that the form of the Hamiltonian depends on whether or not  constants of the motion associated with neutral modes are in involution. The normal mode form of the Hamiltonian was also derived for an unstable system. 
	
	In applying this formalism to determine the normal  modes of a  magnetized 2-ion Coulomb cluster, we found that for  spherically-symmetric confinement the rotational inertia of the cluster is undefined, and we related this surprising result to the fact that constants of the motion associated with rotations of this system are not in involution.  In this case, a change in angular momentum produces a change in crystal orientation rather than a change in rotation frequency.
	
	Thermal fluctuations were also considered, and in particular fluctuations in axial position were evaluated in order to make contact with ongoing experiments which depend, in part,  on keeping these fluctuations below the level set by Lamb-Dicke confinement\cite{bollinger,bollinger2,Sawyer2014,Bohnet2016,Gilmore2017}. A somewhat surprising result of our analysis is the dominant contribution of low-frequency $E\times B$ modes to these axial fluctuations in 3D Coulomb crystals. 
	
This paper focussed on linear modes, but nonlinear interactions between modes is also important. One simple example of this coupling is the modulation of axial mode frequencies as ions slowly shift positions due to low-frequency $E\times B$ modes. This nonlinear behavior will be the subject of future investigations.\cite{Athreya}

The fact that modes separate into three disparate frequency groups in a strong magnetic field also has consequences for the energy equilibration in such a system.\cite{chen} When cooling and heating is applied, modes with disparate frequencies can come to quite different equilibrium energies, depending on the details of the driving.\cite{Tang} In particular, laser cooling of $E\times B$ modes may not be as efficient as for axial and cyclotron modes because the velocities associated with these modes are low, so their energies may not be well-equilibrated with the other mode branches. In 3D crystals this problem may be somewhat alleviated by the axial motion associated with some $E\times B$ modes, which can increase coupling between these modes and axial modes. These effects will be subjects of further study.

The general theory presented in Sec. II kept forcing terms $\bf f$ in the oscillator equations. This is because time-dependent forcing terms are often present in ion crystal experiments, although they were not considered in Sec. III. In some experiments oscillating fields are applied in order to purposely interact with and excite certain modes, such as the axial center of mass mode. Forcing can also be caused  by non-axisymmetric field errors in the external potential that are static in the laboratory frame, and by  non-axisymmetric forces from laser beams (also stationary in the lab frame). In the frame of the rotating crystal these forces oscillate at the rotation frequency $\omega_r$, which can resonate with normal modes.  For example, for the $\beta=3/4$ crystal of Fig. 2, when $\Omega=0$ (the Brillouin limit) the rotation rate is $\omega_r = \sqrt{5/4}\omega_z=1.118\omega_z$, which is within the spectrum of mode frequencies (see Fig. 3). It is therefore possible for  forcing that is static in the lab frame to resonantly excite normal modes in this crystal to large amplitude. On the other hand, when $\Omega=20\omega_z$, $\omega_r$ is either $20.0623\omega_z$ or $0.0623 \omega_z$ in the fast and slow rotation branches respectively, and both values lie just beyond the frequency spectrum limits shown in Fig. 4. Resonant interactions with modes will lead to nonlinear effects, plasma heating, and mode damping which are beyond the scope of this paper's linear analysis, and probably require a simulation approach.  The effect of interaction of modes with resonant external forcing will be further examined in future work. 

The data that support the findings of this study are available from the corresponding author upon reasonable request. The author acknowledges useful discussions with Dr. Matt Affolter, Prof. Dan Arovas,  and Prof. Scott Parker. This work is supported by AFOSR contract FA 9550-19-1-0999, DOE Grant No. DE-SC0018236 and NSF Grant No. PHY1805764.

\appendix

\section{The potential matrix}
In this appendix we work out the form of the $3\times 3$ potential matrix ${\bf V}_{j k }=\partial^2 \Phi/\partial{\bf R}_j\partial {\bf R}_k$ for charges $j$ and $k$ with equilibrium positions ${\bf R}_j$ and ${\bf R}_k$ respectively. The system potential energy $\Phi$ is given by Eq.~\eqref{a57n}, and substituting for this  we obtain
\begin{equation}
{\bf V}_{jk} = \delta_{j k}\frac{\partial^2}{\partial {\bf R}_j^2}\left( \phi_j(R_j,Z_j) +  \sum_{\tiny \begin{array}{c} l=1 \\  l\ne j \end{array}}^N\frac{q_j q_l} {|{\bf R}_j-{\bf R}_l|}\right)
+ (1-\delta_{j k})\frac{\partial^2}{\partial {\bf R}_j\partial {\bf R}_k}\frac{q_j q_k} {|{\bf R}_j-{\bf R}_k|}.
\end{equation}
Evaluating the derivatives yields the following dyadic form for the matrix:
\begin{align}
{\bf V}_{jk} =&\delta_{j k}\left( \hat r \hat r \frac{\partial^2}{\partial R_j^2}+ \hat z \hat z \frac{\partial^2}{\partial Z_j^2}+ (\hat r \hat z+ \hat z \hat r)\frac{\partial^2}{\partial R_j \partial Z_j}\right) \phi_j(R_j,Z_j)\notag \\
+ &\delta_{j k}\sum_ {\tiny \begin{array}{c} l=1 \\  l\ne j \end{array}}^N q_j q_l\left(\frac{3 ({\bf R}_j - {\bf R}_l)({\bf R}_j - {\bf R}_l)}{|{\bf R}_j-{\bf R}_l|^3}- \frac{\bf 1}{|{\bf R}_j-{\bf R}_l|^5}\right) \notag \\
-  &(1-\delta_{j k})q_j q_k\left(\frac{3 ({\bf R}_j - {\bf R}_k)({\bf R}_j - {\bf R}_k)}{|{\bf R}_j-{\bf R}_k|^3}- \frac{\bf 1}{|{\bf R}_j-{\bf R}_k|^5}\right),
\end{align}
where here ${\bf 1}$ is the $3\times 3$ unit matrix.

\section{Diagonalizing the energy using non-canonical variables} 
	In order to evaluate the normal modes of oscillation of a Coulomb crystal  we employed a  Hamiltonian approach using canonical variables in Sec. III. An alternate approach instead uses particle velocities rather than canonical momenta, but employs most of the same techniques as in the Hamiltonian method. In this Appendix we outline this non-canonical method. It is closer to the Lagrangian approach used in Ref.~\onlinecite{freericks}, and may be easier to apply provided that canonical coordinates are not required.

	The linearized equations of motion in the rotating frame of the crystal equilibrium, when written in terms of the perturbed velocities $\delta{\bf v}_j, j=1,...,N$, are
	
	\begin{align}
\delta \dot{\bf r}_i &= \delta {\bf v}_i, \label{f2} \\
m_i \delta \dot{\bf v}_i&=  -\sum_j {\bf V}_{i j}\cdot\delta{\bf r}_j - m_i\Omega_i \delta {\bf v}_i\times\hat z,\label{f3}
\end{align} 
	where theas before the $3\times3$ symmetric tensor ${\bf V}_{i j} = \nabla_i\nabla_j\Phi$.  

In order to solve for the linear normal modes of oscillation of this system, we combine Eqs.~\eqref{f2}and \eqref{f3} into a single vector equation for the $6N$ dimensional  configuration vector ${\bf \eta} = (\delta{\bf r}_1, ...,\delta{\bf r}_N,\delta {\bf v}_1,...,\delta{\bf v}_N)$,as was done in the Hamiltonian method. Then  the linearized equations of motion can be written, in analogy to Eq.~\eqref{a2},  as
\begin{equation}
\dot{\bf \eta} = {\bf D}'\cdot{\bf \eta},
\end{equation}
where the $6N\times6N$ dynamical matrix ${\bf D}'$ consists of four $3N\times3N$ blocks,
\begin{equation}\label{f4}
{\bf D}'=\bigg (
\begin{array}{c c}
 {\bf 0}&  {\bf 1} \\
-{\bf M}^{-1}\cdot {\bf V} & -2{\bf\Omega} 
\end{array}
\bigg ),
\end{equation}
where the Lorentz tensor $\bf \Omega$, the mass tensor $\bf M$, and the potential tensor is $\bf V$ are the same as in Eq.~\eqref{a119n1}.

The system energy $E$ is a conserved quantity, given by
\begin{equation}\label{f5}
E = \frac{1}{2} {\bf \eta}\cdot{\bf E}\cdot{\bf \eta},
\end{equation}
where the symmetric energy matrix $\bf E$ is
\begin{equation}\label{f6}
{\bf E} = \left(
\begin{array}{c c}
{\bf V} & {\bf 0} \\
{\bf 0} & {\bf M}
\end{array}
\right).
\end{equation}

Eigenmodes of the form ${\bf \eta}(t)= \exp(-i\omega t) {\bf \psi}_\omega$  satisfy the eigenvalue problem
\begin{equation}
-i\omega{\bf \psi }_\omega= {\bf D}'\cdot{\bf \psi}_\omega.
\end{equation} 
This eigenvalue problem is essentially identical to the secondary eigenvalue problem used to solve the quadratic eigenvalue problem in Ref.~\onlinecite{freericks}.

This modified non-canonical dynamical matrix still has the property that $i{\bf D}'$ is Hermitian,  with respect to a modified
inner product defined by $({\bf a},{\bf b})' = {\bf a}^*\cdot{\bf E}\cdot{\bf b}$ for vectors $\bf a$ and $\bf b$. This can be proven in an analogous manner to the proof for the Hamiltonian problem in Eqs.~\eqref{a7} and \eqref{a8}. 
Consider the matrix ${\bf L} = {\bf E}\cdot{\bf D}'$. One can show that this matrix is antisymmetric by direct calculation using Eqs.~\eqref{f4} and \eqref{f6}:
\begin{equation}
{\bf L} = 
 \left(
\begin{array}{c c}
{\bf 0} & {\bf V} \\
-{\bf V} & -2{\bf M}\cdot{\bf \Omega}
\end{array}
\right).
\end{equation}
The dot product of the diagonal matrix $\bf M$ and the antisymmetric block-diagonal matrix $\Omega$ is clearly antisymmetric, and therefore $\bf L$ is antisymmetric; also, it is real. This implies that 
${\bf a}^*\cdot{\bf E}\cdot{i \bf D}'\cdot{\bf b} = [{\bf b}^*\cdot{\bf E}\cdot{i \bf D}'\cdot{\bf a}]^* $,
so ${i \bf D}'$ is Hermitian. 
Therefore, the eigenmodes satisfy properties 1,2 and 3 of Sec. II. The eigenvectors form an orthogonal set with respect to the inner product $({\bf a},{\bf b})'$, and so on. 

The energy of the system is diagonalized by the eigenmodes. For simplicity we consider only the case where the system is stable with no zero-frequency modes. Then we may write a general phase space configuration $\bf \eta$ in terms of the complete set of orthogonal eigenvectors 
\begin{equation}
\eta = \sum_\omega a_\omega {\bf \psi}_\omega,
\end{equation}
where $a_\omega$ is the complex amplitude of mode $\omega$. As before, property 3 of Sec. II together with the real nature of $\eta$ implies that $a_{-\omega}=a_\omega^*$. 
Applying this to the energy in Eq.~\eqref{f5} and using orthogonality and property 3  then yields the diagonalized energy,
\begin{equation}
E = \sum_{\omega>0} |a_\omega|^2({\bf \psi}_\omega,{\bf \psi}_\omega)'.
\end{equation}

\section{ Rotational inertia for a Coulomb crystal}
In this appendix we evaluate the rotational inertia of a Coulomb crystal consisting of identical ions. We then specialize the result to a quadrupolar trap in which the ions are confined in the $x-y$ plane.  

In order to evaluate the rotational inertia $(\bar{\bf u}_{0 z},\bar{\bf u}_{0 z})$, we require a solution for  the vector $\bar{\bf u}_{0 z}$  of the equation
\begin{equation}\label{d1}
{\bf H}\cdot\bar{\bf u}_{0 z} = {\bf u}_{0 z}\cdot{\bf J},
\end{equation}
where ${\bf u}_{0 z}$ is given by Eq.~\eqref{eig0}, and  the Hamiltonian matrix $\bf H$ is given by Eq.~\eqref{a119n1}. Taking advantage of the block form of  $\bf H$, we write $\bar {\bf u}_{0 z} = (\bar {\bf r},\bar{\bf p})$, which when used in Eq.~\eqref{d1}  yields two coupled equations,
\begin{align}
({\bf V}+{\bf C})\cdot\bar {\bf r} + {\bf \Omega}\cdot\bar{\bf p} &= -{\bf p}_{0 z}=-\frac{1}{2}m\Omega {\bf R}_\bot \label{d2}\\
{\bf \Omega}^{tr}\cdot\bar{\bf r} + m^{-1}\bar {\bf p} &={\bf r}_{0 z}=\hat z\times{\bf R}, \label{d3}
\end{align}
where ${\bf R}_\bot$ is the projection of  the equilibrium positions ${\bf R}$ onto the $x-y$ plane, and where we have also imposed the assumption of a single species plasma for simplicity, but have not yet assumed a single-plane structure to the equilibrium.
Using Eqs.~\eqref{a120n1} and \eqref{a63} for $\bf \Omega$, and the definition of $\bf C$, we can write these equations as
\begin{align} 
{\bf V}\cdot\bar {\bf r}+\frac{1}{4}m \Omega^2\bar{\bf r}_\bot  - \frac{1}{2}\Omega\hat z\times \bar{\bf p} &= -\frac{1}{2}m\Omega {\bf R}_\bot  \label{d4}\\
\frac{1}{2} \Omega \hat z\times \bar{\bf r} + m^{-1}\bar {\bf p} &=\hat z\times{\bf R}. \label{d5}
\end{align}
Taking a cross-product of Eq.~\eqref{d5} then yields
\begin{equation}\label{d6}
\hat z\times \bar{\bf p} = -m{\bf R}_\bot + \frac{1}{2}m\Omega \bar{\bf r}_\bot.
\end{equation}
Applying this result to Eq.~\eqref{d4} implies
\begin{equation}\label{d7}
-{\bf V}\cdot\bar {\bf r} = m\Omega {\bf R}_\bot. 
\end{equation}
 The left hand side is the electrostatic force due to a displacement $\bar{\bf r}$ of the charges. The equation requires that this force must be purely radial. 
 Once a solution is obtained, then $\bar{\bf p}$ is determined by Eq.~\eqref{d5},
 \begin{equation}
 \bar{\bf p} = m \hat z\times {\bf R} -\frac{1}{2} m \Omega \hat z \times \bar{\bf r}.
 \end{equation}
 When these results are used to calculate the rotational inertia $\bar{\bf u}_{0 z}\cdot{\bf H}\cdot\bar{\bf u}_{0 z}$, the result is
 \begin{equation}\label{dinert}
 (\bar{\bf u}_{0 z},\bar{\bf u}_{0 z}) =  m {\bf R}_\bot\cdot{\bf R}_\bot + \bar{\bf r}\cdot{\bf V}\cdot\bar{\bf r}.
 \end{equation}
 The first term is the usual kinetic rotational inertia of a rigid body consisting of identical masses, and the second term is the extra inertia associated with potential energy from compression of the crystal.
 
In general  Eq.~\eqref{d7} must be solved numerically, but for a single plane equilibrium  in a quadrupole trap the solution is available analytically. In this case it is well-known that a purely radial perturbation in the position of each charge produces a radial restoring force, as required (this occurs in the radial breathing mode). \cite{dubinschiffer, freericks} According to Appendix B in Ref.~\onlinecite{dubinschiffer} the restoring force from a radial expansion is $-{\bf V}\cdot{\bf R}_\bot = -3m\omega_\bot^2{\bf R}_\bot$. Therefore, the solution of Eq.~\eqref{d7} is
 \begin{equation}\label{d8}
 \bar{\bf r} = -\frac{\Omega}{3\omega_\bot^2}  {\bf R}_\bot,
\end{equation}
and together with Eq.~\eqref{d5}
 this implies
 \begin{equation}\label{d9}
 \bar {\bf p} = m \hat z\times {\bf R}(1 +\frac{\Omega^2}{6\omega_\bot^2}).
\end{equation}
 These are the results quoted in Eq.~\eqref{a128nn}. The rotational inertia, Eq.~\eqref{a129nn}, follows from substitution of  Eq.~\eqref{d8} into Eq.~\eqref{dinert}, using Eq.~\eqref{d7}.

\section{ The Bosonic Bogoliubov method revisited}
In this appendix we review the Bogoliubov method and show that it is equivalent to the classical Hermitian method used in Sec. II, while describing a version of the Bogoliubov method that more closely follows the Hermitian method.

The Bogoliubov method for diagonalizing a linearized Hamiltonian is  couched in terms of ``creation and annihilation" pairs ${\bf \psi}=({\bf c},{\bf c}^*)$ rather than the phase space coordinates ${\bf z}=({\bf q},{\bf p})$. In terms of these pairs the Hamiltonian for a linearized system is
\begin{equation}\label{dd1}
H=\frac{1}{2}\psi^*\cdot{\mathscr H}\cdot\psi,
\end{equation}
where the matrix $\mathscr H$ is Hermitian, ${\mathscr H}^\dagger = \mathscr H$, and has the symmetric block form
\begin{equation}\label{dd2}
{\mathscr H} = \left(
\begin{array}{c c}
{\bf A}& {\bf B} \\
{\bf B}^* & {\bf A}^*
\end{array}
\right),
\end{equation} 
where the $N\times N$ matrix ${\bf A}={\bf A}^\dagger$ is Hermitian and the $N\times N$ matrix ${\bf B}={\bf B}^{tr}$ is symmetric. 
The creation and annihilation pairs are related to ${\bf z}$ via the linear transformation 
\begin{equation}\label{dd3}
\psi = {\bf T}\cdot {\bf z}
\end{equation}
 where
\begin{equation}
{\bf T}  = \frac{1}{\sqrt{2}}\left(
\begin{array}{c c}
{\bf 1} & i{\bf 1} \\
{\bf 1} & -i {\bf 1}
\end{array}
\right),
\end{equation}
or in component form ${\bf c} = ({\bf q}+ i {\bf p})/\sqrt{2}, {\bf c}^* = ({\bf q}- i {\bf p})/\sqrt{2}$. These creation/annihilation pairs have  Poisson bracket relations that may be succinctly expressed by the equation
\begin{equation}\label{dd5}
[\psi,\psi^*] = -i \bf\sigma, 
\end{equation}
where the matrix $\sigma$ is defined as
\begin {equation}
{\bf \sigma}  =\left(
\begin{array}{c c}

{\bf 1 } & {\bf 0} \\
{\bf 0} & - {\bf 1}
\end{array}
\right).
\end{equation}
Equation~\eqref{dd5} is equivalent to, and follows from,  the Poisson bracket relations $[{\bf z},{\bf z}] = {\bf J}$.  The relation between the Hamiltonian matrix $\bf H$ of Eq.~\eqref{a1} and the matrix $\mathscr H$ is found by substitution of Eq.~\eqref{dd3} into Eq.~\eqref{dd1}, yielding
 \begin{equation}
 {\bf H} = {\bf T}^\dagger\cdot{\mathscr H}\cdot{\bf T}.
 \end{equation}

The standard Bogoliubov approach is to find a transformation to new creation-annihilation pairs $\phi = ({\bf a},{\bf a}^*)$, 
\begin{equation}\label{dd8}
\psi = {\mathscr S}\cdot\phi
\end{equation}
for some matrix $\mathscr S$ such that the Hamiltonian is diagonalized:
\begin{equation}
H= \frac{1}{2}\phi^*\cdot{\mathscr K}\cdot\phi,
\end{equation}
with the new Hamiltonian matrix ${\mathscr K}$ given by
\begin{equation}\label{dd10}
{\mathscr K}= {\mathscr S}^\dagger\cdot{\mathscr H}\cdot{\mathscr S} = diagonal.
\end{equation}
The added requirement that the transformation be canonical requires $[\phi,\phi^*] = -i \bf\sigma$, which using Eqs.~\eqref{dd8} implies that ${\mathscr S}$ must satisfy the symplectic condition in the creation/annihilation representation, 
\begin{equation}\label{dd11}
{\mathscr S}\cdot{\bf\sigma}\cdot{\mathscr S}^\dagger=\bf\sigma.
\end{equation}
Equations~\eqref{dd10} and \eqref{dd11} are the  Bogoliubov equations for $\mathscr S$, whose solution provides the canonical transformation that diagonalizes the Hamiltonian. In what follows we solve for $\mathscr S$  using an approach similar to that used in Sec. II.

Consider the equation of motion for $\psi$ that follows from Hamiltonian \eqref{dd1}:
\begin{equation}\label{dd12}
\dot\psi = [\psi,H] = [\psi,\psi^*]\cdot{\mathscr H}\cdot\psi = -i {\bf \sigma}\cdot{\mathscr H}\cdot\psi =-i {\mathscr D}\cdot\psi,
\end{equation}
where we introduce the quantum dynamical matrix ${\mathscr D} \equiv {\bf \sigma}\cdot{\mathscr H}$.  This dynamical matrix is analogous to the matrix $\bf D$ introduced in Sec. II. Now consider eigenmodes  of the form $\psi = \exp(-i\omega t){\bf w}_\omega$ for a mode of frequency $\omega$, where ${\bf w}_\omega$ is the associated vector. 
Equation~\eqref{dd12} implies that these vectors are eigenvectors of $\mathscr D$ with eigenvalues $\omega$:
\begin{equation}\label{dd13}
{\mathscr D}\cdot{\bf w}_\omega = \omega {\bf w}_\omega.
\end{equation}
These eigenvectors are related to the eigenvectors ${\bf u}_\omega$ of the dynamical matrix $\bf D$ by Eq.~\eqref{dd3}, which implies that
\begin{equation}\label{dd13.5}
{\bf w}_\omega = {\bf T}\cdot{\bf u}_\omega.
\end{equation}
We can now prove three properties of these eigenmodes that are directly analogous to the three properties in Sec. II:
\begin{enumerate}
\item The eigenvectors ${\bf w}_\omega$ form an orthogonal set with respect to a generalized inner product  defined for any complex vectors $\bf a$ and $\bf b$ as $({\bf a},{\bf b})\equiv{\bf a}^*\cdot{\mathscr H}\cdot{\bf b}$:
\end{enumerate}
\begin{center}
$({\bf w}_\omega,{\bf w}_{\bar\omega})= 0 $ provided that $\omega\ne\bar\omega^*$. 
\end{center}
\begin{enumerate}
\setcounter{enumi}{1}
\item A given eigenvalue $\omega$ is real provided that the corresponding eigenvector satisfies $({\bf w}_\omega,{\bf w}_\omega)\ne 0$.
\item For each eigenmode $(\omega,{\bf w}_\omega)$ for which $\omega\ne 0$, there is a second eigenmode $(-\omega^*, {\bf w}_{-\omega^*})$ for which ${\bf w}_{-\omega^*}={\bf \Lambda}\cdot{\bf w}^*_\omega$, where the matrix ${\bf \Lambda}$ is
\begin{equation}
{\bf \Lambda} = \left(
\begin {array}{c c}
{\bf 0} & {\bf 1} \\
{\bf 1} &{\bf  0} 
\end{array}
\right).
\end{equation}
Thus, for real $\omega$ the $\omega\ne 0$ eigenmodes
come in  $\pm\omega$ pairs. 
\end{enumerate}

As  in Sec. II, the first two properties are a consequence of the spectral theorem for Hermitian matrices. Here, as before, the quantum dynamical matrix 
$\mathscr D$ is Hermitian with respect to the above inner product:

\begin{equation}
({\bf a}, {\mathscr D}\cdot{\bf b}) = ({\bf b}, {\mathscr D}\cdot{\bf a})^*.
\end{equation}

This requires that  the matrix
 ${\mathscr L}\equiv {\mathscr H}\cdot{\mathscr D}$ is a Hermitian matrix: ${\mathscr L}={\mathscr L}^\dagger$, which can be proven using the same set of steps as in Eq.~\eqref{a8}:
 
\begin{equation}\label{dd16}
{\mathscr L}_{j i} = {\mathscr H}_{j k}{\bf \sigma}_{k l} {\mathscr H}_{l i}={\mathscr H}^*_{k j} {\bf \sigma}_{l k} {\mathscr H}^*_{i l} =
 {\mathscr H}^*_{i l} {\bf \sigma}_{l k} {\mathscr H}_{k j} = {\mathscr L}^*_{i j}.
\end{equation}

The third property takes a bit more work than in Sec. II.  Here we use the following property of the quantum Hamiltonian matrix $\mathscr H$  that follows from its special form, Eq.~\eqref{dd2}:
\begin{equation}\label{dd17}
{\mathscr H} = {\bf \Lambda}\cdot{\mathscr H}^*\cdot{\bf \Lambda}.
\end{equation}
Acting on both sides of the equation with ${\bf \sigma}$ and using the identity ${\bf \sigma}\cdot{\bf \sigma}={\bf 1}$ gives
\begin{align}\label{dd18}
{\mathscr D} &= {\bf \sigma}\cdot{\bf \Lambda}\cdot{\bf \sigma}\cdot{\bf \sigma}\cdot{\mathscr H}^*\cdot{\bf \Lambda} \notag \\
&=-{\bf\Lambda}\cdot{\mathscr D}^*\cdot{\bf \Lambda},
\end{align}
where we used the identity ${\bf\sigma}\cdot{\bf\Lambda}\cdot{\bf\sigma}=-{\bf 1}$.
Substituting Eq.~\eqref{dd18} into Eq.~\eqref{dd13}, acting on both sides with $\bf\Lambda$ and using 
${\bf\Lambda}\cdot{\bf\Lambda}={\bf 1}$ then yields
\begin{equation}
{\mathscr D}^*\cdot{\bf\Lambda}\cdot{\bf w}_\omega = -\omega {\bf\Lambda}\cdot{\bf w}_\omega.
\end{equation}
The complex conjugate of this equation proves property 3. 

We can now diagonalize the Hamiltonian using these eigenvectors, proceeding as in Sec. II. We will assume for simplicity
 that $({\bf w}_\omega,{\bf w}_\omega)\ne 0$ for all eigenmodes, so that all eigenfrequencies are real and are also nonzero and non-degenerate. First, we write a general vector 
$\psi = ({\bf c},{\bf c}^*)$ in terms of  a linear combination of the eigenvectors ${\bf w}_\omega$:
\begin{equation}\label{dd20}
\psi(t) =\sum_\omega a_\omega(t){\bf w}_\omega.
\end{equation}

This is merely another way to express the Bogoliubov transformation Eq.~\eqref{dd8}, taking the components of the vector ${\bf\phi}(t)$ to be $a_\omega(t)$ and the transformation matrix $\mathscr S$ to have columns given by the eigenvectors:
\begin{equation}
{\mathscr S} = ({\bf w}_{\omega_1}, {\bf w}_{\omega_2}, ..., {\bf w}_{\omega_{2N}} ).
\end{equation}
Substituting Eq.~\eqref{dd20} into Eq.~\eqref{dd1} and using orthogonality of the eigenvectors (property 1) leads immediately to the diagonal form
\begin{equation}
H = \frac{1}{2} \sum_\omega a_\omega a_\omega^* ({\bf w}_\omega,{\bf w}_\omega).
\end{equation}

To make further progress we order the eigenfrequencies such that frequencies $\omega_1$ to $\omega_N$ are greater than zero, and the next set of $N$ frequencies are their paired opposites as per property 3. This implies that $\mathscr S$ takes the form
\begin{equation}\label{dd23}
{\mathscr S} = ({\bf w}_{\omega_1}, ...,{\bf w}_{\omega_N}, {\bf\Lambda}\cdot{\bf w}^*_{\omega_1},...,{\bf\Lambda}\cdot{\bf w}^*_{\omega_N})
\end{equation}
and that we can write Eq.~\eqref{dd20} as
\begin{equation}\label{dd24}
\psi(t) =\sum_{\omega>0}\left( a_\omega(t){\bf w}_\omega + a_{-\omega}(t){\bf\Lambda}\cdot{\bf w}^*_{\omega}\right).
\end{equation}
In order for this equation to match Eq.~\eqref{dd8} with ${\bf \phi } = ({\bf a},{\bf a}^*)$ this requires 
\begin{equation}
a_{-\omega} =  a_\omega^*,
\end{equation}
which implies that $\phi$ has the required form
\begin{equation}
\phi = (a_{\omega_1},...,a_{\omega_N}, a^*_{\omega_1},...,a^*_{\omega_N}).
\end{equation}
We can use this result to simplify the Hamiltonian, summing only over positive frequencies $\omega_1,...,\omega_N$:
\begin{equation}
H = \frac{1}{2} \sum_{\omega>0} a_\omega a_\omega^* \left\{({\bf w}_\omega,{\bf w}_\omega)+({\bf \Lambda}\cdot{\bf w}^*_\omega,{\bf \Lambda}\cdot{\bf w}^*_\omega)\right\}.
\end{equation}
This can be further simplified using the identity 
\begin{equation}\label{dd27.5}
({\bf \Lambda}\cdot{\bf w}^*_\omega,{\bf \Lambda}\cdot{\bf w}^*_\omega)=({\bf w}_\omega,{\bf w}_\omega),
\end{equation} 
which yields the simplified diagonalized Hamiltonian
\begin{equation}\label{dd28}
H = \sum_{\omega>0} a_\omega a_\omega^* ({\bf w}_\omega,{\bf w}_\omega).
\end{equation}
The identity can be proven with the aid of the complex conjugate of Eq.~\eqref{dd17}:
\begin{equation}
({\bf \Lambda}\cdot{\bf w}^*_\omega,{\bf \Lambda}\cdot{\bf w}^*_\omega)={\bf w}_\omega\cdot{\bf \Lambda}\cdot{\mathscr H}\cdot{\bf \Lambda}\cdot{\bf w}^*_\omega ={\bf w}_\omega\cdot{\mathscr H}^*\cdot{\bf w}^*_\omega={\bf w}^*_\omega\cdot{\mathscr H}\cdot{\bf w}_\omega
\end{equation}
where we used ${\bf\Lambda}^{tr} = \bf\Lambda$ in the first step and ${\mathscr H}^\dagger={\mathscr H}$ in the last step.

Next, we ensure that the transformation to the new $\phi$ variables is canonical by requiring that their Poisson brackets satisfy $[\phi,\phi^*]=-i\bf \sigma$. In component form this requires $[a_\omega, a^*_{\bar\omega}] = -i \delta_{\omega,\bar\omega}$, and $[a_\omega, a_{\bar\omega}] =0$  (for $\omega$ and $\bar\omega$ both greater than zero). We satisfy these equations in the same way as in Sec. II. Equation~\eqref{dd20} and orthogonality of the modes implies that
\begin{equation}
a_\omega(t) = \frac{({\bf w}_\omega,\psi(t))}{({\bf w}_\omega,{\bf w}_\omega)}.
\end{equation}
Applying this to $[a_\omega, a^*_{\bar\omega}]$ we use Eq.~\eqref{dd5} to obtain
\begin{align}
[a_\omega, a^*_{\bar\omega}] &= \frac{ {\bf w}^*_\omega\cdot{\mathscr H}\cdot{(-i \bf \sigma) } \cdot {\mathscr H}\cdot{\bf w}_{\bar\omega} }  { ( {\bf w}_\omega,{\bf w}_\omega ) ( {\bf w}_{\bar\omega},{\bf w}_{\bar\omega} ) } = \frac{ {\bf w}^*_\omega\cdot{\mathscr H}\cdot{(-i \bf \mathscr D})\cdot{\bf w}_{\bar\omega} }  { ( {\bf w}_\omega,{\bf w}_\omega ) ( {\bf w}_{\bar\omega},{\bf w}_{\bar\omega} ) } \notag\\
&= \frac{ {\bf w}^*_\omega\cdot{\mathscr H}\cdot{(-i\bar\omega \bf w}_{\bar\omega}) }  { ( {\bf w}_\omega,{\bf w}_\omega ) ( {\bf w}_{\bar\omega},{\bf w}_{\bar\omega} ) } \notag\\
&= \frac{ -i\omega\delta_{\omega,\bar\omega} }  { ( {\bf w}_\omega,{\bf w}_\omega ) },
\end{align}
where in the first step we used ${\mathscr H}^\dagger = {\mathscr H}$, in the third step we used Eq.~\eqref{dd13}, and in the last step we used orthogonality of the eigenmodes (property 1).
 A similar argument (see Eq.~\eqref{16.6} and \eqref{16.7}) shows that $[a_\omega, a_{\bar\omega}]=0$ when $\omega$ and $\bar\omega$ are greater than zero.
Thus, the transformation is canonical provided that we normalize the eigenvectors so that
\begin{equation}\label{dd32}
( {\bf w}_\omega,{\bf w}_\omega )  = \omega,\ \  \omega>0,
\end{equation}
a result analogous to Eq.~\eqref{a20}.
Applying this to Eq.~\eqref{dd28} yields the diagonalized Hamiltonian in canonical form,
\begin{equation}
H = \sum_{\omega>0} a_\omega a_\omega^* \omega.
\end{equation}

We can more directly connect this approach to the standard Bogoliubov approach by applying the vector ${\bf w}^*_{\bar\omega}\cdot\sigma$ to Eq.~\eqref{dd13}:
\begin{equation}
{\bf w}^*_{\bar\omega}\cdot\sigma\cdot{\mathscr D}\cdot{\bf w}_\omega = \omega {\bf w}^*_{\bar\omega}\cdot\sigma\cdot{\bf w}_\omega.
\end{equation}
Using $\sigma\cdot\sigma = {\bf 1}$ and ${\mathscr D}=\sigma\cdot\mathscr H$ then allows the left hand side to be written as an inner product:
\begin{equation}
({\bf w}_{\bar\omega},{\bf w}_\omega) = \omega {\bf w}^*_{\bar\omega}\cdot\sigma\cdot{\bf w}_\omega.
\end{equation}
Orthogonality of the eigenmodes then yields
\begin{equation}\label{dd35}
({\bf w}_{\omega},{\bf w}_\omega)\delta_{\omega,\bar\omega}=\omega{\bf w}^*_{\bar\omega}\cdot\sigma\cdot{\bf w}_\omega.
\end{equation}
For $\omega>0$ we can substitute for the inner product using Eq.~\eqref{dd32}, while for $\omega<0$ we can employ Eq.~\eqref{dd27.5} to see that $( {\bf w}_\omega,{\bf w}_\omega )  = -\omega,\ \ \omega<0$, which when used in Eq.~\eqref{dd35} yields
\begin{equation}
{\bf w}^*_{\bar\omega}\cdot\sigma\cdot{\bf w}_\omega={\text {sign}} (\omega) \delta_{\omega,\bar\omega}
\end{equation}
However, since the eigenvectors are columns of the transfomation matrix $\mathscr S$ (see Eq.~\eqref{dd23}), this equation is equivalent to 
\begin{equation}
{\mathscr S}^\dagger\cdot\sigma\cdot{\mathscr S}=\sigma.
\end{equation}
Taking the complex conjugate of this equation and using $\sigma^{tr} =\sigma$ yields the symplectic condition, Eq.~\eqref{dd11}. Thus, our transformation matrix $\mathscr S$, given by Eq.~\eqref{dd23} along with eigenvector normalizations \eqref{dd32},  is of the required symplectic form, and also diagonalizes the Hamiltonian as required by the Bogoliubov equations.

Finally, we can rederive the symplectic transformation $\bf S$ between $\bf z$ and $\bf Z$ (see Eq.~\eqref{28}) by employing Eq.~\eqref{dd3} along with ${\bf \phi} = {\bf T}\cdot{\bf Z}$ in Eq.~\eqref{dd8}, which implies
\begin{equation}\label{dd41}
{\bf S} = {\bf T}^{-1}\cdot{\mathscr S}\cdot{\bf T}.
\end{equation}
Since $\bf T$ is a unitary transformation, ${\bf T}^{-1} = {\bf T}^\dagger$ and Eqs.~\eqref{dd13.5} and \eqref{dd23} imply that ${\bf T}^{-1}\cdot{\mathscr S} = ({\bf U},{\bf U}^*)$, which when applied to Eq.~\eqref{dd41} leads back to Eq.~\eqref{29}, ${\bf S}= \sqrt{2}(\text{Re}{\bf U},-\text{Im}{\bf U})$.

\end{document}